\documentclass[twocolumn,twocolappendix,appendixfloats]{aastex63}
\usepackage{apjfonts}
\usepackage{amsmath}
\usepackage{scalerel}

\renewcommand*{\thefootnote}{\fnsymbol{footnote}}

\graphicspath{{./}{figures/}}

\DeclareSymbolFont{matha}{OML}{txmi}{m}{it}
\DeclareMathSymbol{\varv}{\mathord}{matha}{29}

\newcommand\ngc{NGC~3923 }%
\newcommand\msol{$M_{\odot}$}%
\newcommand\rsol{$R_{\odot}$}%
\newcommand{\hei}{\ion{He}{1}}%
\newcommand{\feii}{\ion{Fe}{2}}%
\newcommand{\feiii}{\ion{Fe}{3}}%

\newcommand{\fex}{\ion{Fe}{2}\protect\scaleto{$/III$}{1.4ex}}%

\newcommand{\coiii}{\ion{Co}{3}}%
\newcommand{\caii}{\ion{Ca}{2}}%
\newcommand{\niii}{\ion{Ni}{2}}%
\newcommand{\nai}{\ion{Na}{1}}%

\newcommand{\cii}{\ion{C}{2}}%
\newcommand{\oi}{\ion{O}{1}}%
\newcommand{\siii}{\ion{Si}{2}}%
\newcommand{\sii}{\ion{S}{2}}%
\def\simgt{\lower.5ex\hbox{$\; \buildrel > \over \sim \;$}}%
\def\simlt{\lower.5ex\hbox{$\; \buildrel < \over \sim \;$}}%
\def\tas{Type Ia SN}%
\def\tase{Type Ia SNe}%
\def\he1{${\rm{^{1}He}}$}%
\def\ni56{${\rm{^{56}Ni}}$}%
\def\co56{${\rm{^{56}Co}}$}%
\def\fer56{${\rm{^{56}Fe}}$}%
\def\fe52{${\rm{^{52}Fe}}$}%
\def\chr48{${\rm{^{48}Cr}}$}%
\def\dm15{$\Delta M_{15}(B)$}%
\def\mb{$M_{B}$}%
\def\tas{Type Ia SN}%
\def\tase{Type Ia SNe}%
\def\sn1a{SNe Ia}%

\def\uname{SN~2018aoz}
\def\sbv{$s_{BV}$}%
\def\t0{$t_{\rm 0}$}%
\def\chisq{$\chi^2$}%
\def\chisqr{$\chi^2_{\rm R}$}%
\def\d6s{D$^{\wedge}$6}%
\def\vi{\mbox{$V\!-\!i$}}%

\def\scl{1.2}%

\shorttitle{Infant Red Excess Type~Ia Supernova}
\shortauthors{Ni et al.}

\begin{document}

\title{The origin and evolution of the normal Type Ia SN 2018aoz with infant-phase reddening and excess emission}

\author[0000-0003-3656-5268]{Yuan Qi Ni}
\affiliation{David A. Dunlap Department of Astronomy and Astrophysics, University of Toronto, 50 St. George Street, Toronto, ON M5S 3H4, Canada}

\author[0000-0003-4200-5064]{Dae-Sik Moon}
\affiliation{David A. Dunlap Department of Astronomy and Astrophysics, University of Toronto, 50 St. George Street, Toronto, ON M5S 3H4, Canada}

\author[0000-0001-7081-0082]{Maria R. Drout}
\affiliation{David A. Dunlap Department of Astronomy and Astrophysics, University of Toronto, 50 St. George Street, Toronto, ON M5S 3H4, Canada}

\author[0000-0002-1633-6495]{Abigail Polin}
\affiliation{The Observatories of the Carnegie Institution for Science, 813 Santa Barbara Street, Pasadena, CA 91101, USA}
\affiliation{TAPIR, Walter Burke Institute for Theoretical Physics, Caltech, 1200 East California Boulevard, Pasadena, CA 91125, USA}

\author[0000-0003-4102-380X]{David J. Sand}
\affil{Department of Astronomy/Steward Observatory, University of Arizona, 933 North Cherry Avenue, Rm. N204, Tucson, AZ 85721-0065, USA}

\author[0000-0001-9541-0317]{Santiago Gonz\'alez-Gait\'an}
\affiliation{CENTRA, Instituto Superior T\'ecnico, Universidade de Lisboa, Avenida Rovisco Pais, 1049-001 Lisboa, Portugal}

\author[0000-0001-9670-1546]{Sang Chul Kim}
\affiliation{Korea Astronomy and Space Science Institute, 776 Daedeokdae-ro, Yuseong-gu, Daejeon 34055, Republic of Korea}
\affiliation{Korea University of Science and Technology (UST), 217 Gajeong-ro, Yuseong-gu, Daejeon 34113, Republic of Korea}

\author[0000-0002-6261-1531]{Youngdae Lee}
\affiliation{Department of Astronomy and Space Science, Chungnam National University, 99 Daehak-ro, Yuseong-gu, Daejeon 34134, Republic of Korea}
\affiliation{Korea Astronomy and Space Science Institute, 776 Daedeokdae-ro, Yuseong-gu, Daejeon 34055, Republic of Korea}

\author[0000-0002-3505-3036]{Hong Soo Park}
\affiliation{Korea Astronomy and Space Science Institute, 776 Daedeokdae-ro, Yuseong-gu, Daejeon 34055, Republic of Korea}
\affiliation{Korea University of Science and Technology (UST), 217 Gajeong-ro, Yuseong-gu, Daejeon 34113, Republic of Korea}
\affil{Department of Astronomy/Steward Observatory, University of Arizona, 933 North Cherry Avenue, Rm. N204, Tucson, AZ 85721-0065, USA}

\author[0000-0003-4253-656X]{D. Andrew Howell}
\affiliation{Department of Physics, University of California, Santa Barbara, CA 93106-9530, USA}
\affiliation{Las Cumbres Observatory, 6740 Cortona Dr, Suite 102, Goleta, CA 93117-5575, USA}

\author[0000-0002-3389-0586]{Peter E. Nugent}
\affiliation{Lawrence Berkeley National Laboratory, 1 Cyclotron Road, 1 Berkeley, CA 94720-8197, USA}
\affiliation{Departments of Physics and Astronomy, University of California, Berkeley, Berkeley, CA 94720 USA}

\author[0000-0001-6806-0673]{Anthony L. Piro}
\affiliation{The Observatories of the Carnegie Institution for Science, 813 Santa Barbara Street, Pasadena, CA 91101, USA}

\author[0000-0001-6272-5507]{Peter J. Brown}
\affiliation{Department of Physics and Astronomy, Texas A\&M University, 4242 TAMU, College Station, TX 77843-4242, USA}
\affiliation{George P. and Cynthia Woods Mitchell Institute for Fundamental Physics \& Astronomy, College Station, TX 77843, USA}

\author[0000-0002-1296-6887]{Llu\'is Galbany}
\affiliation{Institute of Space Sciences (ICE, CSIC), Campus UAB, Carrer de Can Magrans, s/n, E-08193 Barcelona, Spain}
\affiliation{Institut d’Estudis Espacials de Catalunya (IEEC), E-08034 Barcelona, Spain}

\author[0000-0003-0035-6659]{Jamison Burke}
\affiliation{Las Cumbres Observatory, 6740 Cortona Dr, Suite 102, Goleta, CA 93117-5575, USA}
\affiliation{Department of Physics, University of California, Santa Barbara, CA 93106-9530, USA}

\author[0000-0002-1125-9187]{Daichi Hiramatsu}
\affiliation{Las Cumbres Observatory, 6740 Cortona Dr, Suite 102, Goleta, CA 93117-5575, USA}
\affiliation{Department of Physics, University of California, Santa Barbara, CA 93106-9530, USA}
\affiliation{Center for Astrophysics \textbar{} Harvard \& Smithsonian, 60 Garden Street, Cambridge, MA 02138-1516, USA}
\affiliation{The NSF AI Institute for Artificial Intelligence and Fundamental Interactions}

\author[0000-0002-0832-2974]{Griffin Hosseinzadeh}
\affil{Department of Astronomy/Steward Observatory, University of Arizona, 933 North Cherry Avenue, Rm. N204, Tucson, AZ 85721-0065, USA}

\author[0000-0001-8818-0795]{Stefano Valenti}
\affiliation{Department of Physics and Astronomy, University of California Davis, 1 Shields Avenue, Davis, CA 95616-5270, USA}

\author[0000-0002-1338-490X]{Niloufar Afsariardchi}
\affiliation{David A. Dunlap Department of Astronomy and Astrophysics, University of Toronto, 50 St. George Street, Toronto, ON M5S 3H4, Canada}

\author[0000-0003-0123-0062]{Jennifer E. Andrews}
\affil{Department of Astronomy/Steward Observatory, University of Arizona, 933 North Cherry Avenue, Rm. N204, Tucson, AZ 85721-0065, USA}

\author[0000-0003-4453-3776]{John Antoniadis}
\affiliation{Institute of Astrophysics, FORTH, Dept. of Physics, University of Crete, Voutes, University Campus, GR-71003, Heraklion, Greece}
\affiliation{Max-Planck Institut für Radioastronomie, Auf dem Hügel 69, 53121, Bonn, DE}
\affiliation{Argelander Institut für Astronomie, Auf dem Hügel 71, 53121, Bonn, DE}

\author[0000-0002-1691-8217]{Rachael L. Beaton}
\altaffiliation{Hubble Fellow}
\altaffiliation{Carnegie-Princeton Fellow}
\affiliation{Department of Astrophysical Sciences, Princeton University, 4 Ivy Lane, Princeton, NJ~08544, USA}
\affiliation{The Observatories of the Carnegie Institution for Science, 813 Santa Barbara Street, Pasadena, CA 91101, USA}

\author[0000-0002-4924-444X]{K. Azalee Bostroem}
\affiliation{DIRAC Institute, Department of Astronomy, University of Washington, 3910 15th Avenue NE, Seattle, WA 98195, USA}

\author[0000-0002-7667-0081]{Raymond G. Carlberg}
\affiliation{David A. Dunlap Department of Astronomy and Astrophysics, University of Toronto, 50 St. George Street, Toronto, ON M5S 3H4, Canada}

\author[0000-0003-1673-970X]{S. Bradley Cenko}
\affiliation{Astrophysics Science Division, NASA Goddard Space Flight Center, MC 661, Greenbelt, MD 20771, USA}
\affiliation{Joint Space-Science Institute, University of Maryland, College Park, MD 20742, USA}

\author[0000-0002-7511-2950]{Sang-Mok Cha}
\affiliation{Korea Astronomy and Space Science Institute, 776 Daedeokdae-ro, Yuseong-gu, Daejeon 34055, Republic of Korea}
\affiliation{School of Space Research, Kyunghee University, 1732 Deogyeong-daero, Giheung-gu, Yongin-si, Gyeonggi-do, 17104, Republic of Korea}

\author[0000-0002-7937-6371]{Yize Dong}
\affiliation{Department of Physics and Astronomy, University of California Davis, 1 Shields Avenue, Davis, CA 95616-5270, USA}

\author[0000-0002-3653-5598]{Avishay Gal-Yam}
\affiliation{Department of Particle Physics and Astrophysics, Weizmann Institute of Science, 234 Herzl St., Rehovot, 76100, Israel}

\author[0000-0002-6703-805X]{Joshua Haislip}
\affiliation{Department of Physics and Astronomy, University of North Carolina at Chapel Hill, 120 E. Cameron Ave, Chapel Hill, NC 27599-3255, USA}

\author[0000-0001-9206-3460]{Thomas W.-S. Holoien}
\affiliation{The Observatories of the Carnegie Institution for Science, 813 Santa Barbara Street, Pasadena, CA 91101, USA}

\author[0000-0001-9487-8583]{Sean D. Johnson}
\affiliation{Department of Astrophysical Sciences, Princeton University, 4 Ivy Lane, Princeton, NJ~08544, USA}

\author[0000-0003-3642-5484]{Vladimir Kouprianov}
\affiliation{Department of Physics and Astronomy, University of North Carolina at Chapel Hill, 120 E. Cameron Ave, Chapel Hill, NC 27599-3255, USA}
\affiliation{Central (Pulkovo) Observatory of Russian Academy of Sciences, 196140 Pulkovskoye Ave. 65/1, Saint Petersburg, Russia}

\author[0000-0001-7594-8072]{Yongseok Lee}
\affiliation{Korea Astronomy and Space Science Institute, 776 Daedeokdae-ro, Yuseong-gu, Daejeon 34055, Republic of Korea}
\affiliation{School of Space Research, Kyunghee University, 1732 Deogyeong-daero, Giheung-gu, Yongin-si, Gyeonggi-do, 17104, Republic of Korea}

\author[0000-0001-9732-2281]{Christopher D. Matzner}
\affiliation{David A. Dunlap Department of Astronomy and Astrophysics, University of Toronto, 50 St. George Street, Toronto, ON M5S 3H4, Canada}

\author[0000-0003-2535-3091]{Nidia Morrell}
\affiliation{Las Campanas Observatory, Carnegie Observatories, Casilla 601, La Serena, Chile}

\author[0000-0001-5807-7893]{Curtis McCully}
\affiliation{Las Cumbres Observatory, 6740 Cortona Dr, Suite 102, Goleta, CA 93117-5575, USA}
\affiliation{Department of Physics, University of California, Santa Barbara, CA 93106-9530, USA}

\author[0000-0003-0006-0188]{Giuliano Pignata}
\affiliation{Departamento de Ciencias Fisicas, Universidad Andres Bello, Avda. Republica 252, Santiago, 8370134, Chile}
\affiliation{Millennium Institute of Astrophysics (MAS), Nuncio Monse\~{n}or Sotero Sanz 100, Providencia, Santiago, Chile}

\author[0000-0002-5060-3673]{Daniel E. Reichart}
\affiliation{Department of Physics and Astronomy, University of North Carolina at Chapel Hill, 120 E. Cameron Ave, Chapel Hill, NC 27599-3255, USA}

\author[0000-0002-5807-5078]{Jeffrey Rich}
\affiliation{The Observatories of the Carnegie Institution for Science, 813 Santa Barbara Street, Pasadena, CA 91101, USA}

\author[0000-0003-4501-8100]{Stuart D. Ryder}
\affiliation{School of Mathematical and Physical Sciences, Macquarie University, 105 Delhi Rd, North Ryde, NSW 2109, Australia}
\affiliation{Macquarie University Research Centre for Astronomy, Astrophysics \& Astrophotonics, Sydney, NSW 2109, Australia}

\author[0000-0001-5510-2424]{Nathan Smith}
\affil{Department of Astronomy/Steward Observatory, University of Arizona, 933 North Cherry Avenue, Rm. N204, Tucson, AZ 85721-0065, USA}

\author[0000-0003-2732-4956]{Samuel Wyatt}
\affil{Department of Astronomy/Steward Observatory, University of Arizona, 933 North Cherry Avenue, Rm. N204, Tucson, AZ 85721-0065, USA}

\author[0000-0002-2898-6532]{Sheng Yang}
\affiliation{The Oskar Klein Centre, Department of Astronomy, Stockholm University, AlbaNova, SE-10691 Stockholm, Sweden}

\correspondingauthor{Yuan Qi Ni}
\email{chris.ni@mail.utoronto.ca}

\begin{abstract}
\object{SN~2018aoz} is a Type Ia SN
with a $B$-band plateau and excess emission in the infant-phase light curves $\lesssim$ 1 day after first light, evidencing an over-density of surface iron-peak elements as shown in our previous study.
Here, we advance the constraints on the nature and origin of \uname\ based on its evolution until the nebular phase. 
Near-peak spectroscopic features show the SN is intermediate between two subtypes of normal Type Ia: Core-Normal and Broad-Line.
The excess emission could have contributions from the radioactive decay of surface iron-peak elements as well as
ejecta interaction with either the binary companion or a small torus of circumstellar material.
Nebular-phase limits on H$\alpha$ and \hei\ favour 
a white dwarf companion, consistent with the small companion size constrained by the low early SN luminosity, while the absence of [\oi] and \hei\ disfavours a violent merger of the progenitor.
Of the two main explosion mechanisms
proposed to explain the distribution of surface iron-peak elements in
\uname, the asymmetric Chandrasekhar-mass explosion is 
less consistent with the progenitor constraints
and
the observed blueshifts of nebular-phase [\feii] and [\niii].
The helium-shell double-detonation explosion is compatible with the observed lack of C spectral features, but current 1-D models are incompatible with the infant-phase excess emission, $B_{\rm max}-V_{\rm max}$ color, and weak strength of nebular-phase [\caii].
Although the explosion processes of \uname\ still need to be more precisely understood, the same processes could produce a significant fraction of \tase\ that appear normal after $\sim$ 1 day.
\end{abstract}


\section{Introduction}\label{sec:intro}

\tase\ are thermonuclear explosions of carbon and oxygen white dwarfs \citep[C+O WDs;][]{Nugent2011nat}.
They are the main source of iron-peak elements in the universe and crucial for measuring extragalactic distances, leading to the discovery of the accelerated cosmological expansion and dark energy \citep{Riess1998aj, perlmutter1999apj}. Despite their fundamental importance, the explosion mechanisms and progenitor systems of \tase\ remain a matter of extensive debate \citep{Maoz2014araa}. Understanding the origins of \tase, particularly of the ``normal'' events comprising $\sim$ 70\% of their population \citep{Blondin2012aj}, will not only clarify the endstates of stellar evolution but will be essential for improving cosmological distance measurements \citep[e.g.,][]{Wang2013sci, Zhang2021mnras}.

There is a broad consensus that \tase\ explode as a result of mass transfer in binary progenitor systems. 
However, uncertainty remains about whether the binary companion involved in normal \tas\ explosions is an evolved non-degenerate star \citep[``single-degenerate scenario'';][]{Whelan&Iben1973apj} or another WD \citep[``double-degenerate scenario'';][]{Iben&Tutukov1984apjs}.
In the latter case, it is unclear whether the explosion would be triggered during WD-WD accretion \citep{Guillochon2010apj, Pakmor2013apj}, or in a complete merger \citep{Pakmor2012apj} or head-on collision of the two WDs \citep{Kushnir2013apj}. 
The ``core-degenerate scenario’’ is a third hypothesis where \tase\ result from mergers of WDs with the cores of asymptotic giant branch stars \citep{Aznar2015mnras}.

The mechanisms responsible for triggering normal \tas\ explosions are also unclear.
Normal \tase\ have long been theorized to be ignited by nuclear burning in the core of a WD when accretion or merger causes its mass to reach the critical Chandrasekhar limit \citep[$\sim$ 1.4 \msol;][]{Mazzali2007sci}.
Alternatively, recent theoretical studies have suggested that the detonation of a thin helium layer on the surface of a sub-Chandrasekhar-mass WD can subsequently ignite carbon in the core, producing normal \tase\ via a ``helium-shell double-detonation'' \citep[He-shell DDet;][]{Polin2019apj, Townsley2019apj, Shen2021apjl, Magee2021mnras}.
One scenario that has been thought to result in a He-shell DDet is the detonation of He-rich material on the WD surface during a double-degenerate accretion process, called ``dynamically-driven double-degenerate double-detonation'' (or \d6s), recently supported by the identification of hyper-velocity Galactic WDs interpreted to be survivors of the scenario \citep{Shen2018apj, Bauer2021apj}.

Multiple explosion and progenitor channels may ultimately contribute to the observed population of \tase. 
In particular, the normal events consist
of two spectroscopically distinct subtypes \citep{Parrent2014apss}: ``Core-Normal/Normal-Velocity'' (CN/NV); and ``Broad-Line/High-Velocity'' (BL/HV).
Events from the two subtypes are nearly indistinguishable in their light curves, with similar peak brightness and decline rate, but differ in their observed spectroscopic features \citep{Branch2006pasp} and ejecta velocities \citep{Wang2009apj}.
Different explosion mechanisms---such as 
Chandrasekhar- and sub-Chandrasekhar-mass 
explosions \citep[e.g.,][]{Polin2019apj, Li2021apj}---have been suggested
to explain the differences between
the two subtypes.
Alternatively, unified origins for the observed spectroscopic diversity in normal events have also been proposed, usually involving an asymmetric explosion mechanism \citep[e.g.,][]{Maeda2010natur}.

Early (e.g., $\lesssim$ 5 days post-explosion) light curves of \tase\ can shed light on their origins by providing critical constraints on the binary companion, circumstellar material (CSM) from accretion or merger, and the distribution of elements in the outer ejecta.
Theoretical models have predicted that the collision between the SN ejecta and a binary companion \citep{Kasen2010apj} or CSM \citep{Piro&Morozova2016apj} can shock heat the ejecta, producing blue excess emission.
Multiple explosion processes, including sub-sonic mixing \citep{Reinecke2002aa} and detonation of surface helium \citep{Polin2019apj, Maeda2018apj, Magee2021mnras}, have also been predicted to lead to over-densities of radioactive iron-peak (Fe-peak) elements, including \ni56, \fe52, and \chr48,
in the shallow layers of the ejecta, leading to excess emission and short-lived color evolution associated with Fe spectroscopic features.
Such color and light curve features within $\sim$ 5 days have been reported in many \tase\ \citep{Jiang2017nat, De2019apj, Hosseinzadeh2017apj, Marion2016apj, Miller2018apj, Li2019apj, Shappee2019apj, Dimitriadis2019apj, Bulla2020apj, Jiang2018apj, Stritzinger2018apj, Miller2020yvq, Tucker2021yvq, Ni2022natas, Deckers2022mnras, Hosseinzadeh2022arxiv}, though there have been recent debates about their interpretation in some normal events \citep[e.g.,][]{Sand2018apj, Shappee2018apj, Ashall2022arxiv}.
However, for the vast majority of \tase\ observed between 1 and 5 days, their light curves match simple power-law profiles in this phase \citep{Bloom2012apj, Foley2012apj, Olling2015nat, Cartier2017mnras, Holmbo2019aa, Yao2019apj, Moon2021apj}.
Such power-law evolution is consistent with an origin that both (1) has a small non-degenerate or WD companion and (2) leads to a \ni56\ distribution in the ejecta that is largely centrally concentrated and
monotonically declining towards the surface.

Another way to critically constrain the explosion mechanism and progenitor system is
to investigate spectral features of \tase\ from 
the so-called ``nebular phase" of
$\gtrsim$ 200 days since $B$-band maximum.
Differences in the Doppler shifts of [\feii] and [\niii] emission lines observed in normal \tase\ have been attributed to the viewing angle effects of asymmetric explosion mechanisms \citep{Maeda2010apj, Maeda2010natur, Maguire2018mnras, Li2021apj}.
Meanwhile, strong [\caii] emission has been associated with incomplete nuclear burning in the core of sub-Chandrasekhar-mass explosions \citep{Polin2021apj, Siebert2020apj}.
For the progenitor, the presence of H$\alpha$ and \hei\ emission by stripped/ablated H and He from a non-degenerate companion
has been predicted by several recent studies as evidence supporting single-degeneracy \citep{Mattila2005aap, Botyanszki2018apj, Dessart2020aa}.
Such H$\alpha$ emission has been observed in the nebular-phase spectra of a few peculiar events \citep[e.g.,][]{Kollmeier2019mnras},
indicating that they may be from single-degenerate progenitors. 
However, systematic searched for H and He emission in the nebular-phase spectra of $>$ 100 \tase\ have failed to find such emission in most of them ($\gtrsim$ 90\%),
disfavouring the single-degenerate scenario as the primary contributor to the \tas\ population \citep{Mattila2005aap, Leonard2007apj, Shappee2013apj, Maguire2016mnras, Tucker2020mnras}.
However, a systematic search for H and He emission in the nebular-phase spectra of 110 \tase\ has failed to find such emission in most of them ($\gtrsim$ 90\%),
disfavouring the single-degenerate scenario as the primary contributor to the \tas\ population \citep{Tucker2020mnras}.
[\oi] emission has also been detected in the nebular-phase spectra 
of two peculiar events and interpreted to be evidence for 
the presence of swept-up unburned O from a double-degenerate merger \citep{Kromer2013apj, Taubenberger2013apj}. 
The identification of such [\oi] emission 
has yet to be made for normal events.

\uname\ is a recent normal \tas\ detected 1.0 hours after its estimated epoch of first light\footnote{First light refers to the epoch when photons first emerge from the ejecta, which may follow the explosion by a few-hours to days in \tase\ depending on the photon diffusion process \citep{Piro&Nakar2013apj, Piro&Nakar2014apj}. In \uname, the epoch of explosion is estimated to be MJD 58205.6 $\pm$ 0.7 based on the observed evolution of photospheric velocity (Paper I).} (MJD~58206.00), the earliest detection for a \tas\ ever made so far \citep[][Paper I hereafter]{Ni2022natas}.
Photometric and spectroscopic observations were obtained 
over the ensuing period of $\sim$ 450 days, 
including light curves of the first 12 hours from the very low brightness of $-$10.5 absolute AB magnitude.
This data set provides the unique opportunity to study 
the entire evolution of a normal \tas\ 
from 1 hour after first light to the nebular phase.
In Paper I, we presented the discovery of two new infant-phase features 
of \tas\ evolution during the first 1.0--12.4 hours: 
a brief $B$-band plateau---which disappears after $\sim$ 0.5 days---and
simultaneous excess emission in the $V$ and $i$ bands.
The subsequent evolution of \uname\ until $\sim$ 110 days is consistent with that of typical normal \tase, with a power-law light curve rise, peak $B$-band absolute magnitude of $-$19.32 mag and \dm15\ of 1.12 mag.
The two infant-phase features result in a rapid reddening of the \bv\ color, which has been associated with line-blanket absorption by an over-density of Fe-peak elements in the outer 1\% of the SN-ejected mass (Paper I). This has important implications for the normal \tas\ explosion mechanism, as such an ejecta composition is primarily predicted by asymmetric Chandrasekhar-mass explosions and He-shell DDets.

Although \uname\ has provided critical information 
on the distribution of surface Fe-peak elements, its evolution to the nebular phase has yet to be explored and additional insights into its origin can be gained by (1) placing constraints on the nature of its companion star (2) examining the physical implications of a range of possible power sources for the infant phase excess emission and (3) assessing its precise subtype among normal \tase.
In this paper, we present new photometric and spectroscopic
observations of the nebular phase of \uname\ in Section~\ref{sec:obs}, as well as detailed modelling
and interpretation of key features to understand its origin
and evolution as follows.
In Section~\ref{sec:lc}, we describe the evolution of the light curves and spectra of \uname, including comparisons of them to those of other \tase\ in order to establish its spectroscopic subtype.
We assess the range of companion stars that are compatible with the luminosity of the observed early light curve in Section~\ref{sec:kasan}.
Sections~\ref{sec:early}, \ref{sec:nebea}, and \ref{sec:hedd} describe our modelling of the infant-phase excess emission, analyses of the nebular-phase observations, 
and comparisons to the predictions of He-shell DDet simulations, respectively.
In Section~\ref{sec:orig}, 
we discuss the implications of our results for the progenitor system and explosion mechanism of \uname, the nature of its infant-phase excess emission, and the origins of normal \tase.
We summarize our results and conclude in Section~\ref{sec:conc}.

\section{Observations and Data} \label{sec:obs}

\uname\ was identified by both the KMTNet Supernova Program \citep[KSP;][]{Moon2016spie, Afsariardchi2019apj, Moon2021apj, Lee2022apj} and Distance Less Than 40 Mpc Survey \citep[DLT40;][]{Tartaglia2018}.
The earliest detection of the SN with signal-to-noise (S/N) $>$ 3 was made by KSP in the $B$ band at 00h54m on 29 March 2018 Universal Time (UT), or MJD 58206.0378.
DLT40 detected the source 1.1 days later in the $r$ band and reported the discovery of \uname\ at 07h25m on 2 April 2018 UT \citep{Sand2018atel}. 
The first spectrum obtained by the Las Cumbres Observatory \citep{Brown2013pasp} at 09h25m on 2 April 2018 UT subsequently classified the source as a \tas\ \citep{Hosseinzadeh2018tnscr}. 
The discovery triggered an extensive campaign of ground- and space-based photometric observations as well as spectroscopic follow-up, obtaining observations in UV to NIR wavebands.
The early observations of \uname\ obtained until $\sim$ 110 days since first light were presented in Paper I.
Here, we present additional KSP photometry continuing from $>$ 250 days since first light, covering the nebular phase (Section~\ref{sec:neblc}), as well as new nebular-phase spectroscopy of the SN (Section~\ref{sec:nebspec}).

\begin{deluxetable*}{cccc}
\tabletypesize{\footnotesize}
\tablecolumns{4} 
\tablewidth{0.99\textwidth}
 \tablecaption{Nebular-phase magnitudes of \uname}
 \tablehead{
 \colhead{Time [MJD]} & \colhead{Band} & \colhead{Magnitude$\rm ^a$ [mag]} & \colhead{Error [mag]}
 } 
\startdata 
58471.68652  & $B$ &    19.195  &       0.065  \\
58471.68830  & $B$ &    19.267  &       0.061  \\
58471.68979  & $V$ &    19.562  &       0.128  \\
58471.69130  & $i$ &    19.430  &       0.177  \\
58472.68728  & $B$ &    19.302  &       0.074  \\
58472.68862  & $V$ &    19.374  &       0.092  \\
58472.69015  & $i$ &    19.585  &       0.285  \\
58473.68083  & $B$ &    19.346  &       0.082  \\
58473.68233  & $V$ &    19.554  &       0.096  \\
58473.68382  & $i$ &    19.420  &       0.177  \\
\enddata
\tablenotetext{{\rm a}}{The $BV$-band magnitudes are in the Vega system,
while the $i$-band magnitudes are in the AB system (see text).}
\tablecomments{Sample of the observed magnitudes of \uname\ during its nebular phase.
The entire observed magnitudes of \uname\ are available in the electronic edition.} 
\end{deluxetable*} 
\label{tab:neblc}

\subsection{Nebular-Phase Photometry} \label{sec:neblc}

We used the three 1.6m telescopes of the Korea Microlensing Telescope Network \citep[KMTNet;][]{Kim2016jkas} in Chile, South Africa, and Australia to conduct photometric observations of \uname\ during its nebular phase, $>$ 200 days since $B$-band maximum.
Each telescope of the network is equipped with an identical wide-field CCD camera with 4 square degree field-of-view and multiple filters in the visible band.
Between 2018 December and 2019 June, we conducted high-cadence monitoring of a 2\degr$\times$ 2\degr\ field containing the source, obtaining $\sim$ 500 images of the field with 60-s exposure times at a mean cadence of $\sim$ 9 hours in each of the $BVI$ bands.
The $B$, $V$, and $I$ bands are observed nearly simultaneously at each epoch with a time difference of $\sim$ 2 minutes between adjacent filters.
The typical limiting magnitude for a point source in these images is 21$-$22 mag at a S/N of 3.
Note that the source was not observed between July and November due to its proximity to the Sun. 

\begin{figure}[t!]
\epsscale{\scl}
\plotone{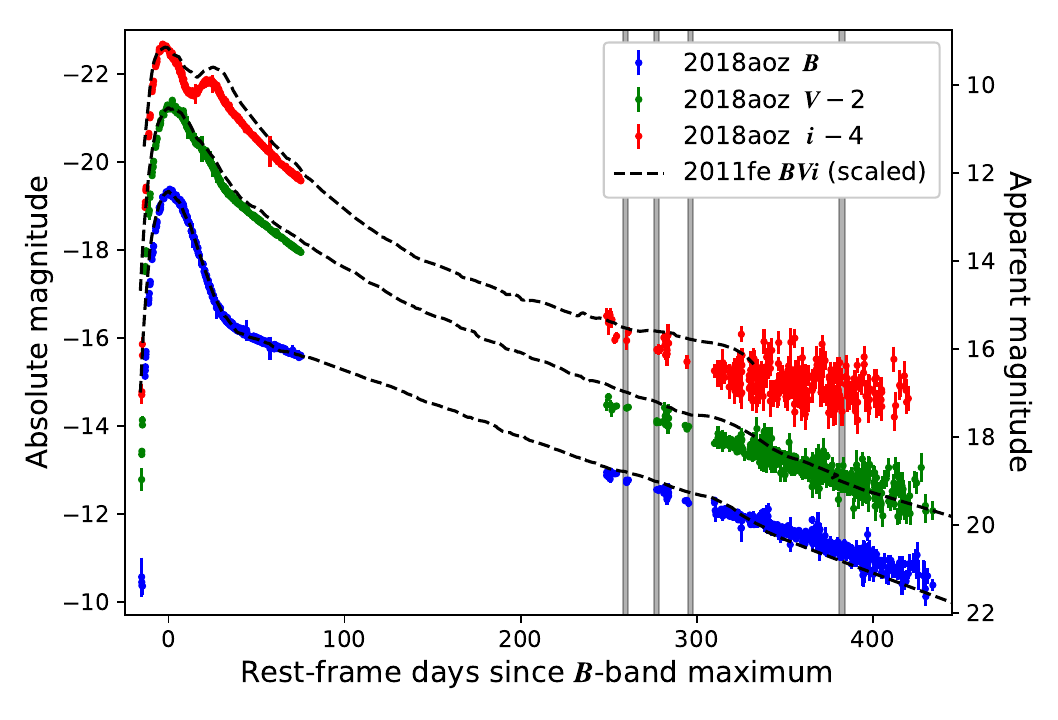}
\caption{The dereddened $BVi$-band light curves of \uname\ (colored circles) relative to the epoch of $B$-band maximum light in rest frame covering its nebular phase are compared to those of SN~2011fe \citep[dashed lines;][]{Munari2013newa,Tsvetkov2013coska} that have been scaled so that they match the \mb\ and \dm15\ values of \uname.
The errorbars represent the 1-$\sigma$ uncertainty level in this figure and all of the following. The vertical grey lines mark the four epochs with nebular-phase spectroscopy (see Table~\ref{tab:nebspec} and Figure~\ref{fig:nebspec}).
\label{fig:neblc}}
\end{figure}

We performed point-spread function (PSF) photometry of \uname\ using the SuperNova Analysis Package (SNAP\footnote{\url{https://github.com/niyuanqi/SNAP}}), a custom python-based pipeline for supernova photometry and analysis.
A local PSF was obtained by fitting a Moffat function \citep{Moffat1969aap, Trujillo2001mnras} to nearby reference stars and simultaneously fitting sky background emission with a first-order polynomial function. 
The fluxes of \uname\ in the $B$ and $V$ bands were obtained by fitting the local PSF near the source location.
Paper I reported the presence of a faint background source $\sim$ 0\farcs8 north-west of the position of \uname\ with apparent magnitudes of 24.90$\,\pm\,$0.27, 24.02$\,\pm\,$0.20, and 22.39$\,\pm\,$0.08 mag in the $BVi$ bands, respectively, that mainly affects the $i$ band.
Therefore, we measure the $i$-band SN flux in the nebular phase by using a Kron aperture containing both sources and subtracting the known flux of the background source from the combined flux in the aperture.
Since the brightness of the background source is significantly fainter than that of the 1-$\sigma$ noise level in $B$- and $V$-band images ($\lesssim$ 23.4 mag) and the SN at any epoch ($<$ 22.0 mag for $B$ band and $<$ 22.1 mag for $V$ band), it is incapable of meaningfully affecting the PSF photometry of the SN in those bands.

Photometric flux calibration was performed against 6--9 standard reference stars within 10\arcmin\ of the source from the AAVSO Photometric All-Sky Survey\footnote{\url{https://www.aavso.org/apass}} database whose apparent magnitudes are in the range of 15--16 mag. The observations in the $BVI$ KMTNet filters were calibrated against reference stars in the nearest AAVSO filters (Johnson~$BV$, and Sloan $i'$; or $BVi$).
For the AAVSO reference stars, their KSP $BVI$ instrumental magnitudes were transformed to standard $BVi$ filters using the equations from \citet{Park2017apj}.
For the SN, since its nebular-phase spectra are significantly different from the AAVSO standard stars used to derive the \citet{Park2017apj} equations, we applied linearly interpolated spectrophotometric (S)--corrections \citep{Stritzinger2002aj}.
These are photometric corrections between instrument and standard filters derived by performing synthetic photometry on spectra obtained at approximately the same epoch.
The calibrated and S-corrected nebular-phase photometry is presented in Table~\ref{tab:neblc} and shown in Figure~\ref{fig:neblc}.

\subsection{Nebular-Phase Spectroscopy} \label{sec:nebspec}

We obtained four low-resolution nebular-phase optical spectra of \uname\ at 259.4, 277.3, 296.4, and 382.5 days since $B$-band maximum with a combination of the Gemini Multi-Object Spectrograph \citep[GMOS;][]{Hook2004} on Gemini-South, the Low Resolution Imaging Spectrometer \cite[LRIS;][]{Oke1995} on Keck, and the Low Dispersion Survey Spectrograph-3 \citep[LDSS-3;][]{Allington1994} on Magellan-Clay.
The spectroscopic observations are summarized in Table~\ref{tab:nebspec}.

\begin{deluxetable*}{lccccc}
\tabletypesize{\footnotesize}
\tablecolumns{5} 
\tablewidth{0.99\textwidth}
 \tablecaption{Nebular-phase spectroscopy of \uname}
 \tablehead{
 \colhead{Date (UT)} & \colhead{Phase} & \colhead{Telescope} & \colhead{Instrument} & \colhead{R} & \colhead{Wavelength [\AA]}
 } 
\startdata 
2018 December 30.28 & $+$259.4 & Gemini S & GMOS & 1690 & 4050--10000\\
2019 January 17.31 & $+$277.3 & Gemini S & GMOS & 1690 & 5000--10000\\
2019 February 5.53 & $+$296.4 & Keck & LRIS & 2000 & 3200--10000 \\
2019 May 3.17 & $+$382.5 & Magellan-Clay & LDSS-3 & 860 & 4250--10000 \\
\enddata
\tablecomments{Phase is observer frame days since $B$-band maximum light (MJD 58221.41).}
\end{deluxetable*} 
\label{tab:nebspec}

\begin{figure}[t!]
\epsscale{\scl}
\plotone{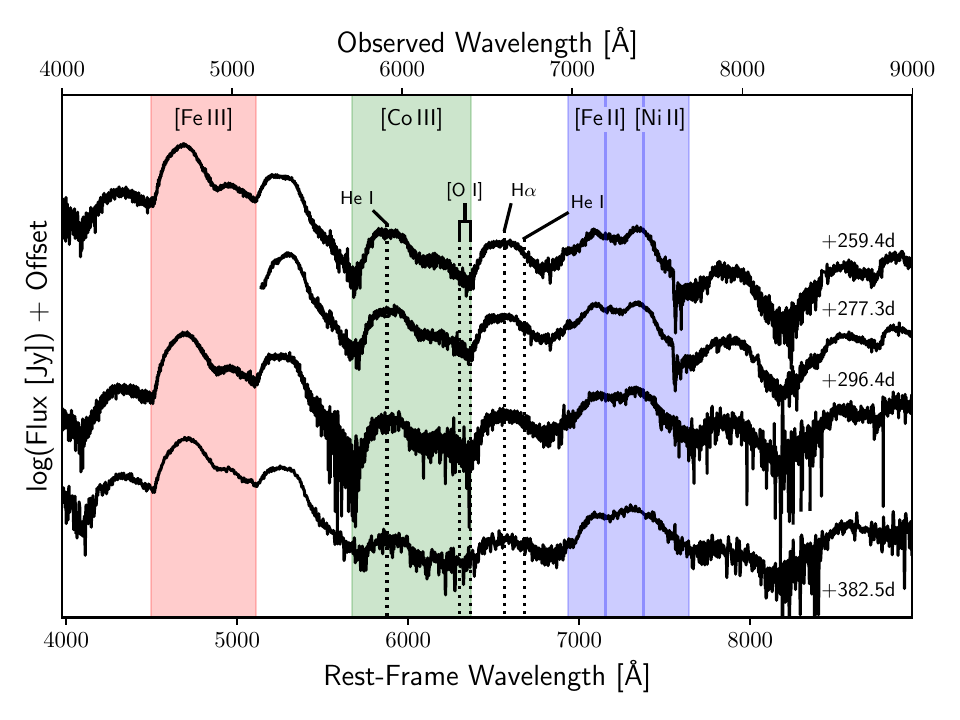}
\caption{The dereddened spectra of \uname\ obtained from four epochs during the nebular phase as labelled on the right side of the figure in days since $B$-band maximum are shown. The spectra are vertically offset for display clarity. The vertical shaded colored regions show the locations of the broad emission features of [\feiii] (red), [\coiii] (green), as well as [\feii] and [\niii] (blue) that are visible. 
While the [\feiii] and [\coiii] features are produced by a blend of several broad emission lines, the [\feii] and [\niii] features are thought to be primarily due to transitions of [\feii]~$\lambda$7155~\AA\ and [\niii]~$\lambda$7378~\AA\ (vertical solid lines), respectively \citep{Maeda2010apj}.
The dotted vertical lines show the expected locations of narrow emission lines associated with non-degenerate companions and circumstellar material (CSM) in \tase: H$\alpha$, \hei, and [\oi]. None of these narrow emission lines are detected.
\label{fig:nebspec}}
\end{figure}

The spectrum from the Magellan Telescope was reduced using standard tasks within IRAF. Bias and flat-field corrections were performed on the two-dimensional frames, one-dimensional spectra were extracted, and wavelength calibration was performed using calibration lamps taken immediately after target exposures. Flux calibration and telluric corrections were peformed with a set of custom IDL scripts \citep{Matheson2008,Blondin2012aj} using spectrophotometric standards observed on the same night. The GMOS spectra were reduced in a similar manner, but using the custom \texttt{gmos} suite of IRAF tasks. Initial flux calibration for GMOS spectra was performed using the IRAF tasks \texttt{standard} and \texttt{calibrate}, and final scaling was performed based on matching to the observed $V$-band photometry from the same epochs. The Keck-LRIS spectrum was reduced using LPipe, a fully-automated IDL pipeline for the LRIS \citep{Perley2019}.
The reduced and dereddened nebular-phase spectra are shown in Figure~\ref{fig:nebspec}.

\subsection{Host Galaxy, Distance, and Reddening}\label{sec:extn}

\uname\ is located at (RA, decl.) = ($\rm 11^h51^m01^s.80$, $-28\degr44\arcmin38\farcs.5$) (J2000), in the halo of its host galaxy NGC~3923 (Paper I).
We adopt the host galaxy redshift of $z$ = 0.0058, distance modulus (DM) of 31.75 $\pm$ 0.08~mag
based on normal \tas\ template fitting,
and extinction correction of $E(B-V)\sim$ 0.09 mag, consistent with the observed Na~I~D lines in the spectrum of \uname\ as well as the expected Galactic extinction towards the source (Paper I).
The extinction towards the source is also confirmed by fitting the observed color evolution of \uname\ during the Lira law phase as detailed in Appendix~\ref{apx:color}.

\section{Early Evolution and Classification}\label{sec:lc}

\subsection{Early Light Curves and the Characteristics of the Infant-Phase Excess Emission}\label{sec:gaus}

\begin{figure}[t!]
\epsscale{\scl}
\plotone{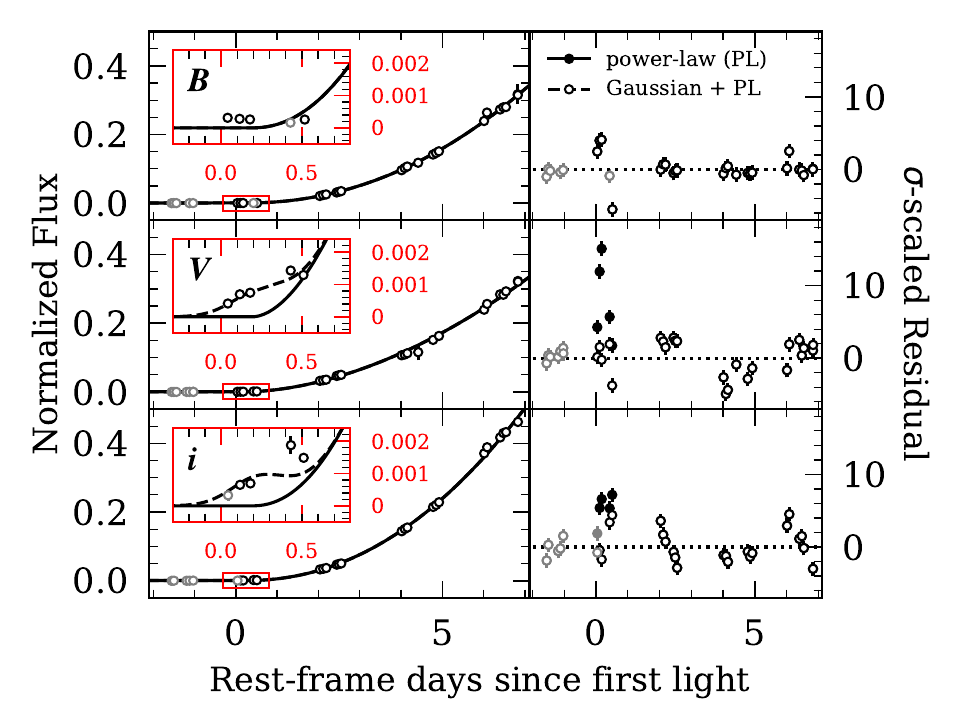}
\caption{(Left) The dereddened early $BVi$-band (top to bottom) forced photometry light curves (circles) of \uname\ up to 40\% of maximum light in rest frame are compared to the best-fit power-law $+$ Gaussian (dashed curves) and its power-law component alone (solid curves). 
The data points with S/N $<$ 3 are greyed out.
The inset zooms in on the infant phase ($\lesssim$ 1 day).
(Right) The $\sigma$-scaled residual of each data point for the best-fit power-law $+$ Gaussian (open circles) and its power-law component alone (closed circles) are shown over the same time interval as the left panel.
\label{fig:gaus}}
\end{figure}

The infant-phase light curves of \uname\ contain the lowest luminosity detected signals from an early \tas\ to date, reaching a depth of $-$10.5 absolute AB magnitude. 
In Paper I, we reported that the dominant source of its early luminosity appears to follow a power-law evolution.
The observed $BVi$-band light curves over 1--7 days since first light (or up to $\sim$ 40\% of peak brightness) follows $L_{\nu} \sim t^{\alpha_{\nu}}$, consistent with the majority of other \tase\ that have been observed in these phases \citep{Nugent2011nat, Foley2012apj, Olling2015nat, Cartier2017mnras, Holmbo2019aa, Miller2020apj, Moon2021apj, Ni2022natas}.
The measured power-law indices for \uname, $\alpha_{(B,V,i)}$ = (2.24, 1.99, 2.26), are also close to the Type Ia population average \citep[$\alpha$ = 2.01;][]{Miller2020apj}. 
In principle, power-law rise is expected for SNe powered by a smooth, centrally-concentrated \ni56\ distribution with a power-law-like tail towards the ejecta surface, where $\alpha$ depends on the steepness of the tail \citep{Piro&Nakar2014apj}.
However, in addition to this component, we also found evidence for excess emission over the power-law in the $V$ and $i$ bands during the first 0--1 days since first light.
This infant-phase excess emission is present during the same epochs as the $B$-band plateau, which has been attributed to line-blanket absorption by an over-density of Fe-peak elements near the ejecta surface.
While, in Paper I, we highlighted that excess radioactive heating by those same Fe-peak elements is one possible explanation for the excess emission, a range of other possible explanations and their implications remains to be thoroughly explored.

Here, we characterize the properties and statistical significance of the infant-phase excess emission in \uname.
Figure~\ref{fig:gaus} (left panels) shows the results of fitting the early light curves of \uname\ during 0--7 days with a power-law $+$ excess emission, where the infant-phase excess emission is modelled by a Gaussian in each of the $V$ and $i$ bands. 
(Note that the $B$-band light curve during 0--1 days is excluded from the fit since it is affected by $B$-band suppression.)
The $V$- and $i$-band infant-phase light curves share the same Gaussian central epoch, $\mu$, and width, $\sigma$,
but each Gaussian is scaled independently.
In each of the $BVi$ bands, the power-law component has the form $L_{\nu} \propto (t - t_{\rm PL})^{\alpha_{\nu}}$, where the onset of the power-law $t_{\rm PL}$ is shared among the bands while the power-law indices $\alpha_{\nu}$ and scalings are independent parameters in each band.
The best-fit power-law $+$ excess emission (dotted curves in Figure~\ref{fig:gaus}) is obtained with $\mu$ = 0.25 days since first light and $\sigma$ = 0.17 days for the Gaussian component, and $t_{\rm PL}$ = 0.19 days since first light and $\alpha_{(B,V,i)}$ = (2.1, 1.8, 2.1) for the power-law component (represented by the solid curves), which appears to adequately fit the observed early light curves (minus the excluded $B$-band light curve during 0--1 days).
The reduced $\chi$-squared statistic (= \chisq\ normalized by the number of degrees of freedom; \chisqr) of 4.0 for this fit is significantly better than the one obtained by fitting a pure power-law to the same light curves (\chisqr\ = 9.2; Paper I), indicating that the $Vi$-band excess emission component is required to explain the observed light curves.

The statistical significance of the $Vi$-band excess emission
is displayed in Figure~\ref{fig:gaus} (right panels), showing the $\sigma$-scaled residual of the best-fit power-law $+$ excess emission (open circles) compared to that of the power-law component alone (closed circles).
The residuals of the power-law component appear to be dominated by the data points from the infant phase.
Note that this is consistent with the \chisqr\ analysis of the power-law fitting in Paper I, where the \chisqr\ error from fitting a pure power-law (\chisqr\ = 9.2) was found to be predominantly from the infant-phase data points (with $\Delta$\chisqr\ = 6.0) than from all subsequent data points (with $\Delta$\chisqr\ = 3.2).
Meanwhile, the power-law $+$ excess emission model significantly reduces the residuals of the $Vi$-band data points from the infant phase, which now provide similar residuals as the data points from later phases.
Thus, the early light curves of \uname\ appear to require the distinct excess emission component peaked between $\sim$ 0.08 and 0.42 days since first light.
During this phase, excess emission is the dominant component of the SN light curve, emitting a total of $\sim$ 2.4 $\times$ 10$^{-9}$ ergs~cm$^{-2}$ into the $V$ and $i$ bands along the line of sight (or $\sim$ 1.4 $\times$ 10$^{44}$ ergs, assuming spherically symmetric emission).
In Section~\ref{sec:early}, we examine potential mechanisms that can produce the observed excess emission.

\subsection{Color Evolution}\label{sec:color}

\begin{figure}[t!]
\epsscale{\scl}
\plotone{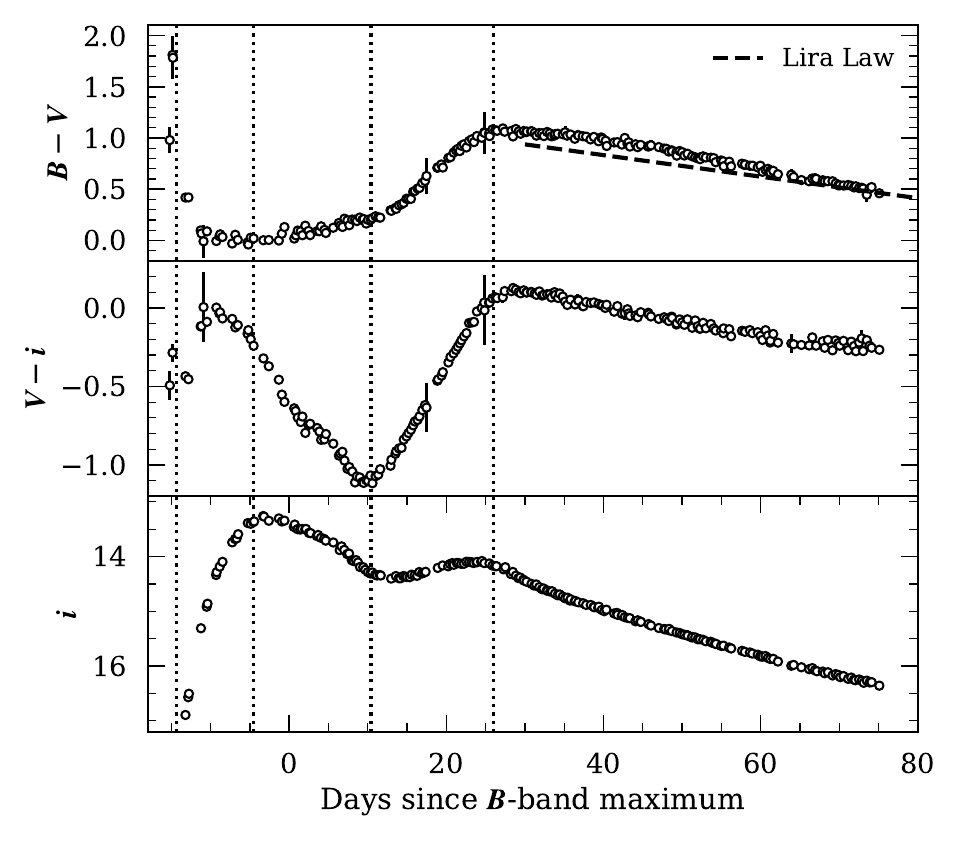}
\caption{The observed (non-dereddened) optical colors of \uname\ (black circles) in \bv\ (top) and \vi\ (middle) aligned with its $i$-band light curve (bottom). The data are binned over 0.3 day intervals.
The vertical dotted lines mark the epochs of $-$14.4, $-$4.6, 10.4, and 26.0 days since peak where the optical colors undergo notable phase transitions in their evolution \citep{Moon2021apj}. The dashed line is the Lira law from \citet{Burns2014apj}. Note that a zoomed-in plot of the un-binned early color evolution focused on the early phases before $\sim -$8 days is shown in Figure~\ref{fig:shockmod}.
\label{fig:optcolor}}
\end{figure}

Figure~\ref{fig:optcolor} presents high-cadence KMTNet color curves of \uname\ in \bv\ (top) and \vi\ (middle) aligned with its $i$-band light curve (bottom).
The observations, which are nearly simultaneous among different filters,
were linearly interpolated to the union of the two sets of epochs for each pair of adjacent filters during subtraction.
The four vertical dotted lines in the figure mark four epochs, $-$14.4, $-$4.6, 10.4, and 26.0 days since $B$-band maximum, where the colors undergo notable phase transitions in their evolution.

The \bv\ color evolution of \uname\ prior to the first color transition epoch, corresponding to the infant phase, was discussed extensively in Paper I. The simultaneous plateau in the $B$-band and rapid rise in the $V$- and $i$-band light curves at these early times lead to an abrupt redward evolution wherein the \bv\ color changes by 1.5 mag between 1.0 and 12.4 hours after first light.
We refer to this redward color evolution as the ``natal red bump'' (NRB), hereafter,
while the ``NRB phase'' refers to the epochs ($\sim$ 1.0--12.4 hours) where the NRB is observed.
The NRB is also identifiable in the \vi\ color,
though with a smaller color change of 0.23 mag between 2.8 and 12.2 hours.
During the NRB phase, the average \bv\ color is $\sim$ 1.7 mag redder than the average \vi\ color, consistent with the presence of Fe absorption lines that selectively suppress the $B$ band.

The entire color evolution after the first color transition epoch is largely consistent with those of other normal \tase, and is best described in relation to the $i$-band light curve \citep{Moon2016spie} as detailed Appendix~\ref{apx:color}.

\begin{figure}[t!]
\epsscale{\scl}
\plotone{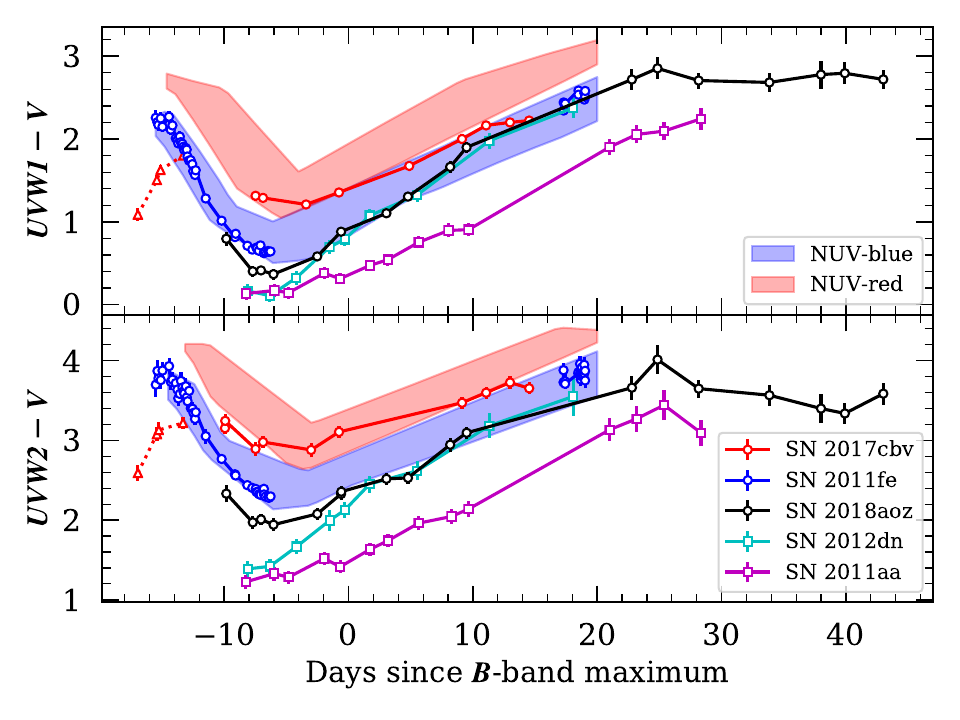}
\caption{The observed (non-dereddened) UV-optical colors of \uname\ (black open circles) in $UVW1-V$ (top) and $UVW2-V$ (bottom) compared to those of SN~2017cbv \citep[red circles;][]{Hosseinzadeh2017apj}, SN~2011fe \citep[blue circles;][]{Brown2012apj}, the NUV-red/blue groups of normal \tase\ \citep[colored shaded areas,][]{Milne2013apj}, and the super-Chandrasekhar-mass \tase~2012dn and 2011aa \citep[cyan and magenta squares, respectively;][]{Brown2014apj}. The red triangles represent the color of SN~2017cbv during its early excess emission. 
\label{fig:uvcolor}}
\end{figure}

Figure~\ref{fig:uvcolor} presents the Swift UV-optical color curves of \uname\ compared to those of other \tase.
The near-peak UV-optical colors of normal \tase\ have been grouped into two categories \citep{Milne2013apj}: ``NUV-red'' \citep[e.g., SN~2017cbv;][]{Hosseinzadeh2017apj} and ``NUV-blue'' \citep[e.g., SN~2011fe;][]{Brown2012apj}.
\citet{Brown2018at} initially reported that \uname\ displayed blue UV-optical colors that are similar to \tase\ with super-Chandrasekhar ejecta masses \citep{Brown2014apj}. 
Indeed, prior to peak, the colors are bluer than SN~2011fe, which is one of the bluest events in the NUV-blue group \citep{Brown2017apj}. 
However, subsequent evolution shows that while lying on the blue edge of the group, \uname\ overall appears to follow the NUV-blue group. 
In particular, the observed colors near peak are not as extreme as those of the super-Chandrasekhar-mass events---for instance, SNe 2012dn and 2011aa \citep[Figure~\ref{fig:uvcolor};][]{Brown2014apj}.

\subsection{Classification}\label{sec:class}

\begin{figure}[t!]
\epsscale{\scl}
\plotone{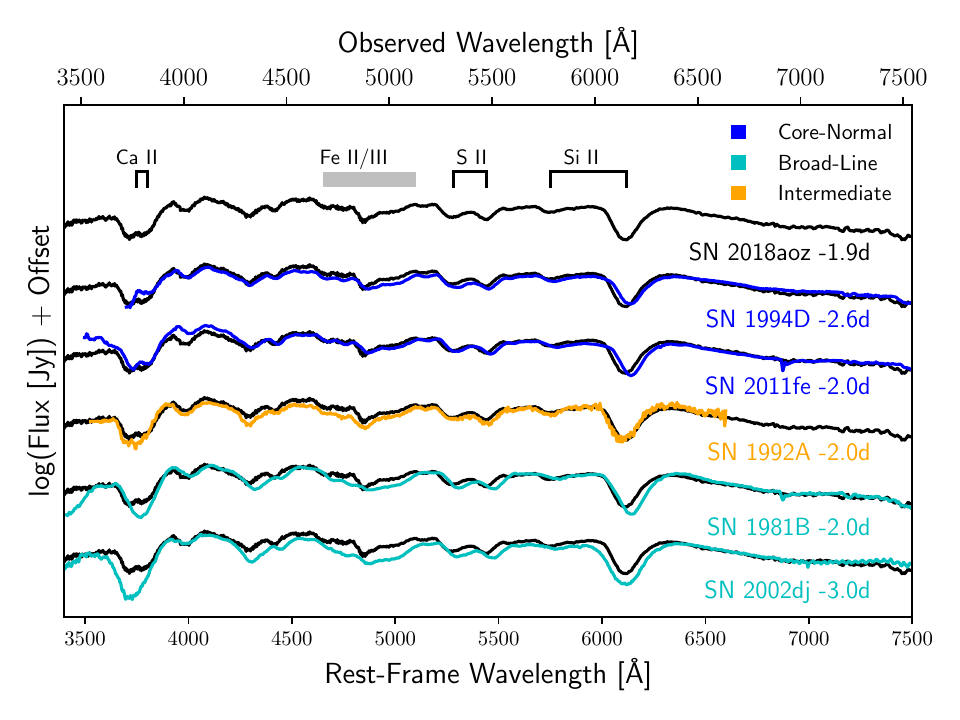}
\caption{The dereddened spectrum of \uname\ (black solid line; Paper I) taken 1.9 days before $B$-band maximum is 
compared to spectra of \tase\ of different subtypes obtained at comparable epochs:
SN~1994D \citep{Meikle1996mnras} and
2011fe \citep{Parrent2012apj}
(Core-Normal subtype; blue);
SN~1981B \citep{Branch1983apj} and SN~2002dj \citep{Pignata2008mnras} (Broad-Line subtype; cyan);
SN~1992A \citep{Kirshner1993apj} (intermediate type; orange).
Observed absorption features of \caii, \fex, \sii, and \siii\ are labelled at the top of the panel.  
\label{fig:class-spec}}
\end{figure}

We classify \uname\ as a normal \tas\ that is intermediate 
between the CN/NV and BL/HV subtypes based on its spectral properties as follows.
(Note that the SN light curves also support this classification as detailed in Appendix~\ref{apx:class}.)
Figure~\ref{fig:class-spec} compares the spectrum of \uname\ taken 1.9 days before $B$-band maximum (Paper I) to spectra of normal \tase\ from the CN and BL subtypes of \citet{Branch2006pasp}: SNe~1994D \citep[CN subtype;][]{Meikle1996mnras},
2011fe \citep[CN subtype;][]{Parrent2012apj}, 
1981B \citep[BL subtype;][]{Branch1983apj}, 2002dj \citep[BL subtype;][]{Pignata2008mnras} and 1992A \citep[intermediate between CN and BL;][]{Kirshner1993apj} from a similar phase.
The spectrum of \uname\ is consistent with 
those of the other normal \tase\ overall, whereas the detailed shapes of key absorption features seem to be intermediate between CN and BL events.
Sharp \fex\ absorption features seen in the spectrum of \uname\ are typical of CN events (e.g., SNe~1994D and 2011fe; blue spectra);
however, the \caii\ and \siii\ absorption features of \uname\ are relatively strong, which is a step in the direction of typical BL events such as SNe~1981B and 2002dj (cyan spectra).
SN~1992A (orange spectrum), classified as marginally BL while bordering CN \citep{Branch2006pasp}, 
is the closest spectroscopic analogue to \uname\ with nearly identical features
in the figure, suggesting that \uname\ is also intermediate between CN and BL.

\begin{figure}[t!]
\epsscale{\scl}
\plotone{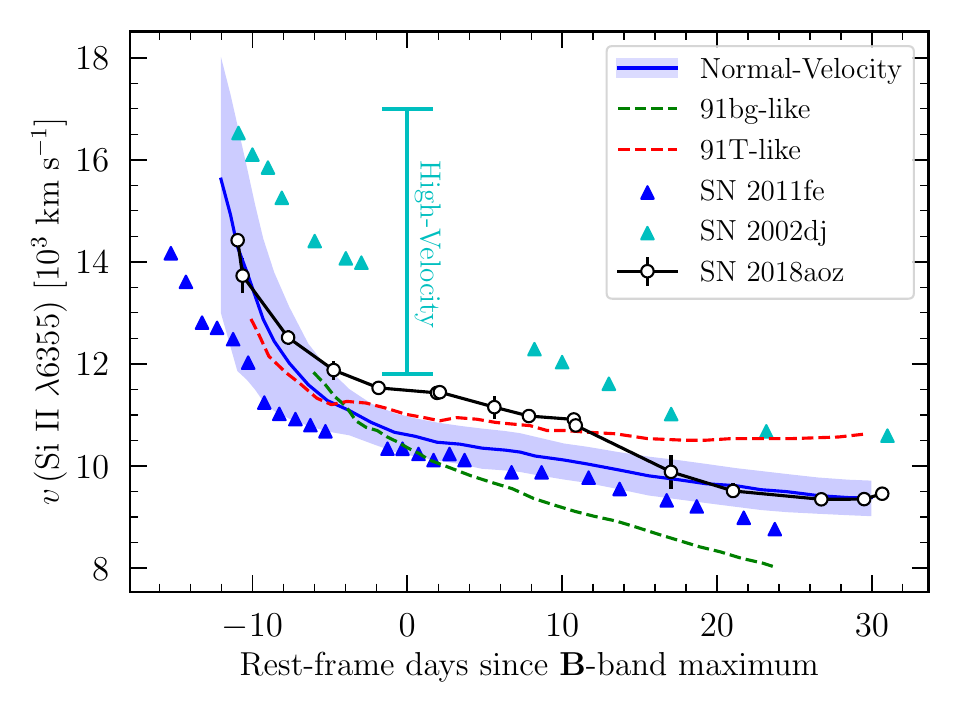}
\caption{The observed velocity evolution of the \siii\ spectral feature of \uname\ (orange circles; Paper I) is compared to the average velocity evolution for NV \tase\ \citep[blue solid curve with shaded 1-$\sigma$ error region;][]{Wang2009apj} as well as that of 91bg-like (green dashed curve) and 91T-like (red dashed curve) events in rest frame. SNe~2011fe \cite[blue triangles;][]{Pereira2013aa} and 2002dj \citep[cyan triangles;][]{Pignata2008mnras} are examples of NV and HV events, respectively, with similar \dm15\ as \uname.
\label{fig:class-si}}
\end{figure}

\begin{figure}[t!]
\epsscale{\scl}
\plotone{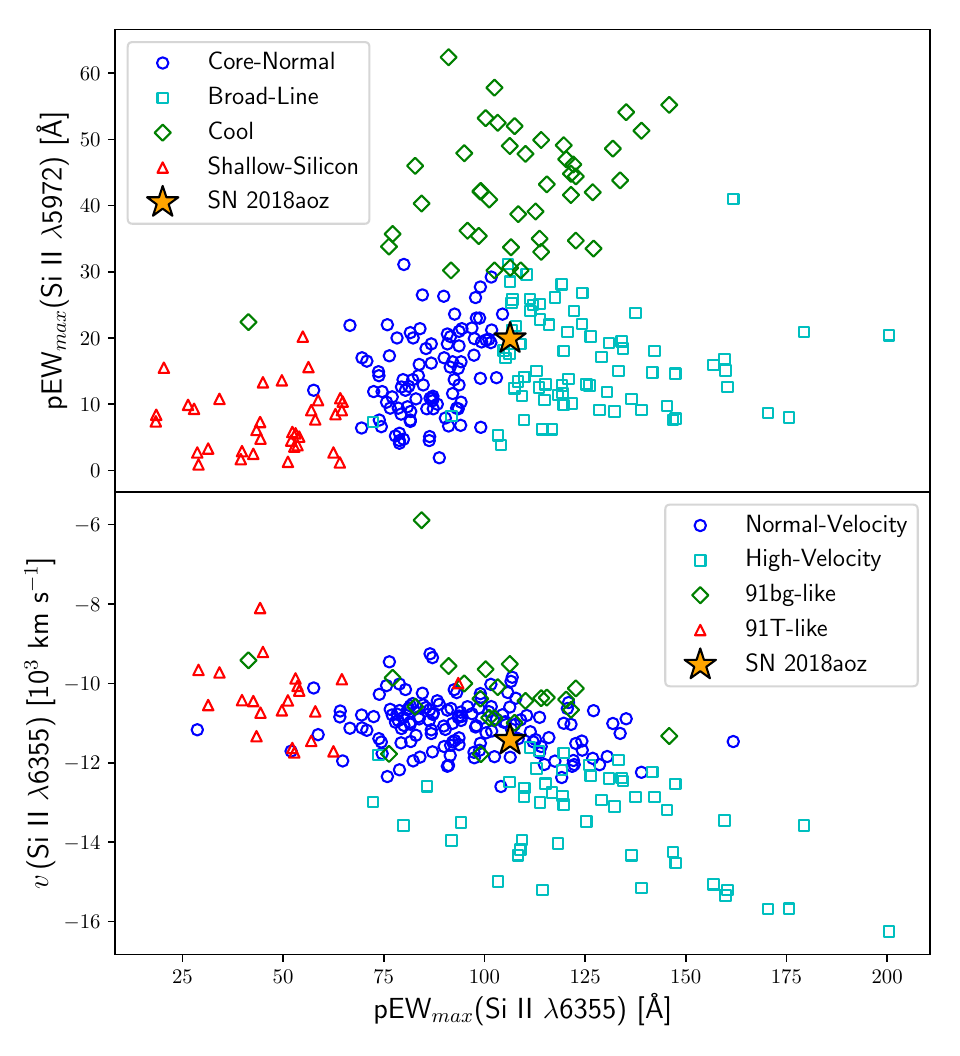}
\caption{Comparison of pseudo-equivalent widths of \siii\ lines (Top) and \siii\ velocity (Bottom) of \uname\ (orange star) with those of other \tase\ \citep{Blondin2012aj}. The colored symbols represent events from the four main subtypes of \tase: CN/NV (blue circles), BL/HV (cyan squares), Cool/91bg-like (green diamonds), and Shallow-Silicon/91T-like (red triangles). Note that BL/NV and CN/BL are both subsets of normal \tase, while Cool/91bg-like and Shallow-Silicon/91T-like are considered peculiar.
\label{fig:class-blondin}}
\end{figure}

Figure~\ref{fig:class-si} compares the evolution of the velocity 
of the \siii~$\lambda$6355~\AA\ feature (``\siii\ velocity'', hereafter) 
of \uname\ (Paper I) to what is expected for the NV and HV subtypes of \tase.
Note that the NV events (e.g., SN~2011fe; blue triangles) are characterized by 
near-peak \siii\ velocities of about $(10.6\pm0.4)\times 10^{3}$~km~s$^{-1}$, 
while the HV events (e.g., SN~2002dj; cyan triangles) have higher near-peak \siii\
velocities in the range of $\sim (11.8-17.0) \times 10^{3}$ km~s$^{-1}$ \citep[vertical cyan interval;][]{Wang2009apj}. 
The NV and HV subtypes largely overlap with CN and BL, respectively \citep{Parrent2014apss}. 
The \siii\ velocity evolution of \uname\ (orange circles) 
during early ($<-$5 days since $B$-band maximum) and late ($>$ 15 days) 
evolutionary phases appears to follow the NV subtype (blue curve with shaded area). 
Around the peak between $\sim$ $-$5 and 15 days, however, its velocity becomes 
significantly higher than the NV population and approaches those of HV events.
The peak \siii\ velocity of $(11.4\pm0.1)\times 10^{3}$~km~s$^{-1}$ in \uname\ 
is about 2-$\sigma$ higher than the NV population average and
near the lower boundary of the HV subtype.
Following the peak, the average \siii\ velocity gradient of \uname\ is $-$68~km~s$^{-1}$~day$^{-1}$, which is on the boundary between the High Velocity Gradient (HVG) and Low Velocity Gradient (LVG) sub-classes of \tase\ \citep{Benetti2005apj} that roughly correspond to HV and NV, respectively \citep{Parrent2014apss}.
The expected \siii\ velocity evolutions of 91bg-like (green dashed curve) 
and 91T-like (red dashed curve), the two most common peculiar types of \tase,
are apparently different from that of \uname\ in  Figure~\ref{fig:class-si} during late ($\gtrsim$ 15 days since $B$-band maximum) evolutionary phases. 
Thus, the \siii\ velocity evolution of \uname\ also supports its intermediate
nature between NV/CN and HV/BL, 
while it is clearly incompatible with those of the prototypical peculiar subtypes.

The intermediate nature of \uname\ between the normal subtypes of CN and BL is confirmed by the pseudo equivalent widths (pEWs) of \siii\ lines from its spectrum taken 1.9 days prior to $B$-band maximum. 
We measure pEWs of 20.22 and 106.4 for the \siii~5972~\AA\ and 6355~\AA\ lines, respectively, using the method of \citet{Branch2006pasp}.
Figure~\ref{fig:class-blondin} compares the  peak \siii\ pEWs and \siii\ velocity of \uname\ to those of a sample of \tase\ \citep{Blondin2012aj} from the CN/NV (blue circles) and BL/HV (cyan squares) subtypes, as well as the peculiar 91bg-like (or ``Shallow-Silicon''; red triangles) and 91T-like (or ``Cool''; green diamonds) subtypes.
The parameters of \uname\ are located at the boundary between the CN/NV and BL/HV subtypes of normal \tase\ in both panels.
The \siii\ pEWs of \uname\ (top panel) are consistent with BL events with \siii~6355~\AA\ pEW $>$ 105 \citep{Blondin2012aj}, while the \siii\ velocity of \uname\ (bottom panel) is consistent with NV events with \siii\ velocity $< 11.8\times 10^{3}$~km~s$^{-1}$, leading to the intermediate classification between the BL (/HV) and NV (/CN) subtypes.

\section{Early Light Curve Constraints on the Companion}\label{sec:kasan}

Early observations of \tase\ have been used to search for excess emission due to ejecta collision with companions \citep[e.g.,][]{Bloom2012apj, Olling2015nat, Marion2016apj, Hosseinzadeh2017apj, Hosseinzadeh2022arxiv, Li2019apj, Dimitriadis2019apj, Shappee2019apj, Moon2021apj}.
With early light curves from the low brightness of $-$10.5 absolute AB magnitudes, observations of \uname\ probe the luminosities expected not only for non-degenerate, but also WD companions for the first time. It therefore provides a unique opportunity to search for such emission and place strict constraints on the nature of the companion star.
Here, we compare the light curves of \uname\ with the analytic ejecta-companion interaction model of \citet[][``K10'' hereafter]{Kasen2010apj} that has been widely adopted for this type of analysis.
The luminosity ($\Gamma$) and effective temperature of the interaction emission in the model depend on the size of the companion (related to the binary separation distance in Roche overflow), as well as the opacity, mass, and kinetic energy of the ejecta.
When observed with a viewing angle $\theta$, the luminosity is $\Gamma\times S(\theta)$, where
\begin{equation}
	S(\theta)\simeq 0.982 \times \exp{[-(\theta/99.7)^2]} + 0.018
\label{eq:kasangle}
\end{equation}
\noindent
describes the angle dependence of the observed luminosity \citep{Olling2015nat}.
Note that the emission is strongest when the progenitor system is observed from the side of the companion star (0\degr; $S=1$) and it is weakest from the side of the progenitor star (180\degr; $S = 0.056$).

\subsection{Comparison to Fiducial Models}\label{sec:kasfid}

Figure~\ref{fig:kasmcVi} (left panels) compares the early $Vri$-band light curves
of \uname\ (black filled circles)
during 0--3 days since first light with what is predicted by the K10 model 
for three cases of non-degenerate binary companions at $\theta$ = 0\degr\ 
in Roche overflow: 1$\,$\msol\ red giant (1RG; red solid curve), 6$\,$\msol\ 
main sequence subgiant (6MS; blue solid curve), 
and 2$\,$\msol\ main sequence subgiant (2MS; indigo solid curve).
In the K10 model, we adopt the electron scattering opacity of $\kappa$ = 0.2~cm$^2$~g$^{-1}$ for H-poor \tas\ ejecta and the ejecta mass and kinetic energy of 0.80~\msol\ and 0.63 $\times$ 10$^{51}$~ergs, respectively, for \uname\ (Paper I).
For all three cases, the predicted emission is brighter than the observed luminosity, disallowing those configurations for the progenitor system under the K10 model.
The $B$-band light curve during 0--1 days was excluded from our comparisons because it is affected by $B$-band suppression while the K10 model assumes a pure blackbody spectral energy distribution.
Note that the values of ejecta mass and kinetic energy we adopted
are the lower limits of the ranges---$\sim$ 0.8--1.0~\msol\ and $\sim$ (0.6--0.8) $\times$ 10$^{51}$ ergs, respectively---that have
been considered for \uname\ (Paper I).
Since larger ejecta mass and kinetic energy both lead to brighter emission 
in the K10 model, the constraints provided in Figure~\ref{fig:kasmcVi} against the companion are conservative with respect to ejecta mass and kinetic energy.
(While $B$-band light curves have usually been used in the search for ejecta-companion interaction emission, we show in Appendix~\ref{apx:kasBsup} that model comparisons with the suppressed $B$-band light curve in the infant phase over-constrains the companion in the case of \uname.)

The K10 model is based on the assumption of local thermodynamic equilibrium (LTE) 
between the shock-heated ejecta and its radiated emission.
According to \citet[][``KS15'' hereafter]{Kutsuna2015pasj}, 
the matter-radiation coupling may not be strong enough to reach LTE due to the low
gas density in the ejecta-companion interaction, indicating that the K10 model may over-estimate
the emission temperature (and luminosity).
Figure~\ref{fig:kasmcVi} (left panels) also compares the observed light curves with the predictions of the two cases of companions from KS15:
1RG (red dotted curve) and 1MS (1~\msol\ main sequence subgiant companion; magenta dotted curve),
both at $\theta$ = 0\degr\ and in Roche overflow. 
While the 1RG case clearly over-predicts the observed emission at $\theta$ = 0\degr,
the case of 1MS is at a very similar brightness to what 
is observed during 0--0.5 days.
We note, however, that KS15 excludes free-free emission 
and Compton scattering---two processes known to accelerate
equilibrium \citep{Weaver1976apjs, Katz2010apj}---in their estimation of
the strength of matter-radiation coupling, likely leading to
under-prediction of emission temperature and luminosity.
Furthermore, no underlying radioactive SN emission is included
in the luminosity calculations by KS15 (and also by K10).
Therefore, it is highly likely that the $\theta$ = 0\degr\ 1MS case is also disallowed
given the close similarity between its prediction and the observed brightness,
though it is difficult to precisely quantify
the effects of excluding the two radiation processes 
and the underlying SN emission in the predicted luminosities.
For the predicted luminosities of ejecta-companion 
interaction alone, the KS15 and K10 models may be regarded as providing upper and lower bounds, respectively.

\subsection{Companion Constraints from Generalized Modeling}\label{sec:kasgen}

\begin{figure*}[t!]
\plotone{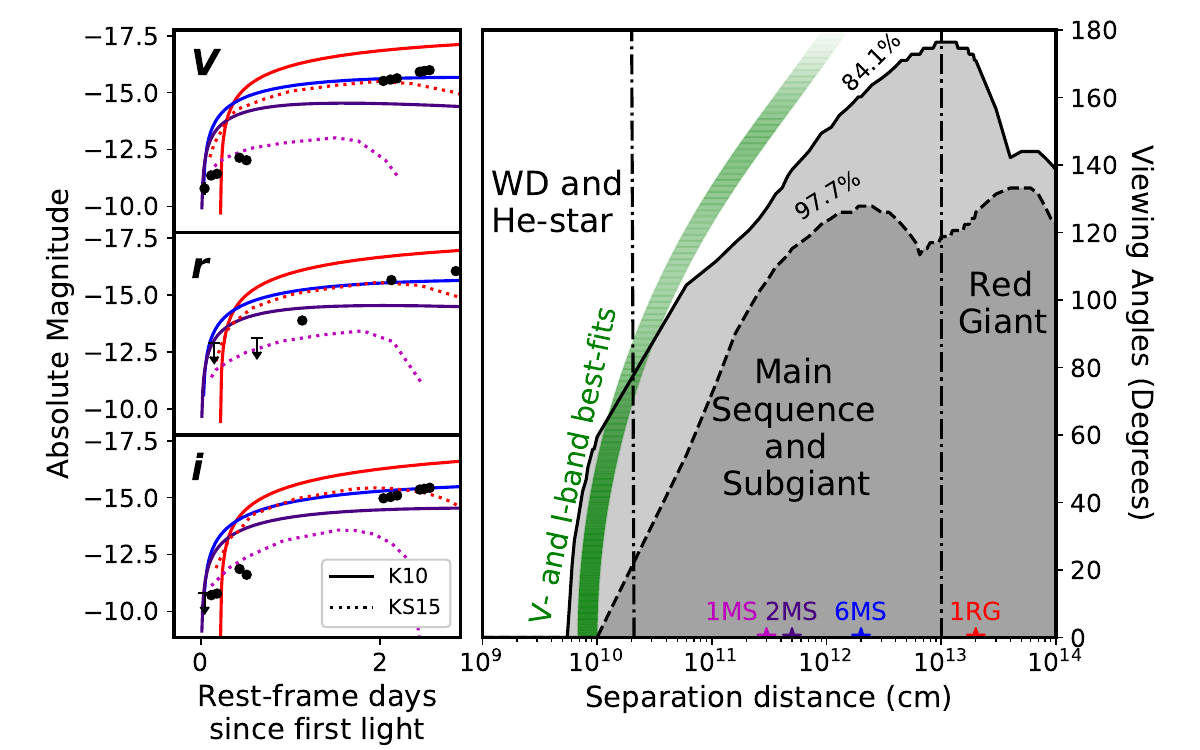}
\caption{(Left) The dereddened early $Vri$-band (from top to bottom) light curves of \uname\ (black circles) within 3 days after first light in rest frame are compared to ejecta-companion interaction models with 0\degr\ viewing angle. The models are of 2MS (indigo solid curves), 6MS (blue solid curves), and 1RG (red solid curves) companions from \citet{Kasen2010apj} as well as 1MS (magenta dotted curves) and 1RG (red dotted curves) companions from \citet{Kutsuna2015pasj}. The black inverted arrows are 3-$\sigma$ detection limits. (Right) The parameter space of separation distances and viewing angles of possible progenitor systems is shown. The vertical dot-dashed lines divide the x-axis (binary separation distance) into WD and He star, Main Sequence and Subgiant, and Red Giant regimes. 
The parameters in the shaded area underneath the solid and dashed black curves are ruled out at 84.1$\%$ and 97.7$\%$ confidence levels, respectively, by the early light curves of \uname\ and the model predictions of \citet{Kasen2010apj}.
The magenta, indigo, blue, and red stars at the bottom of the panel show the parameters for the correspondingly colored models in the left panels.
The green shaded region shows the best-fit separation distances obtained by fitting power-law $+$ \citet{Kasen2010apj} ejecta-companion interaction models for a set of viewing angles between 0--180\degr (see Section~\ref{sec:kasfit}). The transparency of the green shaded region is related to the goodness of the fit (\chisqr), where darker shading corresponds to a better fit.
\label{fig:kasmcVi}}
\end{figure*}

We generalize our analysis using the K10 model to allow for ejecta-companion 
interactions from all possible viewing angles between 0\degr\ and 180\degr\ 
and binary separation distances in the range of 10$^{9}$--10$^{14}$~cm,
following the methods of \citet{Moon2021apj}.
The range of separation distances corresponds to those of companions 
as small as WDs and as large as red supergiants at the Roche limit.
The right panel in Figure~\ref{fig:kasmcVi} shows the extent of this parameter space, where the separation distances are divided into the regimes of WD and He star, Main sequence and Subgiant (MS), and Red Giant (RG)
with two vertical dot-dashed lines approximating the lower bounds
for the MS \citep{Boetticher2017aa} and late-phase RG cases \citep{Seeds1984}.
By comparing the models represented by pairs of these parameters (i.e., viewing angle and separation distance) with the observed luminosities and pre-detection upper limits, we obtain the solid and dashed curves in the figure,
representing the lower limits of acceptable viewing angles as a function of separation distance (i.e., the area under the curve is ruled out) for the 84.1$\%$ and 97.7$\%$
confidence levels, respectively.\footnote{Note that 84.1$\%$ and 97.7$\%$ correspond to the 
the 1- and 2-$\sigma$ levels, respectively, of a Gaussian distribution in one direction.}
The confidence levels account for photometry errors as well as those of the model 
parameters, including redshift, explosion epoch, ejecta mass, 
and ejecta kinetic energy,
estimated using bootstrap assuming Gaussian error distribution.
Note that there are additional systematic uncertainties in the model comparison as mentioned in Section~\ref{sec:kasfid} above:
those associated with (1) the adoption of lower limits for the ejecta mass and kinetic energy of \uname\ and (2) the exclusion of the radioactive SN emission, that are not included in our analysis.
However, both of these uncertainties only allow for stronger 
constraints against the companion (see Section~\ref{sec:kasfid}).
The assumption of Roche overflow may also break if there is a long spin-down phase before the SN explosion during which the binary separation distance can evolve \citep{Meng2019mnras}.

Based on the comparison in Figure~\ref{fig:kasmcVi} (right panel), 
a low-mass ($\lesssim$ few solar mass) main sequence star or subgiant at a high ($\gtrsim$ 80\degr) viewing angle, He-star, or WD are the most likely binary companions in \uname.
\emph{Note that these results are independent of whether ejecta-companion interaction emission has really been detected in \uname.}
Separation distances from $\sim$ 5 $\times$ 10$^{11}$ to $\sim$ 10$^{14}$~cm, corresponding to companions larger than 2MS, are disallowed (at 84.1\% confidence level) for most ($\gtrsim$ 80\% of) viewing angles because the expected luminosity from their ejecta-companion interaction emission would exceed the observed
luminosity of \uname\ in the first three days for $\theta <$ 140--175\degr.
Thus, under the K10 model, if \uname\ had a large main sequence or red giant companion, it would need to have been located within a small range of viewing angles behind the SN.
The ejecta-companion interaction luminosity can be significantly lower than the K10 model predicts if LTE is not reached, as mentioned above, with KS15 providing a lower bound.
However, if we adopt the KS15 model for the 1RG case, the luminosity of 1RG (red dotted curve in Figure~\ref{fig:kasmcVi}, left panels) would be similar to that of 6MS in the K10 model (blue solid curve), for which $\sim$ 90\% of viewing angles are still ruled out (Figure~\ref{fig:kasmcVi}, right panel).
Therefore, the presence of a red giant companion is very unlikely even if the LTE assumption of K10 is not satisfied.
Although there is a small region in the upper-right corner (i.e., large separation and viewing angle) 
of Figure~\ref{fig:kasmcVi} that is not directly ruled out by the comparison, the separation distances correspond to short-lived companions (e.g. red supergiants) which are very unlikely to be found in the halo region of an elliptical galaxy---where \uname\ is located---due to the lack of recent star formation (see Section~\ref{sec:prog}).

\section{Infant-Phase Excess Emission Modelling}\label{sec:early}

\uname\ shows significant excess emission over the power-law rise during 0--1 days since first light (Section~\ref{sec:gaus}).
An over-density of \ni56\ near the ejecta surface can produce excess thermal emission in this phase (Paper I),
but other possibilities---such as ejecta shock interaction---and their subsequent implications for the progenitor system remain unexplored.
We examine the origin of the infant-phase excess emission by fitting the early light curves of \uname\
using a model combining the underlying SN emission (which is represented by a power-law; see Section~\ref{sec:gaus}) and excess emission. 
We compare the fits obtained using models of four conceivable mechanisms for the excess emission: surface \ni56\ heating (Section~\ref{sec:nipl}), ejecta-companion interaction (Section~\ref{sec:kasfit}), ejecta-CSM interaction (CSM; Section~\ref{sec:csm}), and shock breakout (Section~\ref{sec:sbo}).
Note that the characteristic ejecta velocity of \uname, estimated using its observed peak \siii\ velocity of 11400~km~s$^{-1}$ (Paper I), broadly constrains the possible sources of infant-phase emission from ejecta shock interactions to be within $\lesssim$ 10$^{14}$~cm of the progenitor, which includes only the binary companion, 
nearby CSM, and the shock-heated progenitor surface. 

For all of the four excess emission mechanisms, we adopt blackbody spectral energy distributions, because they are based on thermal processes.
We fit the light curves of \uname\ during 0--8 days, but exclude the $B$-band light curve during 0--1 days since it is affected by $B$-band suppression and incompatible with a pure blackbody process (Paper I).
The results obtained using the four models are compared below,
followed by detailed descriptions of each model and the fitting process in the subsequent subsections.

\begin{figure*}[t!]
\epsscale{\scl}
\plotone{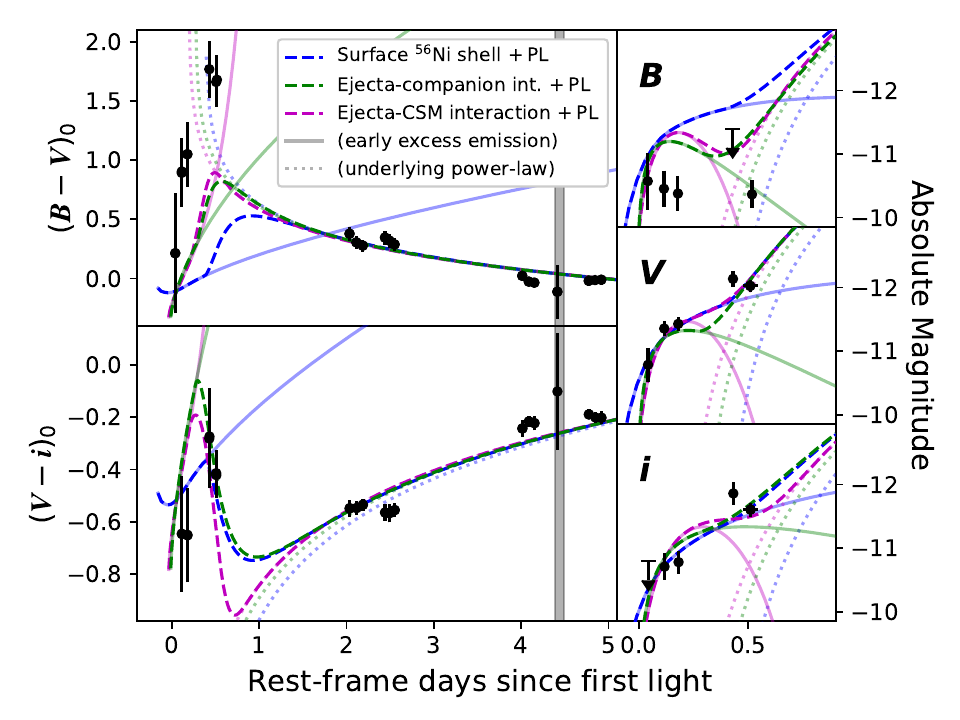}
\caption{(Left) The dereddened \bv\ (top) and \vi\ (bottom) colors of \uname\ in rest frame (circles) are compared with what is expected from power-law (PL) $+$ three models of early excess emission in \tase: (1) surface \ni56\ heating (blue dashed curves; Section~\ref{sec:nipl}), (2) ejecta-companion interaction (green dashed curves; Section~\ref{sec:kasfit}),
and (3) ejecta-CSM interaction (magenta dashed curves; Section~\ref{sec:csm}).
The transparent solid and dotted curves show the early excess emission and underlying power-law components, respectively, of each correspondingly colored model.
The vertical grey line marks the epoch when the first spectrum was taken (4.4 days since first light or $-$11.0 days since peak). (Right) The dereddened $BVi$-band (from top to bottom) light curves of \uname\ in rest frame are compared with those predicted by the same models from the left panels. The inverted arrows are detection limits at a S/N of 3.
\label{fig:shockmod}}
\end{figure*}

Figure~\ref{fig:shockmod} compares the dereddened colors, (\bv)$_0$ (top left panel) and (\vi)$_0$ (bottom left panel),
and $BVi$ light curves (right panels) of \uname\ 
with the best-fit model predictions:
blue dashed curves for surface \ni56\ heating, 
green dashed curves for ejecta-companion interaction, 
and red dashed curves for ejecta-CSM interaction.
(Shock breakout is not shown because it is too faint
to be compared for a reasonable set of model input parameters; Section~\ref{sec:sbo}).
The fit quality is not significantly different for the three best-fits, which have similar \chisqr\ values of 3.4, 3.5, and 3.2, respectively.
As seen in the figure, all three models appear to reproduce the observed $Vi$-band light curves 
of \uname\ as well as the \vi\ color curve similarly well; however, these blackbody excess emission models 
all over-predict the $B$-band luminosity by $\sim$ 0.5$-$1.0 mag in 0.1--0.5 days, 
leading to bluer \bv\ color than observed during the period.
Note that the lower infant-phase $B$-band luminosity compared to the $V$ and $i$ bands in \uname, which is incompatible with pure blackbody emission, 
has been attributed to $B$-band suppression caused by surface Fe-peak elements (Paper I).
As detailed in the following subsections, the best-fit parameters of the surface \ni56\ heating, ejecta-companion interaction, and ejecta-CSM interaction models are all compatible with viable physical processes that can produce the observed infant-phase excess emission in \uname.

\subsection{Radioactive Heating by Excess Surface \ni56}\label{sec:nipl}

We first fit the observed early light curves of \uname\ using the combination of power-law emission (for the underlying SN emission) and the emission from a \ni56 shell distribution (for surface \ni56\ heating).
For the power-law component, we use the power-law described in Section~\ref{sec:gaus} with onset $t_{\rm PL}$ and indices $\alpha_{(B,V,i)}$.
For the infant-phase \ni56\ shell emission, we developed the following model in three steps: 
\begin{enumerate}
    \item We adopt the luminosity calculation for \ni56-powered SNe from \citet[][PN14 herafter]{Piro&Nakar2014apj} based on \ni56\ decay and photon diffusion. In the model, the SN luminosity is determined by the \ni56\ distribution and the ``diffusion depth'', defined as the deepest layer in the ejecta that is visible via photon diffusion. 
    For the evolution of the diffusion depth, we adopt the following equation from Paper I (based on Equation 1 in PN14) describing the fractional mass of the ejecta ($\Delta M/M_{\rm ej}$) in the layers above the diffusion depth at $t-t_0$ days since explosion:
    \begin{equation}
        \frac{\Delta M}{M_{\rm ej}} \approx 1.3 
        \left(\frac{t-t_0}{\tau_m}\right)^{1.76}\ M_{\odot}
    \label{eq:mdiff}
    \end{equation}
    \begin{equation}
        \tau_m = \left( \frac{\kappa}{13.8\, c}\right)^{1/2} \left( \frac{6\,M_{\rm ej}^3}{5 \,E_{\rm ej}}\right)^{1/4}
    \label{eq:taum}
    \end{equation}
    where $\tau_m$ is the geometric mean of the diffusion and expansion timescales \citep{Arnett1982apj, Moon2021apj} related to the ejecta mass, ejecta kinetic energy, and opacity ($M_{\rm ej}$, $E_{\rm ej}$, and $\kappa$, respectively).
    We use the value of $\tau_m$ = 9.51 $\pm$ 0.26 measured from the bolometric light curve of \uname\ (Paper I).
    Note that the explosion epoch, $t_0$, can be different than the onset of the power-law component of the light curve, $t_{\rm PL}$, due to the possibility of a few-hours to days delay \citep[or ``dark phase'';][]{Piro&Nakar2013apj, Piro&Nakar2014apj} before the diffusion depth reaches the underlying main distribution of centrally-concentrated \ni56\ in the ejecta, which is responsible for the power-law rise (Section~\ref{sec:gaus}).
    
    \item The \ni56\ distribution in PN14 (described by a logistic function; see Equation 11 therein) is replaced by the \ni56\ shell distribution with the following functional form:
    \begin{equation}
        X_{56}(t) = \begin{cases}
        X_s, & t - t_0 < t_s\\
        0\ , & t - t_0 > t_s
    \label{eq:nishell}
    \end{cases}
    \end{equation}
    where $X_{56}(t)$ is the mass fraction of \ni56\ at the diffusion depth at $t - t_0$ days since explosion and $t_s$ is the time when the diffusion depth reaches the inner radius of the shell. (Note that the time coordinate $t$ is related to the radial mass coordinate of the diffusion depth by Equation~\ref{eq:mdiff}).
    In the fitting below, we represent the distribution using two physical parameters, $t_s$ and $M_s$, where $M_s = X_s \Delta M(t_s)$ is equal to the total mass of \ni56\ above the diffusion depth at time $t_s$ in the ejecta.
    
    \item The \ni56\ shell emission, originating from radioactive heating of the high-density SN ejecta in infant phases, is assumed to be blackbody distributed (i.e., we assume that the emission is fully thermalized and that gamma-rays are fully trapped) in order to fit the multi-band SN light curves.
    We estimate the blackbody temperature, or ``color temperature'' ($T_c$), of the \ni56\ shell emission using the following equation \citep[based on Equation 12 from][]{Piro&Nakar2013apj}:
    \begin{equation}
        T_c^4 = \frac{L \tau_s}{4\pi \sigma_{SB} r_{ph}^2}\ \ , \ \ \tau_s = \tau_c \left(\frac{r_{ph}}{r_c}\right)^2
    \label{eq:colorT}
    \end{equation}
    where $L$ is the luminosity of the \ni56\ shell emission; $r_{ph}$ is the radius of the photosphere; and $\tau_s$ is a parameter combining the radius, $r_c$ (or ``color depth''), in the ejecta where the \ni56\ radioactive emission is thermalized and the optical depth, $\tau_c$, at the color depth.
    We estimate $r_{ph}$ based on a polytropic (n = 3) ejecta profile expected for an exploding WD undergoing homologous expansion \citep{Piro&Nakar2013apj}, and assume $\tau_s$ is roughly constant over the $\sim$ 1-day infant phase.
    Note that $\tau_s$ is expected to be close to unity since both $\tau_c$ and $r_{ph}/r_{c}$ are typically not much larger than unity \citep{Piro&Nakar2013apj}, so the assumption can at most contribute error with near-unity order.
\end{enumerate}

Fitting the $BVi$-band light curves of \uname\ up to 8 days since first light, excluding the $B$-band light curve during 0--1 days, we obtain the best-fit power-law $+$ surface \ni56\ heating model with \chisqr\ = 3.4.
The parameters are $t_0$ = $-$0.17 days, $t_s$ = 0.30 days, $t_{\rm PL}$ = 0.38 days, all since the epoch of first light (MJD~58206.00) in rest frame, $M_s$ = 8.3 $\times$ 10$^{-4}$~\msol, $\tau_s$ = 18.2, and $\alpha_{B,V,i}$ = (2.03, 1.74, 2.08).
For the properties of the early excess emission in the fit, we allowed for $t_s$ in the range of 0.01--1.0 days, $X_s$ in 1--99\%, and $\tau_s$ in 2/3--100.
The best-fit (blue dashed curves in Figure~\ref{fig:shockmod}) appears to provides an excellent match to the observed infant-phase excess emission of \uname\ in the $V$ and $i$ bands with a 8.3 $\times$ 10$^{-4}$~\msol\ shell of excess \ni56\ in the outer 0.65\% of the SN-ejected mass.
If this is the origin of the infant-phase excess emission, then the difference between the best-fit $t_0$ and $t_{\rm PL}$ parameters indicates the presence of a $\sim$ 0.55-day dark phase in \uname, similar in length to the one reported in the normal \tas~2011fe \citep[$\sim$ 0.5 days;][]{Piro&Nakar2014apj}. 
We also note that the best-fit indices $\alpha_{B,V,i}$ are slightly lower than those obtained in Section~\ref{sec:gaus}, though they are still consistent with the $\alpha\sim$ 2 expectation for power-law rise that has been found in other normal \tase.

The best-fit mass and location of surface \ni56\ obtained above, 8.3 $\times$ 10$^{-4}$~\msol\ of \ni56\ in the outer 0.65\% of the SN-ejected mass, are larger and deeper in the ejecta, respectively, than 1.8 $\times$ 10$^{-4}$~\msol\ of \ni56\ in the outer 0.31\% of the SN-ejected mass obtained in Paper I by fitting the infant-phase excess emission of \uname\ with a purely \ni56-powered blackbody model (as opposed to the power-law $+$ surface \ni56\ model). 
These numbers are broadly comparable with the location and quantity of Fe-peak elements required to explain the $B$-band suppression associated with the NRB, $\sim$ 10$^{-3}$~\msol\ in the outer $\sim$ 1\% of the SN ejecta (Paper I).
However, radiative transfer simulations that account for both line formation and incomplete gamma-ray trapping are required to determine if any single distribution of Fe-peak elements can reproduce both the infant-phase excess emission and NRB features in \uname\ simultaneously.
We discuss this in the context of thin-shell He-shell DDet simulations in Section~\ref{sec:hedd}, below.

\subsection{Ejecta Interaction with the Companion}\label{sec:kasfit}

We model the
infant-phase emission of \uname\ with a combination
of radioactive SN emission and ejecta-companion interaction emission---the former 
with a power-law (Section~\ref{sec:gaus}) and the latter with the K10 model (Section~\ref{sec:kasan}).
For the K10 model, we use the electron scattering opacity of $\kappa$ = 0.2~cm$^2$~g$^{-1}$ for H-poor \tas\ ejecta following Section~\ref{sec:kasan}.
Fitting the observed $BVi$ light curves during 0--8 days, excluding the $B$-band light curve during 0--1 days, we obtain the green shaded region in Figure~\ref{fig:kasmcVi} (right panel) showing 
the distribution of the best-fit companion separation distances ($a$) for viewing angles ($\theta$) between 0\degr\ and 180\degr.
The upper and lower boundaries of the region were obtained using two cases of relatively small and large ejecta masses and kinetic energies, respectively, derived by modelling the light curves of \uname\ (Paper I) as follows:
(1) $M_{\rm ej}$ = 0.80~\msol\ and $E_{\rm ej}$ = 0.63 $\times$ 10$^{51}$~ergs based on the \citet{Arnett1982apj} model; and (2) $M_{\rm ej}$ = 1.05~\msol\ based on He-shell DDet simulations, corresponding to $E_{\rm ej}$ = 0.82 $\times$ 10$^{51}$~ergs for the characteristic ejecta velocity of 11400 km~s$^{-1}$.

\begin{deluxetable*}{cccll}
\tabletypesize{\footnotesize}
\tablecolumns{5} 
\tablewidth{0.99\textwidth}
 \tablecaption{Ejecta-companion interaction model fit parameters}
 \tablehead{
 \colhead{$M_{\rm ej}$ and $E_{\rm ej}$} & \colhead{Viewing angle} & \colhead{Fit parameters} & \colhead{\chisqr}
 } 
\startdata 
 0.80~\msol\ and 0.63 $\times$ 10$^{51}$ ergs{$\rm ^a$} & $\theta$ = 0\degr\ & $a$ = 1.0 $\times$ 10$^{10}$~cm & 3.48 (lowest) \\
  &    & $t_{0}$ = $-$0.01 days & \\
  &    & $t_{PL}$ = 0.27 days &  \\
  &    & $\alpha_{B, V, i}$ = (2.07, 1.77, 2.10) &  \\
 \hline
  &   $\theta$ = 180\degr & $a$ = 1.5 $\times$ 10$^{12}$~cm & 3.54 \\
  &    & $t_{0}$ = $-$0.23 days &  \\
  &    & $t_{PL}$ = 0.37 days &  \\
  &    & $\alpha_{B, V, i}$ = (2.05, 1.75, 2.08) &  \\
 \hline
 1.05~\msol\ and 0.82 $\times$ 10$^{51}$ ergs{$\rm ^b$} & $\theta$ = 0\degr\ & $a$ = 6.8 $\times$ 10$^{9}$~cm & 3.47 (lowest) \\
  &    & $t_{0}$ = 0.00 days &  \\
  &    & $t_{PL}$ = 0.25 days &  \\
  &    & $\alpha_{B, V, i}$ = (2.08, 1.78, 2.11) &  \\
 \hline
  &   $\theta$ = 180\degr & $a$ = 8.8 $\times$ 10$^{11}$~cm & 3.52 \\
  &    & $t_{0}$ = $-$0.19 days &  \\
  &    & $t_{PL}$ = 0.37 days &  \\
  &    & $\alpha_{B, V, i}$ = (2.04, 1.74, 2.08) &  \\
\enddata
\tablenotetext{{\rm a}}{From applying the \citet{Arnett1982apj} model to the light curves of \uname, as typically done for radioactively-powered SNe \citep[e.g.,][]{Li2019apj, Drout2016apj}, approximating the \ni56-dominated opacity in the photospheric phase as $\kappa\sim$ 0.1~cm$^2$~g$^{-1}$ (Paper I).}
\tablenotetext{{\rm b}}{From He-shell DDet simulations (Paper I).}
\tablecomments{$t_0$ and $t_{\rm PL}$ are in days since the epoch of first light (MJD~58206.00) in rest frame.}
\end{deluxetable*}
\label{tab:kas}

Table~\ref{tab:kas} shows the range of fit parameters obtained using different $M_{\rm ej}$, $E_{\rm ej}$, and $\theta$.
In the fit, we allowed for $a$ in the range of 10$^9$--10$^{13}$~cm, which includes all reasonable Roche separation distances for \tas\ progenitors (see Figure~\ref{fig:kasmcVi}).
The best-fit model light and color curves with $\theta$ = 0\degr, which are nearly identical for the two cases of $M_{\rm ej}$ and $E_{\rm ej}$ (green dashed curves in Figure~\ref{fig:shockmod}), provide a very similar goodness of fit to those of \uname\ as the best-fit surface \ni56\ heating model (Section~\ref{sec:nipl}).
The differences between the onsets of the K10 and power-law components in the models (= $t_0$ and $t_{\rm PL}$, respectively) range from $\sim$ 0.3 days for the lowest-\chisqr\ case of $\theta$ = 0\degr\ to $\sim$ 0.6 days for the case of $\theta$ = 180\degr.
These differences are consistent with $t_{PL} - t_0$ of 0.54 days obtained with the surface \ni56\ heating model, pointing to an approximately half-day post-explosion delay (or dark phase) in \uname\ for the diffusion of the radioactive SN emission responsible for the power-law rise.
As seen in the table, the change in \chisqr\ between $\theta$ = 0\degr\ and 180\degr\ is less than 2\%, indicating that the goodness of fit of the ejecta-companion interaction model does not change significantly with separation distance ($a$) ranging from (0.7--1.0) $\times$ 10$^{10}$~cm for $\theta$ = 0\degr\ to (0.9--1.5) $\times$ 10$^{12}$~cm for $\theta$ = 180\degr.
These separation distances correspond to two types of companions that appear to be nearly equally compatible with the observed infant-phase excess emission of \uname\ under the K10 model: (1) a low-mass ($\lesssim$ few solar mass) main sequence star or subgiant at $\gtrsim$ 80\degr\ viewing angle; or (2) a WD or He-star at $\lesssim$ 80\degr\ viewing angle.
Note that the case of ejecta interaction with a WD companion for the origin of the infant-phase excess emission implies a \d6s\ origin for \uname\ since it is the only scenario that predicts the presence of a surviving WD companion after the SN explosion.

\subsection{Ejecta Interaction with Circumstellar Material}\label{sec:csm}

The interaction between the SN ejecta and CSM near the progenitor can produce excess emission with properties dependent on the mass and spatial distribution of the CSM.
We model the early light curves of \uname\ as a combination of power-law (for the underlying SN emission) and ejecta-CSM interaction emission (for the infant-phase excess emission), adopting the model of \citet[][P15, hereafter]{Piro2015apj} for the latter.
Here, we describe the CSM model and geometries (Section~\ref{sec:csmmod}) considered, and then discuss the results in the context of both H-poor (Section~\ref{sec:merger}) and H-rich CSM (Section~\ref{sec:accrete}).

\subsubsection{Model Description}\label{sec:csmmod}

The observed interaction emission is largely determined by properties of the outermost CSM layer \citep{Piro2015apj, Nakar2014apj}, represented as a uniform-density and spherically-symmetric envelope with mass $M_{\rm env}$ and radius $R_{\rm env}$ in the P15 model.
The luminosity ($L_{\rm CSM}$) is provided by the following equation determined by $M_{\rm env}$, $R_{\rm env}$, ejecta mass ($M_{\rm ej}$), ejecta kinetic energy ($E_{\rm ej}$), and opacity ($\kappa$):
\begin{equation}
    L_{\rm CSM}(t) = \frac{t_{\rm env} E_{\rm env}}{t_p^2} \exp{\left[- \frac{t(t+2t_{\rm env})}{2t_p^2}\right]}
\label{eq:Lcsm}
\end{equation}
where $t_{\rm env} \propto E_{\rm ej}^{-0.5} M_{\rm ej}^{0.35} M_{\rm env}^{0.15} R_{\rm env}$ is the envelope expansion timescale post-explosion, $E_{\rm env} \propto E_{\rm ej} M_{\rm ej}^{0.7} M_{\rm env}^{-0.7}$ is the total energy transferred from the ejecta to the envelope, $t_p \propto \kappa^{0.5} E_{\rm ej}^{-0.25} M_{\rm ej}^{0.17} M_{\rm env}^{0.57}$ is the emission peak epoch, and $t$ is time in seconds since the explosion epoch ($t_0$).
Adopting a blackbody for the spectral energy distribution of the interaction emission, the blackbody temperature follows
\begin{equation}
    T_{\rm CSM}(t) = \left[ \frac{L_{\rm CSM}(t)}{4\pi \sigma_{\rm SB} (R_{\rm env} + \varv_{\rm env} t)^2} \right]^{1/4}
\label{eq:Tcsm}
\end{equation}
where $\varv_{\rm env} \propto E_{\rm ej}^{-0.5} M_{\rm ej}^{0.35} M_{\rm env}^{0.15}$ is the envelope expansion velocity post-explosion.

We also consider ejecta interaction with CSM distributed in an equatorially-concentrated disk or torus as follows.
Such CSM may divert the flow of SN ejecta away from the equatorial plane, where it obscures the ejecta-CSM interaction from viewing angles ($\theta$) above and below the equatorial plane ($\theta$ = 0\degr).
Note that similar obscuration is expected for ejecta-companion interaction due to the diverted flow of SN ejecta around the companion \citep{Kasen2010apj}, resulting in attenuated brightness of the interaction as described by Equation~\ref{eq:kasangle} for viewing angles $\theta$ away from the binary axis towards the companion ($\theta$ = 0\degr\ in Equation~\ref{eq:kasangle}).
We approximate the attenuation of ejecta-CSM interaction brightness for a viewing angle $\theta$ above or below the equatorial plane as similar to that of ejecta-companion interaction for the same angle $\theta$ away from the binary axis towards the companion for a distant observer, assuming similar flow of SN ejecta away from the interaction region.
The brightness of ejecta interaction with equatorially-concentrated CSM would thus be $L_{\rm CSM}\times S(\theta)$ for $\theta$ ranging from the equatorial plane (0\degr) to the poles (90\degr), using $S(\theta)$ from Equation~\ref{eq:kasangle}.
$S(0\degr)$ = 1.0 means the brightness along the equatorial plane is identical to the case of spherically symmetric CSM, $L_{\rm CSM}$, while $\theta$ = 90\degr\ provides the minimum observed brightness of $L_{\rm CSM}\times 0.45$.

\subsubsection{Circumstellar Material from a WD or He-star Companion}\label{sec:merger}

\begin{deluxetable*}{cclcl}
\tabletypesize{\footnotesize}
\tablecolumns{5} 
\tablewidth{0.99\textwidth}
 \tablecaption{Ejecta-CSM interaction model fit parameters for H-poor CSM ($\kappa$ = 0.2~cm$^2$~g$^{-1}$)}
 \tablehead{
 \colhead{$M_{\rm ej}$ and $E_{\rm ej}${$\rm ^a$}} & \colhead{Viewing angle} & \colhead{Fit parameters} & \colhead{\chisqr} & \colhead{CSM properties}
 } 
\startdata 
  0.80~\msol\ and 0.63 $\times$ 10$^{51}$ ergs & $\theta$ = 0\degr & $M_{\rm env}$ = 2.0 $\times$ 10$^{-3}$~\msol\ & 3.29 & $\rho_{\rm env}$ = 22~g~cm$^{-3}$ \\
  &    & $R_{\rm env}$ = 3.5 $\times$ 10$^{9}$~cm & & $M_{\rm CSM}$ ($\rho \propto r^{-3}$) = 0.0046~\msol\ \\
  &    & $t_{0}$ = $-$0.04 days &  \\
  &    & $t_{PL}$ = 0.19 days &  \\
  &    & $\alpha_{B, V, i}$ = (2.12, 1.81, 2.14) &  \\
  \hline
   & $\theta$ = 90\degr & $M_{\rm env}$ = 1.7 $\times$ 10$^{-3}$~\msol\ & 3.29 & $\rho_{\rm env}$ = 0.47~g~cm$^{-3}$ \\
  &    & $R_{\rm env}$ = 1.2 $\times$ 10$^{10}$~cm & & $M_{\rm CSM}$ ($\rho \propto r^{-3}$) = 0.0065~\msol\ \\
  &    & $t_{0}$ = $-$0.06 days &   \\
  &    & $t_{PL}$ = 0.19 days &  \\
  &    & $\alpha_{B, V, i}$ = (2.12, 1.82, 2.14) &  \\
  \hline
  1.05~\msol\ and 0.82 $\times$ 10$^{51}$ ergs & $\theta$ = 0\degr & $M_{\rm env}$ = 2.1 $\times$ 10$^{-3}$~\msol\ & 3.29 & $\rho_{\rm env}$ = 34~g~cm$^{-3}$ \\
  &    & $R_{\rm env}$ = 3.1 $\times$ 10$^{9}$~cm & & $M_{\rm CSM}$ ($\rho \propto r^{-3}$) = 0.0045~\msol\ \\
  &    & $t_{0}$ = $-$0.03 days &   \\
  &    & $t_{PL}$ = 0.19 days &  \\
  &    & $\alpha_{B, V, i}$ = (2.12, 1.81, 2.14) &  \\
 \hline
   & $\theta$ = 90\degr & $M_{\rm env}$ = 1.7 $\times$ 10$^{-3}$~\msol\ & 3.29 & $\rho_{\rm env}$ = 0.61~g~cm$^{-3}$ \\
  &   & $R_{\rm env}$ = 1.1 $\times$ 10$^{10}$~cm & & $M_{\rm CSM}$ ($\rho \propto r^{-3}$) = 0.0065~\msol\ \\
  &    & $t_{0}$ = $-$0.06 days &   \\
  &    & $t_{PL}$ = 0.20 days &  \\
  &    & $\alpha_{B, V, i}$ = (2.12, 1.82, 2.14) &  \\
\enddata
\tablenotetext{{\rm a}}{The two cases of $M_{\rm ej}$ and $E_{\rm ej}$ are the same as the ones used for ejecta-companion interaction in Table~\ref{tab:kas}.}
\tablecomments{$t_0$ and $t_{\rm PL}$ are in days since the epoch of first light (MJD~58206.00) in rest frame.}
\end{deluxetable*}
\label{tab:csm}

We primarily consider the case of H-poor CSM originating from a WD or He-star companion, using the electron scattering opacity of $\kappa$ = 0.2~cm$^2$~g$^{-1}$, since those are the most likely companions for \uname\ based on the constraints derived from the early light curves (Section~\ref{sec:kasan}) and nebular-phase spectra (Section~\ref{sec:neb}).
In this case, the SN explosion could occur after the merger of the binary or during an earlier stage of binary mass transfer \citep{Shen2015apj}.
The distribution of CSM initially after the merger and during earlier stages of mass transfer is expected to be equatorially-concentrated,
rather than spherically symmetric as assumed in P15, 
though the distribution can evolve towards spherical symmetry on a timescale of hours after the merger \citep{Guillochon2010apj, Pakmor2013apj, Schwab2012mnras}.

Table~\ref{tab:csm} presents the best-fit parameters obtained by 
fitting the early light curves of \uname\ during 0--8 days since first light, excluding the $B$-band light curve during 0--1 days, using two extreme cases of viewing angles, $\theta$ = 0\degr\ (equal to spherically symmetric case) and $\theta$ = 90\degr, and two cases of relatively small and large ejecta masses and kinetic energies for \uname\ following Section~\ref{sec:kasfit}.
Note that the reduced $\chi$-squared statistics of the four cases are nearly identical (\chisqr\ $\sim$ 3.3), indicating that the goodness of fit is very similar for the different cases. 
For the properties of the early excess emission in the fit, we allowed for ranges of $M_{\rm env}$ in 10$^{-5}$--10$^{-2}$~\msol\ and $R_{\rm env}$ in 10$^9$--10$^{13}$~cm that easily accommodate the obtained fit parameters.
Figure~\ref{fig:shockmod} compares the light and color curves of the best-fit ejecta-CSM interaction model obtained in the case of $M_{\rm ej}$ = 0.80~\msol, $E_{\rm ej}$ = 0.63 $\times$ 10$^{51}$~ergs, and $\theta$ = 0\degr\ (red dashed curves) to those of \uname\ and the two other best-fit models of surface \ni56\ heating (Section~\ref{sec:nipl}) and ejecta-companion interaction (Section~\ref{sec:kasfit}), where the goodness of fit is very similar for the three models.
As seen in the table, the difference of $t_{PL} - t_0 \sim$ 0.22--0.26 days obtained for the ejecta-CSM interaction model is near the lower extreme of the range obtained for the surface \ni56\ heating (0.54 days) and ejecta-companion interaction (0.28--0.60 days) models, consistent with there being a delay (or dark phase) of $\lesssim$ 1 day between the explosion and the onset of power-law rise in \uname\ (Paper I).

We examine whether the CSM mass, $M_{\rm env}$, required to fit the observed infant-phase excess emission
is compatible with the expectations of CSM after a merger (or ``post-merger CSM'').
The total CSM mass ($M_{\rm CSM}$) is not necessarily equal to $M_{\rm env}$
since the envelope represents only the outermost layer of CSM near $R_{\rm env}$ that dominates the ejecta-CSM interaction emission.
$M_{\rm CSM}\gtrsim M_{\rm env}$ in general, where $M_{\rm CSM}$ = $M_{\rm env}$ is for the case of entirely uniform-density CSM, and $M_{\rm CSM}/M_{\rm env}$ increases with the central-concentration of the CSM density distribution.
Adopting a $\rho \propto r^{-3}$ density distribution expected for post-merger CSM \citep{Piro&Morozova2016apj}, we obtain the following equation for $M_{\rm CSM}$ in terms of $M_{\rm env}$ and $R_{\rm env}$:
\begin{equation}
    M_{\rm CSM} \simeq 4\pi R_{\rm env}^3 \rho_{\rm env} \log{(R_{\rm env}/R_*)}
\label{eq:Mcsm}
\end{equation}
where $\rho_{\rm env} = 3 M_{\rm env} / 4\pi R_{\rm env}^3$ is the CSM density in the outermost layer (= envelope density) and $R_*$ is the progenitor radius.
$R_*$ is taken to be $\sim$ 6 $\times$ 10$^8$~cm, the expected shock breakout radius of \uname\ (Section~\ref{sec:sbo}).

Table~\ref{tab:csm} column 5 provides the derived CSM properties of $\rho_{\rm env}$ 
and $M_{\rm CSM}$ that would be implied by the fit parameters using Equation~\ref{eq:Mcsm}.
Overall, these properties appear to be incompatible with the theoretical expectations for post-merger CSM.
In simulations of violent merger, the post-merger CSM mass can be $\sim$ 0.1--0.7\,\msol\ depending on the companion mass \citep{Dan2014mnras}, which are much larger than $M_{\rm CSM}\lesssim$ 0.007~\msol\ required to fit the observed infant-phase excess emission.
The post-merger CSM radius is also expected to expand on short timescales, beginning from $\sim$ 10$^{10}$~cm during the merger and expanding to $\sim$ 10$^{11}$~cm in only a few hours after the merger \citep{Piro&Morozova2016apj}, becoming less compatible with the fitted CSM radii of $R_{\rm env} \lesssim$ 10$^{10}$~cm on the timescale of the infant-phase excess emission.
Thus, the emission is not likely to be from post-merger CSM.

We instead consider CSM of smaller mass and radius expected in ``pre-merger'' stages of binary mass transfer, before the WD or He-star companion is disrupted, for the origin of the infant-phase excess emission.
For example, in simulations of He-shell DDets from WD-WD mergers, $\lesssim$ 0.1~\msol\ of CSM is expected to be present at the time of explosion, which occurs before the merger is completed, distributed in a torus around the progenitor star.
The outermost layers of the torus are located at $\gtrsim$~10$^9$~cm where the CSM density is expected to be $\lesssim$~10$^3$~g~cm$^{-3}$ \citep{Guillochon2010apj, Pakmor2013apj}.
These pre-merger CSM properties are comparable to $R_{\rm env}$ = (3.1--3.5) $\times$ 10$^{9}$~cm and $\rho_{\rm env}$ = 22--34~g~cm$^{-3}$ obtained
for the cases with $\theta$ = 0\degr\ viewing angle (see Table~\ref{tab:csm}).
If the pre-merger CSM density distribution is similar to the post-merger case ($\rho \propto r^{-3}$), then the corresponding total CSM mass of $M_{\rm CSM}\sim$ 0.005~\msol\ is relatively small compared to the CSM masses expected in He-shell DDet simulations \citep[$\sim$ 0.05--0.10~\msol;][]{Guillochon2010apj}.
However, since $M_{\rm CSM}$ can be larger for steeper pre-merger CSM density distributions, 
ejecta interaction with pre-merger CSM remains possible for the origin of the observed infant-phase excess emission in \uname.

\subsubsection{Circumstellar Material from a Main Sequence or Subgiant Companion}\label{sec:accrete}

\begin{deluxetable*}{cclcl}
\tabletypesize{\footnotesize}
\tablecolumns{5} 
\tablewidth{0.99\textwidth}
 \tablecaption{Ejecta-CSM interaction model fit parameters for solar-composition CSM ($\kappa$ = 0.34~cm$^2$~g$^{-1}$)}
 \tablehead{
 \colhead{$M_{\rm ej}$ and $E_{\rm ej}${$\rm ^a$}} & \colhead{Viewing angle} & \colhead{Fit parameters} & \colhead{\chisqr} & \colhead{CSM properties}
 } 
\startdata 
 0.80~\msol\ and 0.63 $\times$ 10$^{51}$ ergs & $\theta$ = 0\degr\ & $M_{\rm env}$ = 1.3 $\times$ 10$^{-3}$~\msol\ & 3.30 & $\rho_{\rm env}$ = 5.3~g~cm$^{-3}$ \\
  &    & $R_{\rm env}$ = 4.9 $\times$ 10$^{9}$~cm & & $M_{\rm CSM}$ ($\rho \propto r^{-3}$) = 0.0035~\msol\ \\
  &    & $t_{0}$ = $-$0.03 days &  \\
  &    & $t_{PL}$ = 0.19 days &  \\
  &    & $\alpha_{B, V, i}$ = (2.12, 1.81, 2.14) &  \\
 \hline
   & $\theta$ = 90\degr\ & $M_{\rm env}$ = 1.1 $\times$ 10$^{-3}$~\msol\ & 3.28 & $\rho_{\rm env}$ = 0.19~g~cm$^{-3}$ \\
  &    & $R_{\rm env}$ = 1.4 $\times$ 10$^{10}$~cm & & $M_{\rm CSM}$ ($\rho \propto r^{-3}$) = 0.0045~\msol\ \\
  &    & $t_{0}$ = $-$0.05 days &   \\
  &    & $t_{PL}$ = 0.20 days &  \\
  &    & $\alpha_{B, V, i}$ = (2.12, 1.82, 2.14) &  \\
 \hline
 1.05~\msol\ and 0.82 $\times$ 10$^{51}$ ergs & $\theta$ = 0\degr\ & $M_{\rm env}$ = 1.4 $\times$ 10$^{-3}$~\msol\ & 3.30 & $\rho_{\rm env}$ = 7.8~g~cm$^{-3}$ \\
  &    & $R_{\rm env}$ = 4.4 $\times$ 10$^{9}$~cm & & $M_{\rm CSM}$ ($\rho \propto r^{-3}$) = 0.0035~\msol\ \\
  &    & $t_{0}$ = $-$0.03 days &   \\
  &    & $t_{PL}$ = 0.19 days &  \\
  &    & $\alpha_{B, V, i}$ = (2.12, 1.81, 2.14) &  \\
 \hline
   &  $\theta$ = 90\degr\ & $M_{\rm env}$ = 1.2 $\times$ 10$^{-3}$~\msol\ & 3.28 & $\rho_{\rm env}$ = 0.33~g~cm$^{-3}$ \\
  &    & $R_{\rm env}$ = 1.2 $\times$ 10$^{10}$~cm & & $M_{\rm CSM}$ ($\rho \propto r^{-3}$) = 0.0046~\msol\ \\
  &    & $t_{0}$ = $-$0.05 days &   \\
  &    & $t_{PL}$ = 0.20 days &  \\
  &    & $\alpha_{B, V, i}$ = (2.12, 1.82, 2.14) &  \\
\enddata
\tablenotetext{{\rm a}}{The two cases of $M_{\rm ej}$ and $E_{\rm ej}$ are the same as the ones used for ejecta-companion interaction in Table~\ref{tab:kas}.}
\tablecomments{$t_0$ and $t_{\rm PL}$ are in days since the epoch of first light (MJD~58206.00) in rest frame.}
\end{deluxetable*}
\label{tab:csmaccrete}

For the less likely case of a few solar mass main-sequence or subgiant companion in \uname\ (Section~\ref{sec:kasan}), we briefly consider the possibility of ejecta interaction with solar composition CSM from such companions, adopting the electron scattering opacity of $\kappa$ = 0.34~cm$^2$~g$^{-1}$.
Since those companions are mainly expected to trigger \tase\ via accretion \citep{Maoz2014araa}, the CSM is likely to be an equatorially-concentrated disk or toroid.
Table~\ref{tab:csmaccrete} shows the best-fit parameters obtained by 
fitting the early light curves of \uname\ using the aforementioned two extreme cases of viewing angles for equatorially-concentrated CSM, $\theta$ = 0\degr\ for equatorial viewing angle and 90\degr\ for polar viewing angle, and two cases of relatively small and large ejecta masses and kinetic energies for \uname. 
For uniform-density CSM, the total CSM mass is expected to be $M_{\rm CSM}$ = $M_{\rm env}$, ranging in 0.0011--0.0014~\msol, while the total CSM mass expected for the relatively steep CSM density distribution of $\rho \propto r^{-3}$ ranges in 0.0035--0.0046~\msol\ (Table~\ref{tab:csmaccrete} column 5 based on Equation~\ref{eq:Mcsm}).
Overall, if \uname\ was triggered by accretion from a few solar mass main-sequence or subgiant companion, then the observed infant-phase excess emission can be produced by ejecta interaction with a $\sim$ 0.001--0.005~\msol\ accretion disk near $\sim$ (4--14) $\times$ 10$^{9}$~cm.

\subsection{Search for Shock Breakout}\label{sec:sbo}

Shock breakout is expected to occur shortly after a SN explosion 
when the outgoing shockwave breaks through the surface of the progenitor star \citep{Piro2010apj, Nakar2010apj}.
Early observations of \tase\ have been used to search for evidence of shock breakout based on the expected thermal emission from the shock-heated envelope.
However, since the luminosity of the shock breakout emission scales with the radius of the progenitor star, this emission has not yet been observed in \tase\ due to the small size of WDs.
The non-detection of shock breakout emission in early \tase\ has been used to constrain the radius of the SN progenitor, e.g., in the case of SN~2011fe to be $\lesssim$ 0.02~\rsol\ \citep{Bloom2012apj}.

We investigate the origin of the infant-phase excess emission by comparing the observed early light curves of \uname\ with what is expected from shock breakout emission.
Adopting the model of \citet{Piro2010apj}, which assumes an approximately spherically-symmetric explosion and a radial shock acceleration law,
the shock breakout emission luminosity ($L_{\rm SBO}$) and temperature ($T_{\rm SBO}$) are determined by
the ejecta mass ($M_{\rm ej}$) and the radius of the progenitor star at the time of shock breakout ($R_{\rm SBO}$) as follows:
\begin{align}
    L_{\rm SBO}(t) &= 7^{-4/3} \times 2 \times 10^{40} (g_9/K_{13})^{-0.41} V_9^{1.9} \rho_6^{0.36} R_{8.5}^{0.83} t_4^{-0.16}\ {\rm erg s}^{-1}  
\label{eq:Lsbo}\\
    T_{\rm SBO}(t) &= 7^{-1/3} \times 2 \times 10^{4} (g_9/K_{13})^{-0.058} V_9^{0.030} \rho_6^{0.0058} R_{8.5}^{0.11} t_4^{-0.44}\ {\rm K}
\label{eq:Tsbo}
\end{align}
where the first terms are correction factors to fix the improper scalings \citep{Bloom2012apj}, $g_9 \propto M_{\rm ej}/R_{\rm SBO}^2$ represents surface gravity, $K_{13} = K/(10^{13}\ {\rm cgs})$ represents the non-relativistic degenerate equation of state constant ($K$) with $\mu_e \sim 2$ for C+O WDs, $V_9 \sim 0.6$ and $\rho_6 \sim 2 g_9^{0.11}$ represent the shock velocity and density, respectively, $R_{8.5} = R_{\rm SBO}/(10^{8.5}\ {\rm cm})$, and $t_4 = (t-t_0)/(10^4\ {\rm s})$ represents time since the epoch of explosion ($t_0$).

We fit the early light curves of \uname\ during 0--8 days since first light (excluding the $B$-band light curve during 0--1 days) as a combination of power-law (for the underlying SN emission) and shock breakout emission (for the infant-phase excess emission). 
The best-fit shock breakout radius is $R_{\rm SBO}$ = (3.5--3.7) $\times$ 10$^9$~cm, or 0.050--0.053~\rsol, where the lower and upper limits of the range represent the results obtained using two cases of relatively small and large values of $M_{\rm ej}$ (= 0.80 and 1.05~\msol), respectively, for \uname\ following the methods of Sections~\ref{sec:kasfit} and \ref{sec:csm}.

This range of fitted $R_{\rm SBO}$ is larger than the $R_{\rm SBO}$ that can be reasonably expected for the progenitor of \uname, indicating that shock breakout is unlikely to be the origin of the observed infant-phase excess emission.
For a typical C+O WD with mass $\sim$ 1.0~\msol, the shock breakout radius after possible expansion due to a deflagration phase is expected to be $\sim$ 6~$\times$~10$^{8}$~cm \citep{Piro2010apj}.
While explosion asymmetry can lead to a factor of $\lesssim$ 2 difference in the inferred $R_{\rm SBO}$, corresponding to the expected range of angular variation of shock breakout luminosity for a compact progenitor \citep{Afsariardchi2018apj}, an unreasonably intense deflagration phase is still be required to achieve such a large radius as the fitted $R_{\rm SBO}$ = (3.5--3.7)~$\times$~10$^{9}$~cm. 
Moreover, explosion mechanisms dominated by deflagration typically leave substantial amounts of unburnt carbon \citep{Nomoto1984apj}, while no C spectral features are seen in \uname\ (Section~\ref{sec:carbon}).
For the case of a He-shell DDet origin, which lacks a deflagration phase, the radius of the best-fit He-shell DDet model progenitor for \uname---a 1.05~\msol\ C+O WD with a 0.01~\msol\ He-shell---is only 5.14~$\times$~10$^8$~cm (Paper I).
In this case, while the He-shell can generally be expected to undergo some shock-driven expansion during the He-shell DDet process, the fitted $R_{\rm SBO}$ of (3.5--3.7)~$\times$~10$^{9}$~cm is unrealistic for the best-fit He-shell DDet model progenitor due to the extremely small He-shell mass.

\section{Nebular-Phase Evolution: Constraints on the Progenitor System and Explosion Mechanism}\label{sec:nebea}

\subsection{Nebular-Phase Light Curves}\label{sec:neblcev}

Figure~\ref{fig:neblc} compares the evolution of \uname\ light curves in $BVi$ (blue, green, and red circles) bands from the beginning to the nebular phase with those of normal \tas~2011fe\footnote[1]{The $I$-band magnitudes of SN~2011fe were converted to $i$ band by subtracting $-2.5\log_{10}(3631~{\rm Jy} /2416~{\rm Jy})$.} \citep[dashed lines;][]{Munari2013newa,Tsvetkov2013coska} that have been scaled to match 
the peak absolute magnitude, \mb\ = $-$19.32 mag, and post-peak decline rate, \dm15 = 1.12 mag,
of \uname.
The light curves of SNe~2018aoz and 2011fe show a good agreement overall throughout their evolution, especially in the $B$-band,
confirming that the evolution of \uname\ in the nebular phase continues to match those of normal \tase.
The nebular-phase light curves of \uname\ decline linearly at rates of 0.0131 $\pm$ 0.0001, 0.0129 $\pm$ 0.0001, and 0.0083 $\pm$ 0.0003 mags~day$^{-1}$ in the $B$, $V$, and $i$ bands, respectively, with $BVi$-averaged decline rate of 0.0127 mags~day$^{-1}$.
For comparison, we measure the light curve decline rates of SN~2011fe 
during the nebular phase to be 
0.0134, 0.0138, 0.0099 mags~day$^{-1}$
in the $B$, $V$, and $i$ bands, respectively.

\subsection{Nebular-Phase Spectra}\label{sec:nebspecev}

Figure~\ref{fig:nebspec} shows the identification of nebular-phase [\feiii]~$\lambda$4658~\AA\ and [\coiii]~$\lambda$5888~\AA\ features in \uname\ (red and green vertical regions, respectively), whose flux ratio is associated with the evolution of \ni56\ $\rightarrow$ \co56\ $\rightarrow$ \fer56\ radioactive decay in \tase\ \citep{Kuchner1994apj}, as well as that of a double-peaked feature near 7290~\AA\ (blue vertical region).
The nebular-phase 7290~\AA\ feature of \tase\ can be from the [\caii] $\lambda$7291, 7323~\AA\ doublet, [\feii] $\lambda$7155~\AA, [\niii] $\lambda$7378~\AA, or some combination thereof \citep{Polin2021apj, Flors2020mnras}.
The double-peaked 7290~\AA\ feature observed in \uname, as well as most normal events \citep[e.g., SN~2011fe;][]{Mazzali2015mnras}, is most likely dominated by [\feii] and [\niii] emission since the [\caii] feature would not be resolved as a doublet at typical \tas\ velocities \citep{Polin2021apj}.
For each of the four nebular-phase spectra of \uname, Table~\ref{tab:nebflux} provides the measured fluxes of [\feiii], [\coiii], and the 7290~\AA\ feature ([\feii] $+$ [\niii]), as well as the ratios between the fluxes of the 7290~\AA\ feature to those of [\feiii]~$\lambda$4658~\AA, called ``7290~\AA/[\feiii]'' hereafter.
To obtain uncertainties in the fluxes, we estimated noise levels by smoothing each spectrum with a second-order Savitsky–Golay filter with a width of 150~\AA, which is $\lesssim$ 1/4 of the feature widths.
The average 7290~\AA/[\feiii] ratio of 0.149 $\pm$ 0.007 between 120 and 320 days since peak for \uname\ is near the lower extreme of what has been found in normal \tase\ in the range of $\sim$ 0.1--1.0 \citep{Polin2021apj}.

\begin{deluxetable*}{ccccc}
\tabletypesize{\footnotesize}
\tablecolumns{4} 
\tablewidth{0.99\textwidth}
 \tablecaption{Nebular-phase broad emission line fluxes}
 \tablehead{
 \colhead{Phase$\rm ^a$} & \colhead{$[$\feiii$]$} & \colhead{$[$\coiii$]$} &
 \colhead{[\feii] $+$ [\niii]} &
 \colhead{7290~\AA/[\feiii]$\rm ^b$}\\ 
 \colhead{} & \colhead{$\lambda$~4658~\AA} & \colhead{$\lambda$~5888~\AA} &
 \colhead{$\lambda$~7155~\AA, 7378~\AA} &
 \colhead{}\\ 
 \colhead{} & \colhead{($10^{-14}$ erg s$^{-1}$ cm$^{-2}$)} & \colhead{($10^{-14}$ erg s$^{-1}$ cm$^{-2}$)} & \colhead{($10^{-14}$ erg s$^{-1}$ cm$^{-2}$)} & \colhead{}
 } 
\startdata 
 $+$259.4 & 10.934 $\pm$ 0.010 & 1.347 $\pm$ 0.003 & 1.710 $\pm$ 0.003 & 0.1564 $\pm$ 0.0003 \\
 $+$277.3 & N/A & 1.029 $\pm$ 0.003 & 1.443 $\pm$ 0.002 & N/A \\
 $+$296.4 & 7.629 $\pm$ 0.009 & 0.632 $\pm$ 0.003 & 1.086 $\pm$ 0.002 & 0.1424 $\pm$ 0.0003 \\
 $+$382.5 & 2.658 $\pm$ 0.005 & 0.245 $\pm$ 0.002 & 0.580 $\pm$ 0.002 & 0.2181 $\pm$ 0.0008 \\
\enddata
\tablenotetext{{\rm a}}{Phases are measured in observer frame days since $B$-band maximum.}
\tablenotetext{{\rm b}}{The ratio of the flux of the 7290~\AA\ feature to that of [\feiii] \citep{Polin2021apj}}
\end{deluxetable*}
\label{tab:nebflux}

Figure~\ref{fig:nebspec} also shows the expected positions of
narrow emission lines of H, He, and O near H$\alpha$ $\lambda$6563~\AA, \hei~$\lambda$5875, 6678~\AA, and [\oi]~$\lambda$6300, 6364~\AA, respectively.
In \tase, these low-velocity lines may be produced by swept-up material from the companion \citep[e.g.,][]{Kollmeier2019mnras} or CSM, including disrupted companion material following a violent merger \citep{Kromer2013apj, Mazzali2022mnras, Tucker2022apj}.
All of the lines appear to be absent in \uname, similar to most other \tase\ \citep{Mattila2005aap, Leonard2007apj, Shappee2013apj, Maguire2016mnras, Tucker2020mnras}, which argues against the presence of a substantial amount of swept-up material (see below).
By injecting synthetic emission lines of H and He with a FWHM = 1000~km~s$^{-1}$ into the observed nebular spectra, modelling
Doppler shifts from the rest wavelength of up to $\pm$1000~km~s$^{-1}$, and estimating the rms of the local spectral region following the methods of \citet{Sand2018apj, Sand2019apj}, we find 3$\sigma$ flux upper limits for the H$\alpha$ and \hei\ lines.
We do the same for [\oi], but using FWHM = 2000~km~s$^{-1}$ and up to $\pm$2000~km~s$^{-1}$ Doppler shifts that can be expected for [\oi] \citep{Taubenberger2013apj}.
Table~\ref{tab:neblims} presents the measured upper limits and their corresponding luminosities.

\begin{deluxetable*}{cccccc}
\tabletypesize{\footnotesize}
\tablecolumns{6} 
\tablewidth{0.99\textwidth}
 \tablecaption{Nebular-phase emission line flux and luminosity limits}
 \tablehead{
 \colhead{Line} & \colhead{Phase$\rm ^a$} & \colhead{Flux Limit} & \colhead{Luminosity Limit} & \colhead{Mass Limit$\rm ^b$} & \colhead{Mass Limit$\rm ^c$} \\ 
 \colhead{} & \colhead{} & \colhead{($10^{-17}$ erg s$^{-1}$ cm$^{-2}$)} & \colhead{($10^{36}$ erg s$^{-1}$)} & \colhead{($10^{-4}$ \msol)} & \colhead{($10^{-4}$ \msol)}
 } 
\startdata 
H$\alpha$~$\lambda$6563~\AA\ & $+$259.4 & 5.4 & 3.2 & 4 & 10--16  \\
 & $+$277.3 & 5.0 & 3.0 & 4 & 12--18 \\
 & $+$296.4 & 6.3 & 3.8 & 5 & 16--24 \\
 & $+$382.5 & 7.7 & 4.6 & 9 & 100--130\\
\hline
\hei~$\lambda$5875~\AA\ & $+$259.4 & 10.0 & 6.0 & 25 & \\
 & $+$277.3 & 7.8 & 4.7 & 25 & \\
 & $+$296.4 & 9.0 & 5.4 & 33 & \\
 & $+$382.5 & 17.6 & 10.5 & 104 & \\
\hline
\hei~$\lambda$6678~\AA\ & $+$259.4 & 5.4 & 3.2 & 18 &  \\
 & $+$277.3 & 6.7 & 4.0 & 25 &  \\
 & $+$296.4 & 10.5 & 6.3 & 43 &  \\
 & $+$382.5 & 7.7 & 4.6 & 55 &  \\
\hline
$[$\oi$]$~$\lambda$6300~\AA\ & $+$259.4 & 16.9 & 10.1 &  &  \\
 & $+$277.3 & 17.6 & 10.5 &  &  \\
 & $+$296.4 & 21.6 & 13.0 &  &  \\
 & $+$382.5 & 25.8 & 15.5 &  &  \\
 \hline
$[$\oi$]$~$\lambda$6364~\AA\ & $+$259.4 & 16.9 & 10.1 &  &  \\
 & $+$277.3 & 21.1 & 12.6 &  &  \\
 & $+$296.4 & 21.6 & 13.0 &  &  \\
 & $+$382.5 & 25.8 & 15.5 &  &  \\
\enddata
\tablenotetext{{\rm a}}{Phases are measured in observer frame days since $B$-band maximum.}
\tablecomments{All implanted lines have peak fluxes corresponding to three times the local rms with a FWHM = 1000 km s$^{-1}$, except for the \hei~$\lambda$5875~\AA\ line, where we used a peak flux of four times the local rms. We infer upper limits on the mass of the emitting elements based on the luminosity limits and the model predictions of $\rm ^b$\citet{Botyanszki2018apj} and $\rm ^c$\citet{Dessart2020aa}.}
\end{deluxetable*} 
\label{tab:neblims}

\subsection{Constraints on Non-Degenerate Companion and Circumstellar Material}\label{sec:neb}

We now place constraints on the presence of a
non-degenerate companion or CSM from the violent merger case of double-degeneracy in the progenitor of \uname\ based on the absence of predicted emission lines from their unburned swept-up material in the observed nebular-phase spectra.
The ejecta of \tase\ are expected to strip/ablate $\sim$ 0.1~\msol\ of H- or He-rich materials from a non-degenerate companion \citep[e.g.,][]{Botyanszki2018apj}, while in the case of a violent merger, $\sim$ 0.1--0.7~\msol\ of H-poor CSM composed of O or He can be expected depending on the mass and composition of the companion WD \citep{Dan2014mnras}.
Multiple spectral synthesis studies have shown that even trace amounts ($\sim 10^{-3}$\,\msol) of low-velocity H will lead to observable nebular-phase H$\alpha$ emission \citep{Mattila2005aap, Botyanszki2018apj, Dessart2020aa}.
For H-poor material, some recent studies \citep{Dessart2020aa} find that $\lesssim$ 0.2~\msol\ of O or He could be hidden by metal line-blanketing while other studies find that even very small amounts, $\sim$ 0.05~\msol\ of O \citep{Mazzali2022mnras} or $\sim$ 10$^{-3}$~\msol\ of He \citep{Botyanszki2018apj}, would be observable.

In columns 5 and 6 of Table~\ref{tab:neblims}, we use the models of \citet{Botyanszki2018apj} and \citet{Dessart2020aa}, respectively,
to obtain upper limits on the masses of low-velocity H and He based on the observed upper limits on the luminosities of their lines.
The range of values in column 6 correspond to the range of mass upper limits derived for the set of delayed-detonation as well as sub-Chandrasekhar-mass explosion models from \citet{Dessart2020aa}. 
As seen in the columns, even the most conservative upper limits on H and He masses in \uname\ permit less than 1.3 $\times$ 10$^{-2}$~\msol\ and 1.0 $\times$ 10$^{-2}$~\msol\ of each, respectively, making the presence of even a trace amount of H extremely unlikely, as supported by both models, while \citet{Botyanszki2018apj} would also exclude the presence of significant He.
Overall, the nebular-phase spectra of \uname\ disfavour the presence of a non-degenerate companion and, to a lesser extent, the large mass of H-poor CSM expected from a violent merger.
The presence of a main sequence or red giant companion is especially disfavoured by the H constraints based on both \citet{Botyanszki2018apj} and \citet{Dessart2020aa}, while the He constraints based on \citet{Botyanszki2018apj} would even disfavour a naked He-star companion.

\subsection{Constraints on Sub-Chandrasekhar-Mass and Asymmetric Explosion Mechanisms}\label{sec:nebex}

Nebular-phase spectra of \tase\ can also offer constraints on the explosion mechanism and geometry.
In particular, the strength of [\caii] emission near 7290~\AA\ can be linked to the mass of the progenitor WD.
Pure detonations of low-mass WDs are expected to undergo incomplete burning due to their lower density \citep{Polin2021apj}, leading to the production of Ca mixed with other Fe-peak elements, which then cool efficiently via [\caii] in the nebular phase \citep{Polin2021apj, Hoeflich2021apj}.
As explained above, the double-peaked 7290~\AA\ features seen in \uname\ and most normal \tase\ (e.g., SN~2011fe) are likely dominated by [\feii] and [\niii], indicating weak [\caii] emission.
This may imply the explosion of a relatively massive WD.
For instance, a 1.26~\msol\ WD explosion can produce weak and double-peaked emission near 7290~\AA\ as seen in SN~2011fe \citep{Mazzali2015mnras}.
Comparisons to the models of \citet{Polin2021apj} also support a $\gtrsim$ 1.2~\msol\ WD explosion for the weak emission seen in \uname\ (see Figure~\ref{fig:ddetpolin}).
In this case, however, reconciling the relatively large total mass with the estimated ejecta mass of \uname\ based on 1-D modelling of its fast-rising light curves (0.8--1.05~\msol; Paper I) may rely on viewing angle effects and explosion asymmetry, which can be caused by tidal heating or a surface He-shell detonation \citep{Iben1994apj, Iben1998apj}.

\begin{figure}[t!]
\epsscale{\scl}
\plotone{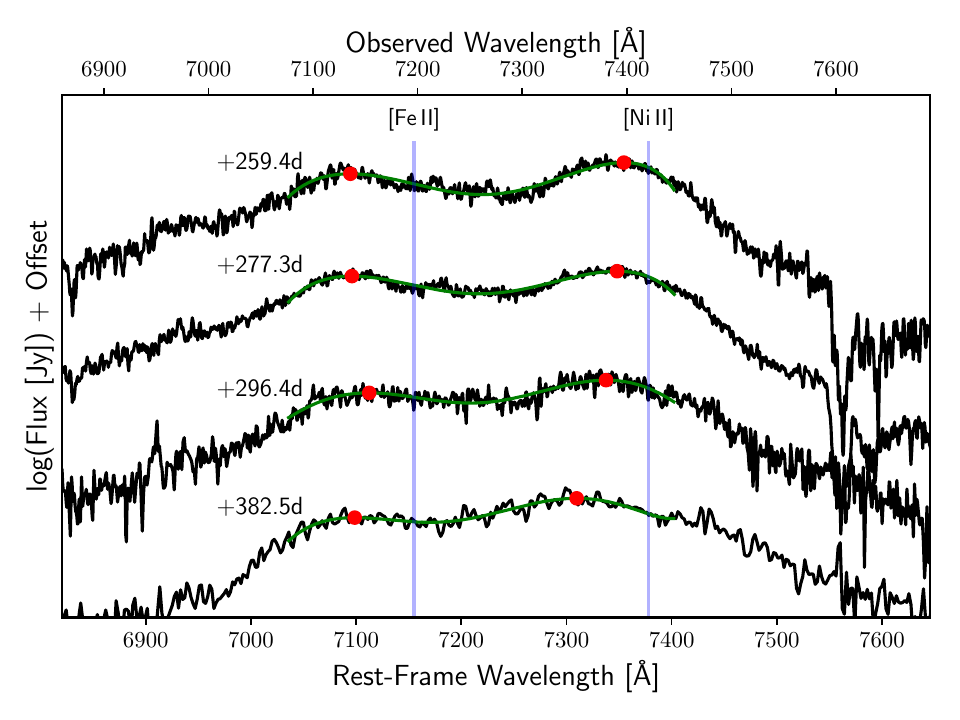}
\caption{The dereddened nebular-phase spectra of \uname\ (black) zoomed in on the [\feii] and [\niii] features whose rest-frame wavelengths are indicated with blue vertical lines.
The red circles are the peaks of them emission features obtained by fitting polynomials (n=12; green curves) to the spectra.
\label{fig:nebvel}}
\end{figure}

Such an explosion asymmetry may leave imprints on other lines in the nebular phase. 
In particular, the [\feii]~$\lambda$7155~\AA\ and [\niii]~$\lambda$7378~\AA\ emission features seen in normal \tase\ are expected to be Doppler shifted as a result of the motion of the ejecta core in an asymmetric explosion mechanism \citep{Maeda2010apj, Maeda2010natur, Li2021apj}.
Note that these two emission features are dominated by single line transitions (with minor contribution from the weak [\feii]~$\lambda$7171~\AA\ line), whereas other emission features of Fe-peak elements, including [\feiii] and [\coiii], have contributions from a blend of broad absorption features with wavelength separations smaller than their typical line widths \citep{Maeda2010apj, Flors2020mnras}.
Thus, shifts in the central wavelengths of [\feiii] and [\coiii] are not solely attributable to Doppler velocity.
Figure~\ref{fig:nebvel} shows that the peak wavelengths of the observed [\feii] and [\niii] features are blueshifted in the nebular-phase spectra of \uname, corresponding to average velocities of ($-$2.3 $\pm$ 0.3) and ($-$1.6 $\pm$ 0.7) $\times$ 10$^3$ km~s$^{-1}$, respectively.
These are among the most blueshifted velocities reported for those features in \tase\ \citep{Li2021apj, Maeda2010natur}.
For both asymmetric Chandrasekhar-mass explosion mechanisms \citep{Maeda2010apj} and double-detonations \citep{Boos2021apj}, the observed velocities of Fe and Ni in \uname\ point to a viewing angle where the primary component of the ejecta core is approaching the observer.
However, we note that asymmetric explosions only provide faster-rising light curves from viewing angles where the ejecta core is \emph{receding} from the observer \citep[see][]{Shen2021apj, Boos2021apj} since those directions provide higher ejecta velocities and lower densities, which leads to shorter diffusion times.
Thus, asymmetric effects alone are unable to reconcile the fast-rising light curves with a relatively large total ejecta mass.
We discuss possible explanations for these features in Section~\ref{sec:asym}.

\section{Comparison to He-shell Double-Detonation Models}\label{sec:hedd}

The $B$-band plateau and rapid redward \bv\ color evolution of \uname\ within the first $\sim$ 1 day post-explosion have been attributed to the presence of an over-density of Fe-peak elements in the outer 1\% of the SN-ejected mass (Paper I).
In addition, the relatively short 15.3-day rise-time of \uname\ among normal \tase\ indicates that \uname\ either (1) was a spherically symmetric explosion with a total ejecta mass of $\sim$0.8--1.0 \msol, which is significantly smaller than the Chandrasekhar mass of $\sim$ 1.4~\msol, or (2) was an asymmetric explosion.
Among the proposed explanations for the distribution of Fe-peak elements in the outer ejecta of \uname---off-center deflagration, gravitationally confined detonation, and He-shell DDet---only He-shell DDet is compatible with a sub-Chandrasekhar total ejecta mass \citep{Kromer2010apj, Woosley2011apj, Polin2019apj}.
Below we examine the compatibility between the other observed properties of \uname\ and a set of 1-D He-shell DDet model predictions.

\subsection{He-shell Double-Detonation Simulations}\label{sec:heddsim}

For our comparisons, we primarily use the set of thin He-shell DDet models from Paper I with core C+O WD and He-shell masses ranging in 1.00--1.10~\msol\ and 0.01--0.012~\msol, respectively, created following the methods of \citet{Polin2019apj}.
The modelling process involves two stages. First we perform hydrodynamic simulations with full nucleosynthesis using Castro, a compressible hydrodynamics code built on the AMReX framework \citep{Almgren2010apj, Zingale2018jphc}. After the SN ejecta has reached homologous expansion we perform radiative transport calculations with Sedona \citep{Kasen2006bapj} in order to produce synthetic light curves and spectra of our models.
The only way our methods differ from the \citet{Polin2019apj} study is that we begin our radiative transport simulations earlier than the previously published models (beginning at 0.1 days instead of 0.25 days) in order to model the natal epochs observed in \uname.
In Paper I, we found that the model with a 1.05 M$_\odot$ WD $+$ 0.01 M$_\odot$ He-shell provided the best-fit to the early (0--8 days since first light) \bv\ color and near-peak $BVi$ luminosity evolution of \uname. Here, we provide an expanded comparison between these models and both the infant-phase and near-peak properties of the SN. We supplement these with comparisons to the predictions for the nebular-phase emission line ratios of the sub-Chandrasekhar-mass He-shell DDet models from \citet{Polin2021apj}.

\subsection{Comparison to Infant-Phase Evolution}\label{sec:heddearly}

\begin{figure*}[t!]
\epsscale{\scl}
\plotone{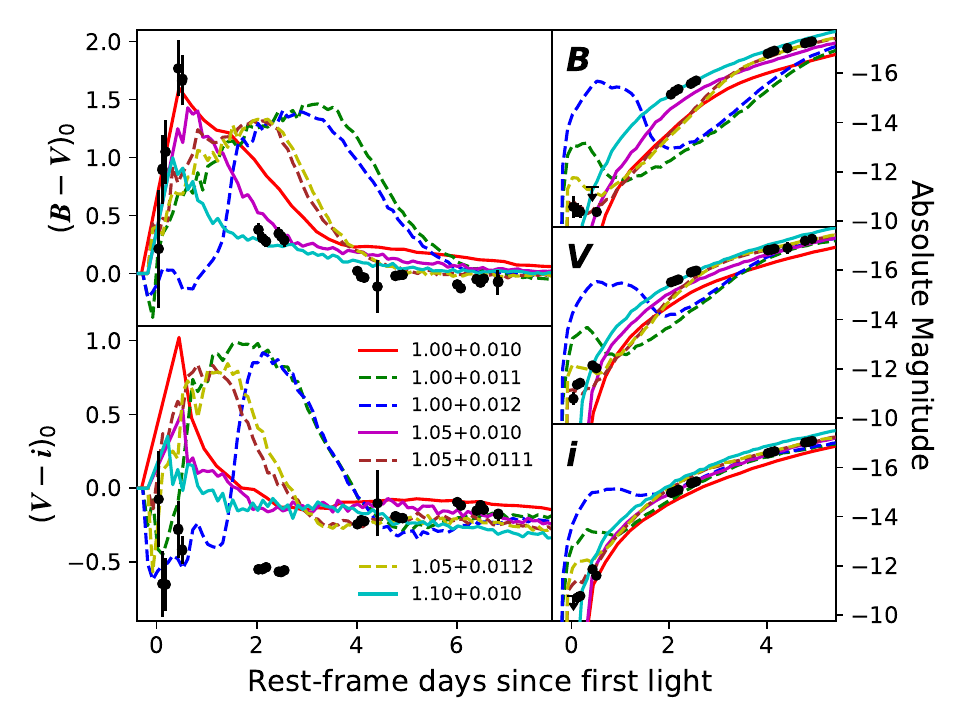}
\caption{The early observations of \uname\ (black circles), including the dereddened color (left panels) and light (right panels) curves in rest frame, are compared to the outcomes of the He-shell DDet simulations (colored curves) labelled with WD mass $+$ He-shell mass in \msol.
The inverted arrow (top-right) is a detection limit at a S/N of 2.
\label{fig:ddetearly}}
\end{figure*}

Figure~\ref{fig:ddetearly} compares the observed early color 
and light curves (filled black circles) of \uname\ with those (colored curves)
predicted by the He-shell DDet models during the first 5 days.
As noted in Paper I, the 1.05~\msol\ WD + 0.010~\msol\ He-shell model (magenta curves) provides the best-fit to the early \bv\ color evolution of \uname, including the timing and magnitude of the rapidly evolving NRB-phase color.
All of the models, on the other hand, 
poorly fit the observed \vi\ color, significantly over-predicting the amount of reddening observed. 
This discrepancy could be due to either (1) a line effect, such as differences in the modelled and observed strength of \caii\ features in the $I$-band around 8000~\AA\ due to Ca produced by the initial He-shell detonation \citep{Polin2021apj}, or (2) a continuum effect, such as differences in the color temperature that are influenced by the radioactive heating rate (Section~\ref{sec:heddexp}).

In addition, none of the current suite of models can fit the early light curves entirely during the first 5 days (right panels), although some can reproduce the observed light curve evolution of the SN at various phases. 
From 0.5 days onward, the cyan and magenta models (1.10 and 1.05~\msol\ WDs with 0.010~\msol\ He-shells) provide the best-fits, although both models significantly under-predict the observed luminosity over the earliest $\lesssim$ 0.5 days.
On the one hand, as noted in Paper I, it is only necessary to add an additional 1.1 $\times$ 10$^{-3}$~\msol\ of He to the best-fit model to match the $\lesssim$ 0.5 day luminosity, as shown by the brown curves (1.05~\msol\ WD + 0.0111~\msol\ He-shell), demonstrating the high sensitivity of model predictions to the He-shell mass.
However, this model provides a worse fit to the subsequent light curves during 0.5--5 days, the timing and duration of the early color evolution (left panels), and near-peak observations (see Section~\ref{sec:heddpeak} below) of \uname\ than the magenta model.
The implications of these discrepancies are discussed in Section~\ref{sec:heddexp} below.

\subsection{Comparison to Maximum-Light Properties}\label{sec:heddpeak}

\begin{figure}[t!]
\epsscale{\scl}
\plotone{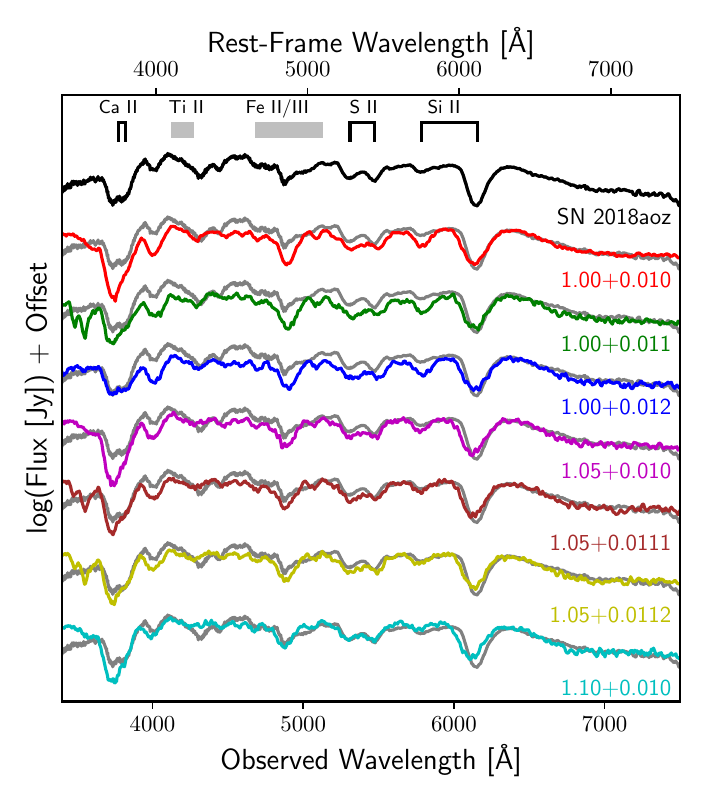}
\caption{The dereddened spectrum of \uname\ (black and grey curves) from 1.9 days before $B$-band maximum in the rest frame ($\sim$ 13.9 days since explosion) is compared to the outcomes of the same-colored
He-shell DDet simulations of Figure~\ref{fig:ddetearly} from the nearest post-explosion phase for various WD and He-shell masses as labelled (WD mass $+$ He-shell mass in \msol).
For clear comparison, the simulated spectra have been smoothed by box-car convolution, resulting in effective spectral resolutions of $R$ = 300.
\label{fig:ddetlate}}
\end{figure}

Figure~\ref{fig:ddetlate} compares the near-peak spectrum of \uname\ those of the He-shell DDet models.
All of the models predict clear absorption features in the vicinity of the observed \siii, \sii, \fex, and \caii\ features in \uname, as labelled at the top of the left panel.
The models with 1.05~\msol\ WD mass provide the best match to the observed spectral features.

\begin{figure}[t!]
\epsscale{\scl}
\plotone{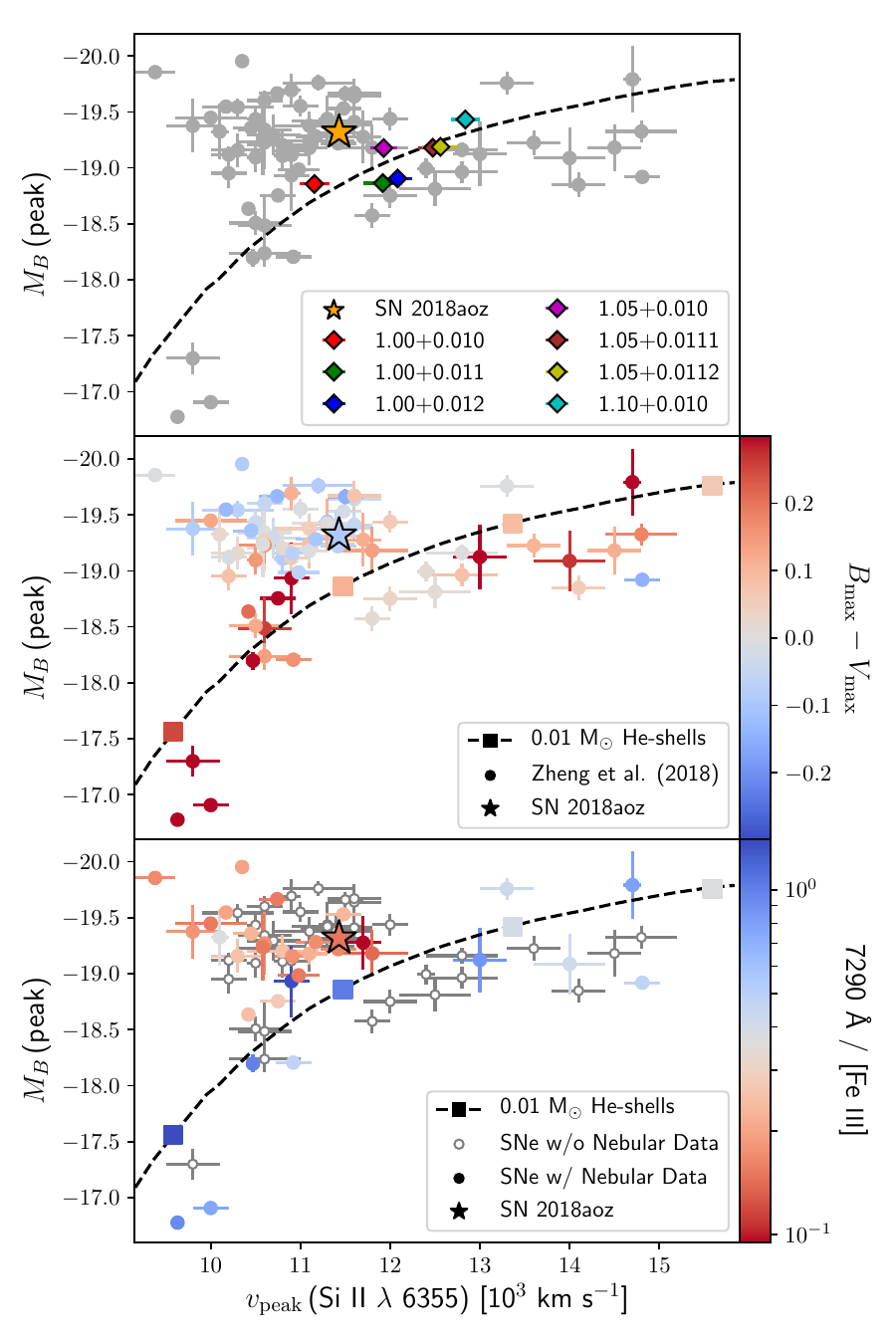}
\caption{
The distribution of $M_B$ (peak) vs. $v_{\rm peak}$ (\siii) of \uname\ (filled stars) and the outcomes of our He-shell DDet simulations (filled diamonds with the same colors as the models from Figures~\ref{fig:ddetearly} and \ref{fig:ddetlate}) compared with that of other SNe (circles) and He-shell DDet models from \citet[][top and middle panel]{Polin2019apj} and \citet[][bottom panel]{Polin2021apj},
where $M_B$ (peak) and $v_{\rm peak}$ (\siii) represent the peak $B$-band absolute magnitude and \siii\ velocity, respectively.
(Note that the two parameters of the other SNe used in \citealt{Polin2019apj,Polin2021apj} 
are originally from  \citealt{Zheng2018apj}.)
The dashed curves represent the relationship between the two parameters predicted by the He-shell DDet models (filled squares in the middle and bottom panels) of \citet{Polin2019apj} with 0.01~\msol\ He-shells.
The colors of the data points in the middle panel correspond to $B_{\rm max} - V_{\rm max}$ magnitudes (right-hand y-axis), while those of the filled data points in the bottom panel do the nebular-phase 7290\AA/[\feiii] line ratio. 
The SNe with large $B_{\rm max} - V_{\rm max}$ magnitudes (red-colored circles in middle panel)
and large 7290\AA/[\feiii] ratios (blue-colored circles in bottom panel) 
scattered near the dashed lines have been suggested to be produced by He-shell DDets \citep{Polin2019apj, Polin2021apj}.
\label{fig:ddetpolin}}
\end{figure}

Figure~\ref{fig:ddetpolin} \citep[top panel; adapted from][Figure 11]{Polin2019apj} compares both the peak luminosity and ejecta velocity, measured by the peak $B$-band magnitude and \siii\ velocity, of \uname\ to those of the He-shell DDet models along with other \tase\ and previously published He-shell DDet models from \citet{Polin2019apj}.
The cyan and red models provide the best matches to the observed peak $B$-band magnitude and ejecta velocity of \uname\ (orange star), respectively, while the magenta model with 1.05~\msol\ WD $+$ 0.010~\msol\ He-shell provides the closest match when both features are simultaneously considered.
However, we note that there is some separation between \uname\ and the models.
In particular, \citet{Polin2019apj} identified two broad populations of SNe within this plot: an apparent clustering at $v_{\rm peak}$~(\siii) $\sim$ 11,000~km~s$^{-1}$ and high peak magnitudes, and a non-clustered population with a tail extending to higher velocities. 
The former is mainly composed of CN and 91T-like \tase, while the latter is composed of BL and 91bg-like events (see Figure~\ref{fig:class-parrent}, bottom panel).
As noted by \citet{Polin2019apj}, the He-shell DDet models exhibit a peak brightness and velocity relationship (black dashed line) that generally follows the BL/91bg-like tail.
Several key predicted features of He-shell DDet are also prevalent among SNe from the BL/91bg-like tail, 
including sub-Chandrasekhar inferred ejecta masses \citep{Scalzo2019mnras} and lack of C spectral features \citep{Maguire2014mnras}, consistent with a He-shell DDet origin for them \citep{Polin2019apj}.
For \uname, the measured values of $M_B$~(peak) = $-$19.319 $\pm$ 0.009 mag and $v_{\rm peak}$~(\siii) = (11.43 $\pm$ 0.12) $\times$ 10$^3$ km~s$^{-1}$ place it close to the boundary between the CN/91T-like cluster and the BL/91bg-like tail populations, consistent with its intermediate nature between CN and BL (Section~\ref{sec:class}), though it is more similar to events from the CN/91T-like cluster overall.

While \uname\ could simply be an edge case between these two populations, in some other maximum-light features there is even less agreement between \uname\ and the He-shell DDet models.
In Figure~\ref{fig:ddetpolin} (middle panel), we add \uname\ to Figure 12 from \citet{Polin2019apj}, which plots peak $B$-band magnitude, \siii\ velocity, and \bv\ ``color'', $B_{\rm max} - V_{\rm max}$, for a population of \tase\ (circles) and He-shell DDet models (squares).
\citet{Polin2019apj} noted that most objects from the BL/91bg-like tail population in this plot of peak $B$-band magnitude versus peak \siii\ velocity exhibit red $B_{\rm max} - V_{\rm max}$ consistent with the models, further suggesting a common He-shell DDet origin.
In contrast, \uname\ with a relatively blue $B_{\rm max} - V_{\rm max}$ of $-$0.093 $\pm$ 0.013 is more consistent with the clustered CN events.

\subsection{Comparison to Nebular-Phase Properties}\label{sec:heddneb}

The nebular-phase flux ratio of the 7290~\AA\ emission feature to [\feiii] $\lambda$~4658~\AA, (``7290~\AA/[\feiii]"; Section~\ref{sec:nebspecev})
from $\sim$ 120--320 days since peak has also been suggested to distinguish He-shell DDet events from other normal \tase\ \citep{Polin2021apj}.
He-shell DDet models typically produce substantially more Ca than what is predicted in Chandrasekhar-mass explosions as a result of incomplete nuclear burning in the core of sub-Chandrasekhar-mass WDs, which may be observed as strong [\caii] emission near 7290\AA\ in the optically-thin nebular phase (Section~\ref{sec:nebex}).
In the bottom panel of Figure~\ref{fig:ddetpolin} \citep[adapted from][Figure 9]{Polin2021apj}, we compare \uname\ to the same population of \tase\ and set of He-shell DDet models as in the top panel with the color scale now representing their nebular-phase 7290\AA/[\feiii] ratios.
As noted by \citet{Polin2021apj}, SNe from the BL/91bg-like tail show stronger contributions from [\caii], leading to larger 7290\AA/[\feiii] ratios consistent with the He-shell DDet model predictions, while those from the CN/91T-like cluster have smaller ratios.
For \uname, its relatively small 7290\AA/[\feiii] ratio of 0.149 $\pm$ 0.007 once again identifies it with the SNe from the CN-subtype cluster, which are inconsistent with the He-shell DDet models.

\subsection{Search for Carbon}\label{sec:carbon}

Another key prediction of He-shell DDet models is efficient carbon burning, which leaves $\lesssim 10^{-5}\,$\msol\ of unburnt carbon in the SN ejecta \citep{Polin2019apj} in contrast to the substantial amount ($\sim 0.03\,$\msol) typically predicted by some other explosion models such as pure deflagration \citep{Nomoto1984apj} and pulsating delayed detonation \citep{Hoeflich1995apj}.
We search for the \cii~$\lambda$6580~\AA\ absorption feature near \siii~$\lambda$6355~\AA\ that has been used to examine the presence of carbon in \tas\ ejecta \citep{Parrent2011apj, Blondin2012aj, Maguire2014mnras}.
As detailed below, although \cii~$\lambda$6580~\AA\ is expected to be visible if the carbon mass fraction in the photosphere is greater than $\sim 0.005$ \citep{Heringer2019apj}, we detect no such feature in \uname\ throughout its evolution. This indicates that the carbon mass fraction is below this value in most layers of the ejecta of \uname, compatible with the He-shell DDet prediction. 
We note, however, that the absence of unburnt carbon is also possible for some non-DDet explosion models \citep[e.g., pulsating delayed detonations with very slow deflagration speeds;][]{Hoeflich1995apj}. 

\begin{figure}[t!]
\epsscale{\scl}
\plotone{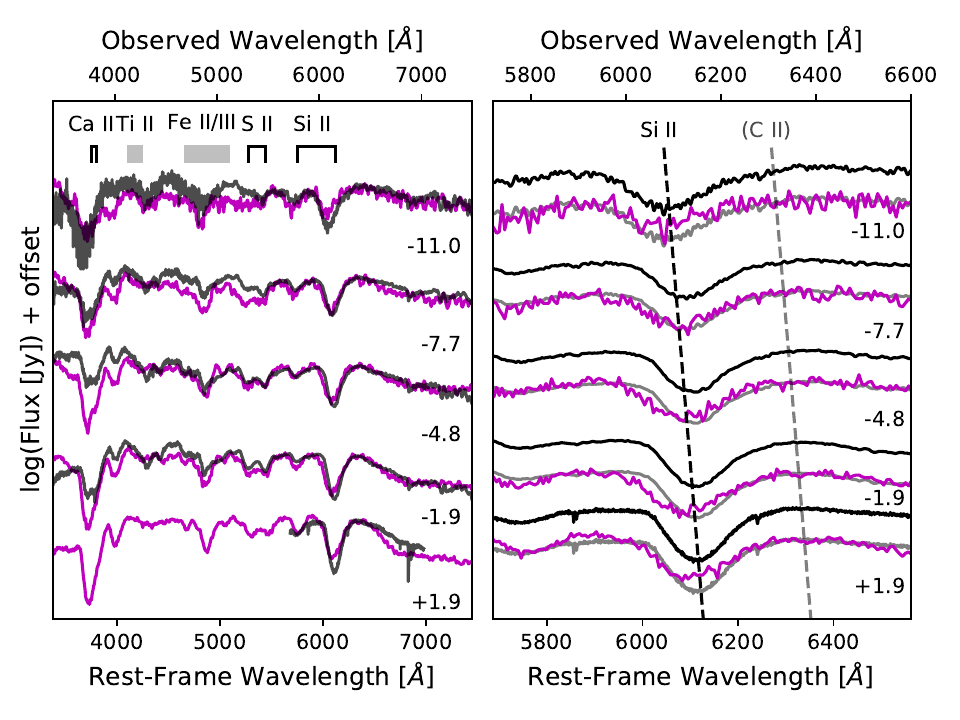}
\caption{(Left) The dereddened early spectra of \uname\ (black) are compared to the predictions of the best-fit He-shell DDet model (1.05~\msol\ WD + 0.010~\msol\ He-shell; magenta curves) in rest frame, as labeled in days since $B$-band maximum.
For clear comparison, the model spectra have been filtered by box-car convolution, resulting in effective spectral resolutions of $R$ = 300.
(Right) Same as the left panel, but showing unfiltered model spectra zoomed in on the vicinity of the observed \siii\ absorption feature.
The observed (black) spectra are translated downwards (grey) by subtracting a constant value for better comparison with the best-fit He-shell DDet model predictions (magenta). 
The approximate minima of the observed \siii\ features and the expected relative positions of the \cii\ absorption feature are shown with black and grey dashed lines, respectively. 
Note that \cii\ is not visible in any of the spectra.
\label{fig:ddetspec}}
\end{figure}

Figure~\ref{fig:ddetspec} compares the predicted spectral evolution (magenta curves) of the best-fit He-shell DDet model (1.05~\msol\ WD + 0.010~\msol\ He-shell) to the observed spectra of \uname\ until approximately $B$-band maximum.
We find an absence of the \cii~$\lambda$6580~\AA\ feature in all the spectra starting from as early as 4.4 days since first light ($-$11.0 days since $B$-band maximum), 
consistent with a lack of unburnt carbon throughout most layers of the ejecta.
However, the lack of earlier spectroscopic observations before 4.4 days probing 
carbon in the fast-expanding outer ejecta potentially allows 
a substantial amount of unburnt carbon to be
hidden in the outer $\sim$ 30\% of the ejected mass (Equation~\ref{eq:mdiff}).
Note also that \cii\ in \tase\ has been detected as early as $-$15 days
since $B$-band maximum \citep{Parrent2011apj}, earlier than our first spectroscopic observations.
In some cases, the \cii\ feature fades to become undetectable long before $B$-band maximum \citep{Brown2019apj}; however,
in NUV-blue events \citep[e.g., SN~2011fe;][]{Pereira2013aa}, the \cii~$\lambda$6580~\AA\ feature is almost always visible until roughly $B$-band maximum \citep[see][and references therein]{Milne2013apj}.
Since \uname\ is an extremely NUV-blue event (Section~\ref{sec:color}), the absence of \cii\ from $-$11 days since $B$-band maximum in the source appears to be an exceptional case.

\section{The Nature of SN~2018aoz and Implications for the Origins of Type Ia Supernovae}\label{sec:orig}

\subsection{Nature of the Companion Star}\label{sec:prog}

Our analyses of the early light curves (Section~\ref{sec:kasan}) and nebular-phase spectra (Section~\ref{sec:neb}) of \uname\ indicate that the binary companion of its progenitor is most likely to be a secondary WD.
First, our analysis of the early light curves disfavours binary companions larger than low-mass (few solar mass) main sequence stars based on the absence of their ejecta-companion interaction emission, leaving low-mass main sequence stars at large viewing angles ($\gtrsim$ 80\degr), naked He-stars, and WDs as the most likely possibilities for the companion.
Note that all three possibilities have been predicted to be involved in \tas\ explosions \citep{Maoz2014araa}.
Second, our modelling of the nebular-phase spectra of \uname\ further disfavours low-mass main sequence stars and naked He-stars as follows.
In the single-degenerate scenario, the SN ejecta is expected to strip/ablate $\sim$ 0.1--0.5~\msol\ of H-/He-rich material from the companion \citep{Dessart2020aa}, and most models predict that this leads to H emission in the nebular phase \citep{Mattila2005aap, Botyanszki2018apj, Dessart2020aa} while one model also predicts He emission \citep{Botyanszki2018apj}.
Our modelling of the nebular-phase spectra permits $\lesssim$ 10$^{-2}$~\msol\ of each element, disfavouring the single-degenerate scenario for \uname, consistent with what has been found in 94\% of \tase\ \citep{Tucker2020mnras}.
We note, however, that disagreements between current model predictions for the emission from early ejecta-companion interactions \citep{Kasen2010apj, Kutsuna2015pasj} and from stripped/ablated materials in the nebular phase \citep{Botyanszki2018apj, Dessart2020aa} precludes a definitive determination of the companion nature based on these analyses for now.

Although we cannot rule out a single-degenerate progenitor for \uname, we can almost completely rule out the case for a red giant companion, as predicted in the classical single-degenerate scenario.
Such a companion likely requires an extreme viewing angle in order to hide the ejecta-companion interaction in the early phase and emission from stripped/ablated H in the nebular phase, as supported by multiple models \citep{Kasen2010apj, Kutsuna2015pasj, Mattila2005aap, Botyanszki2018apj, Dessart2020aa}.
The He-star channel for \tase\ is also a less favourable progenitor for \uname\ due to the short ($\lesssim$ 0.2 Gyr) delay-time of the channel after star formation \citep{Wang2010aa}.
This delay-time is difficult to reconcile with the immediate environment of the SN due to the lack of recent star formation therein.
Metal abundance ratios $\gtrsim$ 40\arcsec\ from the center of \ngc reflect an old ($\sim$ 8--14 Gyr) stellar age in those regions \citep{Kim2012apj}, while recent star formation in the halo of \ngc where 
\uname\ was found is even less feasible.
The lack of recent star formation at the SN location is also supported by the lack of local dust extinction \citep{Sakurai2013eps}, as evidenced by the absence of \nai\ doublet lines at the host galaxy redshift (Paper I).

\subsection{Origin of the Infant-Phase Excess Emission}\label{sec:orgex}

We have shown that three mechanisms are capable of producing the observed infant-phase excess emission in \uname\ (Section~\ref{sec:early}): (1) radioactive heating by surface Fe-peak elements; (2) ejecta interaction with the binary companion; or (3) ejecta interaction with CSM.
Since the presence of surface Fe-peak elements is also required to explain the observed $B$-band suppression, it is likely that at least \emph{some} of the emission is due to radioactive heating by those same Fe-peak elements. 
However, the luminosity produced will depend sensitively on both the specific nucleosynthetic products in the outer ejecta (which will vary with explosion mechanism) as well as the degree of gamma-ray trapping. Indeed, as shown in Section~\ref{sec:heddearly}, it is possible for physical models to produce the level of $B$-band suppression needed to explain the NRB, while under-predicting the infant-phase luminosity. 
Thus, ejecta interactions with the binary companion and/or CSM may also contribute to---and potentially dominate---the luminosity at early times.

Within this context, we note that
both the ejecta-companion and ejecta-CSM interaction cases for the origin of the infant-phase excess emission in \uname\
are compatible with its favoured double-degenerate progenitor (Section~\ref{sec:prog}). 
In the case of ejecta-companion interaction, the observed infant-phase excess emission requires a small binary companion size, consistent with either a WD, He-star, or low-mass (few solar mass) main sequence star (Section~\ref{sec:kasfit}).

In the case of ejecta-CSM interaction, CSM with small mass and radius is required, consistent with what is expected from the companion accretion process (Section~\ref{sec:csm}).
The mass of CSM needed for the observed infant-phase excess emission ($\gtrsim$ 10$^{-3}$~\msol) and our strongest constraints on swept-up H from the nebular-phase spectra ($\lesssim$ 4 $\times$ 10$^{-4}$~\msol; Table~\ref{tab:nebflux}) further
requires the fractional mass of H in the total CSM mass to be $\lesssim$ 50\%.
These mass requirements are compatible with H-poor CSM originating from the accretion process of either a WD or He-star companion.

\subsection{Explosion Geometry and Progenitor Mass}\label{sec:asym}

As noted in Paper I and Section~\ref{sec:hedd}, the relatively short 15.3-day rise-time of \uname\ indicates that \uname\ was either (1) a spherically symmetric explosion with a low total ejecta mass of $\sim$ 0.8--1.0; or (2) an asymmetric explosion. 
The weak strength of [\caii] emission in the nebular phase of \uname\ disfavours the former (Section~\ref{sec:nebex}). 
However, in the remaining case of a high ejecta mass asymmetric explosion, the extreme blueshifts of the observed nebular-phase [\feii] and [\niii] features in \uname\ conflict with the observation of a short rise time (Section~\ref{sec:nebex}).

We examine three possible scenarios that can explain the weak [\caii] emission and blueshifts of [\feii] and [\niii] emission in the nebular-phase 7290~\AA\ feature of \uname\ simultaneously with its short rise time.
\begin{enumerate}
    \item \uname\ may originate from the explosion of a relatively high-mass ($\gtrsim$ 1.2~\msol) WD, where complete nuclear burning in the core results in weak nebular-phase [\caii] emission.
    The explosion can be moderately asymmetric, resulting in blueshifted [\feii] and [\niii] lines in the nebular phase from a viewing angle where the ejecta core is approaching the observer. 
    In this case, the fast-rising light curves of \uname\ can only be explained by the presence of a preceding dark phase \citep{Piro&Nakar2013apj}.
    However, this scenario is disfavoured for two reasons. First, we found no evidence for a long ($>$ 1 day) dark phase in \uname\ (Paper I) based on its observed ejecta velocity evolution \citep{Piro&Nakar2014apj}.
    Second, for an asymmetric Chandrasekhar-mass explosion, such a viewing angle would be less compatible with the presence of surface Fe-peak elements, which is primarily predicted in the direction where the ejecta core is receding from the observer \citep[e.g.,][]{Maeda2010apjCh}.
    
    \item Alternatively, the explosion of a high-mass WD can be compatible with the fast-rising light curve if the ejecta core is receding from the observer. In this case, blueshifted [\feii] and [\niii] lines in the nebular phase may require them to be optically thick, causing the shielding of the receding part of the ejecta by the approaching part.
    Note that Fe-peak elements have been suggested to act as an “Fe-curtain”, blocking the radiation from obstructed regions \citep{Leonard2007apj, Dessart2020aa}.

    \item \uname\ may originate from the asymmetric explosion of a lower-mass WD.
    Since nuclear burning is more complete in the densest regions of the ejecta, which is near the off-center point of carbon ignition, Ca production increases towards the low-density opposing direction \citep[e.g.,][]{Boos2021apj}. 
    For a highly asymmetric explosion, there could be limited overlap between the distributions of Ca and Fe-peak elements in the core, resulting in weak [\caii] emission in the nebular phase \citep{Polin2021apj, Hoeflich2021apj}.
    For a moderately asymmetric explosion where the core is approaching the observer, shielding of the most Ca-rich regions by parts of the intervening core may cause blueshifted [\feii] and [\niii] to dominate the 7290~\AA\ feature if their lines are optically thick.
\end{enumerate}

\subsection{Implications for the Asymmetric Chandrasekhar-Mass Explosion Mechanism}\label{sec:chexp}

One possible origin of surface Fe-peak elements associated with the observed $B$-band suppression in \uname\ is subsonic mixing in an asymmetric Chandrasekhar-mass explosion (Paper I), which has long been theorized to produce normal \tase\ and their observed properties.
In particular, the relationship between the observed light curves of \tase\ and \dm15\ \citep[i.e., Phillips relation][]{Phillips1999aj}, as well as the residual differences in their peak luminosities, peak colors, and nebular-phase [\feii] and [\niii] line shifts, have been found to be attributable to viewing angle effects and the details of the deflagration-to-detonation transition in the model \citep{Kasen2009nat, Maeda2010natur, Maeda2010apj, Maeda2011mnras}.
A Chandrasekhar-mass explosion is also the main scenario that is thought to produce \tase\ with weak [\caii] emission in the nebular phase \citep{Polin2021apj, Mazzali2015mnras} such as \uname, since complete nuclear burning in the core of high-mass WDs produces little Ca.
However, as explained in Sections~\ref{sec:nebex} and \ref{sec:asym}, both the short rise-time and presence of surface Fe-peak elements in \uname\ point to a viewing angle where the ejecta core is receding from the observer under the asymmetric Chandrasekhar-mass explosion scenario.
Reconciling the receding motion of the ejecta core with the observed blueshifts of nebular-phase [\feii] and [\niii] in \uname\ may thus require those lines to remain optically thick until $\sim$ 380 days post-peak.

Between low-mass (few solar mass) main sequence and WD companions for the progenitor system of \uname\ (Section~\ref{sec:prog}), the former is more compatible with the standard Chandrasekhar-mass explosion model, though it would require modifications to models that predict material will be stripped/ablated from the companion and visible at late times \citep{Dessart2020aa, Botyanszki2018apj}. 
In contrast, if the companion is a WD, then this scenario faces a number of constraints due to the accretion process between WDs often being dynamically unstable \citep{Shen2015apj} and tending to result in either a He-shell DDet or a violent merger.
The former (= He-shell DDet) leads to a different explosion mechanism as discussed below (Section~\ref{sec:heddexp}), while the latter (= violent merger) is disfavoured by the absence of unburnt O and He signatures in the nebular phase (Section~\ref{sec:neb}).
To avoid dynamically evolving towards He-shell DDet or violent merger, the case of a Chandrasekhar-mass explosion for \uname\ under double-degeneracy may require a relatively massive and rare primary WD that is
already near the Chandrasekhar mass ($\sim$ 1.4~\msol) at the start of accretion,
significantly larger than that of most WDs in the range of
0.5--0.8~\msol\ \citep{Tremblay2016mnras}.
Alternatively, violent merger is still possible if the nebular-phase O and He emission lines are hidden by metal line-blanket absorption.
In this case, the observed infant-phase excess emission would need to be from surface radioactive heating since a pre-merger explosion is required for both (1) ejecta-companion interactions to occur and (2) ejecta-CSM interaction properties to be compatible with the infant-phase observations (Section~\ref{sec:csm}).

\subsection{Implications for the He-shell Double-Detonation Explosion Mechanism}\label{sec:heddexp}

He-shell DDet is another explosion mechanism that naturally explains the presence of surface Fe-peak elements associated with the observed $B$-band suppression in \uname\ (Paper I).
1-D simulations of thin He-shell DDets (Section~\ref{sec:hedd}) are able to reproduce the rapid \bv\ color evolution of the NRB phase in \uname, as well as its overall spectroscopic features and light curves, with the 1.05~\msol\ WD + 0.010~\msol\ model providing the best-fit.
Thus, a He-shell DDet origin for \uname\ would confirm the predictions of recent theoretical models indicating that detonations of He-shells as thin as 0.01~\msol\ can successfully trigger \tase, including normal events.
In addition to the presence of surface Fe-peak elements, \uname\ also exhibits a number of other features that may be explained by a He-shell DDet origin.
This includes (1) the absence of C spectral features (Section~\ref{sec:carbon}),
(2) the short observed rise time, 
which can be explained by a sub-Chandrasekhar ejecta mass (Paper I), 
and (3) the small inferred companion size (Section~\ref{sec:prog}), 
as the typical progenitor channels for He-shell DDets involve accretion from a He-star, He-WD, or He/CO hybrid companion \citep{Shen2014apj, Guillochon2010apj}.

However, 
\uname\ fails several additional diagnostic criteria proposed by \citet{Polin2019apj,Polin2021apj} that are used to recognize this explosion mechanism.
First, the current suite of He-shell DDet models cannot entirely reproduce the infant-phase features of \uname\ (Section~\ref{sec:heddearly}).
The 1.05~\msol\ WD + 0.010~\msol\ He-shell model is the only He-shell DDet model that can match the early ($\lesssim$ 5 days since first light) \bv\ color evolution of \uname\ associated with surface Fe-peak elements; however, this model under-predicts the observed luminosity of the infant-phase excess emission and produces early ($\lesssim$ 4 days) \vi\ colors that are redder than observed.
Second, although \uname\ exhibits properties in common with both CN and BL 
subtypes of Type Ia SNe (Section~\ref{sec:class}), its blue near-peak color 
appears to be more compatible overall with the bulk of CN events
as opposed to the swath of BL (and also 91bg-like) events that 
show reddened near-peak colors 
predicted to be caused by ashes of 
the He-shell detonation (Section~\ref{sec:heddpeak}).
Third,
compared to the He-shell DDet models of \citet{Polin2021apj} and other Type Ia SNe, the nebular-phase 7290~\AA/[\feiii] flux ratio observed in \uname\ is much lower than what is predicted in the models as well as what is observed in the BL/91bg-like events suspected to be from He-shell DDet (Section~\ref{sec:heddneb}).

Thus, the observed properties of \uname\ appear less compatible with the model predictions overall and show a closer resemblance to SNe that are not suspected of being thin He-shell DDets than SNe that are.
This disfavours the He-shell DDet explosion mechanism for the origin of \uname, or at least requires modifications to the standard scenario of thin He-shell DDets as described by \citet{Polin2019apj,Polin2021apj}.
The first two discrepancies may be attributable to differences in the ashes of the initial He-shell detonation between the models and \uname, which would impact both the radioactive heating rate in the infant phase and spectroscopic features near maximum light \citep{Magee2021mnras}. 
Different nucleosynthetic yields may be possible if the evolutionary path leads to a He-shell with different initial conditions (e.g., composition, density) from those adopted by the models. 
The under-prediction of early luminosity by the best-fit He-shell DDet model may also indicate that an additional source of luminosity beyond radioactive heating is required at early times (e.g., shock interaction).

In contrast, production of nebular-phase [\caii] emission in sub-Chandrasekhar-mass explosions primarily depends on the total mass of the progenitor WD and the relative distribution of Ca and radioactive Fe-peak elements in the ejecta core.
Recent multi-dimensional He-shell DDet simulations have found that the explosion mechanism is inherently non-spherically symmetric \citep[e.g.,][]{Boos2021apj}, and its viewing angle effects have been suggested to explain the different ejecta velocities of \tase\ from the CN and BL subtypes as well as their differences in peak colors and nebular-phase [\feii] line shifts \citep{Boos2021apj, Li2019apj}.
The asymmetric explosion can also shift the distributions of Ca and Fe-peak elements.
As noted in Section~\ref{sec:asym}, nuclear burning in an asymmetric sub-Chandrasekhar-mass explosion is more complete near the ignition point of central carbon, which results in the distribution of Ca being offset towards the opposite side of the ejecta core.
Thus, a viewing angle where the Ca-rich region is shielded by the core may result in [\caii] being hidden if [\feii] and [\niii] lines remain optically thick in the nebular phase.
Weak [\caii] emission may also result from a sub-Chandrasekhar-mass explosion with a
higher total mass \citep[e.g., 1.26~\msol;][]{Mazzali2015mnras}.
However, as with the case of a Chandrasekhar-mass explosion (Section~\ref{sec:chexp}), reconciling the short rise time of \uname\ with the high ejecta mass in this case may still require optically thick [\feii] and [\niii] lines.
More detailed multi-dimensional modelling is necessary to ascertain if such effects can explain the observations of \uname.

\subsection{The \d6s Scenario}\label{sec:d6exp}

One specific He-shell DDet scenario that may yield initial conditions varying from the hydrostatic models of \citet{Polin2019apj,Polin2021apj} and also arises from the favoured double-degenerate progenitor of \uname\ is a ``dynamically-driven double-degenerate double-detonation'', or D$^{\wedge}$6 \citep{Shen2018apj}, scenario.
D$^{\wedge}$6 is a proposed origin for \tase\ wherein dynamic (unstable) accretion during the coalescence of a double-degenerate binary composed of two WDs leads to a \tas\ triggered by He-shell DDet.
While detailed models would be necessary to assess the overall consistency of \d6s\ with observations of \uname,
motivated by the possible requirement of additional emission sources beyond radioactive heating at early times (Section~\ref{sec:heddexp}), we show below that \d6s\ also naturally provides infant-phase emission at the level observed in \uname\ via ejecta interactions with CSM and/or the companion.

First, due to the dynamical nature of the accretion, a torus of CSM is expected to be present around the primary WD at the time of explosion \citep[e.g.,][]{Guillochon2010apj,Pakmor2013apj}. 
As detailed in Section~\ref{sec:csm}, the small mass and radius ($\lesssim$ 0.007~\msol\ and $\lesssim$ 10$^{10}$~cm) of CSM required to fit the observed infant-phase excess emission is compatible with CSM properties predicted in hydrodynamic simulations of this accretion process. 
Note that the fitted CSM properties were obtained by assuming an ejecta mass of 1.05~\msol, which is favoured by He-shell DDet models (Section~\ref{sec:heddsim}).

Second, models have shown that when all nuclear reactions are considered \citep{Shen&Moore2014apj} the He-shell DDet of the primary can occur during the early phases of dynamical accretion, before the companion WD has been fully disrupted \citep{Pakmor2013apj}. Thus, for the D$^{\wedge}$6 scenario, ejecta interaction with the companion should occur.
Dynamically unstable mass transfer between two WDs is expected for mass ratios $\gtrsim$ 0.2 \citep{Shen2018apj}, corresponding to companion masses of $\gtrsim$ 0.2 M$_\odot$ for a 1.05 M$_\odot$ primary.
Adopting the temperature of $\sim$ 3.0$\times$ 10$^4$~K for a tidally-heated He-WD in Roche overflow and the corresponding mass-radius relationship \citep{Panei2007mnras}, the expected separation for a 0.2 M$_\odot$ He-WD is $\sim$ 1.2 $\times$ 10$^{10}$~cm while larger companion masses lead to smaller separation distances.
These separations overlap with the lower end of the binary separation distances (6.8 $\times$ 10$^9$~cm; Section~\ref{sec:kasfit}) that can fit the observed infant-phase excess emission for an assumed ejecta mass of 1.05~\msol. 
In addition, rapid mass loss during dynamical accretion is expected to both widen the binary and inflate the donor WD \citep{Kremer2015apj}, indicating that both higher mass He-WDs and He/CO hybrids could also provide non-negligible contribution to the infant-phase excess emission of \uname\ in the D$^{\wedge}$6 scenario.

We note that if either ejecta-companion or ejecta-CSM interaction are the origin of the observed infant-phase excess emission under \d6s, two distinct physical processes would be required to
produce the infant-phase features of \uname: line-blanket absorption by surface Fe-peak elements produced in the He-shell DDet; and shock interaction from the ejecta colliding with either the companion or CSM.
While these processes are naturally predicted together in the D$^{\wedge}$6 scenario at the low luminosity level probed by the infant-phase observations of \uname, we emphasize that there are currently no theoretical models that consider the observational outcomes of both processes simultaneously.

\subsection{Implications for the Explosion Mechanisms of Normal Type Ia SNe}\label{sec:tase}

The exact explosion mechanism of \uname\ remains uncertain---as neither asymmetric Chandrasekhar mass explosion nor He-shell DDet models are currently capable of explaining all of the observations. 
However, whatever its nature, the explosion mechanism of \uname\ appears to produce a \tas\ with normal properties after the infant phase, indicating that it is a potentially prevalent explosion mechanism among \tase. 
As shown in Section~\ref{sec:class}, \uname\ is intermediate between
the CN and BL subtypes of normal \tase---corresponding to 38\% and 30\% of the entire \tas\ population \citep{Blondin2012aj}, respectively---and
shares spectroscopic similarities with both groups.
Thus, assuming that the reported infant-phase features first identified in \uname\ are found among spectroscopically similar SNe, 
then an explosion mechanism capable of producing normal \tase\ with surface Fe-peak elements may be responsible for up to 68\% of \tase\ from these two normal subtypes.

\section{Summary and Conclusion}\label{sec:conc}

The observations of \uname\ starting from the infant phase ($<$ 1 day since first light) and continuing to the late nebular phase ($\gtrsim$ 200 days since peak) have provided one of the most extensive sets of clues for understanding the origin and evolution of a \tas.
We summarize our main results and conclusions as follows.

\begin{itemize}

    \item The near-peak light curves and spectroscopic features of \uname\ 
    show that it is intermediate between the CN/NV and BL/HV subtypes of normal \tase, manifesting its nature as a normal event.
    The evolution of its \bv\ and \vi\ colors after the infant
    phase are also consistent with those of other normal \tase, 
    while the infant-phase color evolution
    is revealed for the first time, showing the rapid reddening of both colors over the first $\sim$ 0.5 days (or ``NRB'').
    \uname\ belongs to the NUV-blue group of normal \tase\ based on its UV-optical
    colors, with some of the bluest UV-optical colors reported in the group prior to $B$-band maximum. 
    No C spectral features are detected throughout the SN evolution beginning from the first spectrum $\sim$ 4.4 days since first light, which is exceptional among NUV-blue events while similar to typical BL events.
    
    \item The early $BVi$-band light curves of \uname\ during 0--7 days consist of 
    three components wherein two infant-phase features are embedded in an underlying
    power-law component that rises overall during the period.
    The two infant-phase features are
    (1) $B$-band plateau during $\sim$ 0--1 days (Paper I) and 
    (2) excess emission during 0.08--0.42 days, together resulting in the NRB color evolution.
    
    \item The $B$-band plateau feature has been attributed to $B$-band suppression by surface Fe-peak elements (Paper I), while we find that three mechanisms can contribute to the observed infant-phase excess emission: (1) radioactive heating by the surface Fe-peak elements; (2) ejecta shock interaction with the binary companion; and (3) ejecta shock interaction with CSM.
    
    \item Shock breakout is unlikely to be a significant contributor to the infant-phase excess emission.
    
    \item A small companion---such as a WD, He-star, or low-mass (few solar mass) 
    main sequence star---is required to attribute the infant-phase excess
    emission to ejecta-companion interaction,
    and the absence of H and He emission lines throughout 
    the nebular phase favours the WD companion.
    The presence of a red giant companion is particularly incompatible with the observed luminosity over the first few days, while the environment of the SN in the halo of the NGC~3923 elliptical galaxy argues against short delay-time companions, including He-stars as well as high-mass giants.
    
    \item Attributing the infant-phase excess emission to ejecta-CSM interaction requires a CSM distribution with small total mass ($\lesssim$ 0.007~\msol) and radius ($\lesssim$ 10$^{10}$~cm) at the time of explosion, 
    more consistent with what is expected during the binary accretion process 
    than after a violent merger.
    The presence of CSM from a violent merger is further disfavoured by the absence of He and O lines in the nebular phase.
    
    \item The weak strength of nebular-phase [\caii] emission and the observed blueshifts of [\feii] and [\niii] are not well explained by either explosions of high- or low-mass WDs. Both cases may require [\feii] and [\niii] lines to remain optically thick until $\sim$ 380 days since peak in addition to explosion asymmetry.
    
    \item Our 1-D thin He-shell DDet simulations are capable of explaining the observed \bv\ NRB color evolution associated with the $B$-band suppression by surface Fe-peak elements, overall evolution of optical luminosity and spectra, and absence of C spectral features in \uname. However, the model that best matches the observed \bv\ color evolution of the SN fails to reproduce its infant-phase excess emission and early \vi\ color. In addition, in a number of observed properties that have been suggested to identify the explosion mechanism, including $B_{\rm max} - V_{\rm max}$ and nebular-phase [\caii]/[\feiii] line ratio, \uname\ is more similar to the bulk of CN \tase, as opposed to the population of BL/91bg-like SNe that closely resemble the He-shell DDet models. Modifications to the standard thin He-shell DDet scenario (e.g., explosion asymmetry) may ameliorate some of these discrepancies.
    
    \item Both asymmetric Chandrasekhar-mass explosion and the \d6s\ scenario accommodate the presence of surface  Fe-peak elements and the observed infant-phase excess emission in \uname. However, neither model is currently capable of explaining all of the observations.
    
    \item The normal Type Ia nature of \uname\ and its spectroscopic similarity with a significant fraction of the \tas\ population indicates that \uname\ shares a common origin with at least some fraction of normal events, assuming that the reported infant-phase features first identified in \uname\ are found among spectroscopically similar SNe.
\end{itemize}

Our analyses highlight the importance of (1) deep, high-cadence survey observations that are capable of probing the low-luminosity signals of \tase\ in their earliest phases and (2) follow-up observations of light curves and spectra over the entire evolution of the SN until the nebular phase.
As the only \tas\ to date with sufficiently early (= $\sim$ 0--0.5 days) and deep (= $\sim -$10 to $-$12 absolute AB magnitudes) multi-band observations to detect the infant-phase $B$-band suppression and excess emission, \uname\ can provide an important point of reference for future efforts to model crucial physical processes in the infancy of Type Ia SN explosions.

\acknowledgments

This research has made use of the KMTNet system operated by the Korea Astronomy and Space Science Institute (KASI) and the data were obtained at three host sites of CTIO in Chile, SAAO in South Africa, and SSO in Australia.
This research is also based on observations obtained at the international Gemini-S Observatory, a program of NSF’s NOIRLab, which is managed by the Association of Universities for Research in Astronomy (AURA) under a cooperative agreement with the National Science Foundation on behalf of the Gemini Observatory partnership: the National Science Foundation (United States), National Research Council (Canada), Agencia Nacional de Investigaci\'{o}n y Desarrollo (Chile), Ministerio de Ciencia, Tecnolog\'{i}a e Innovaci\'{o}n (Argentina), Minist\'{e}rio da Ci\^{e}ncia, Tecnologia, Inova\c{c}\~{o}es e Comunica\c{c}\~{o}es (Brazil), and Korea Astronomy and Space Science Institute (Republic of Korea).
The Gemini-S observations were obtained under the K-GMT Science Program (PID: GS-2018A-Q-117 and GS-2018B-Q-121) of KASI and acquired through the Gemini Observatory Archive at NSF’s NOIRLab.
This paper includes data gathered with the 6.5 meter Magellan Telescopes located at Las Campanas Observatory, Chile.
This work makes use of observations from the Las Cumbres Observatory (LCO) global telescope network. The LCO team is supported by NSF grants AST-1911225 and AST-1911151, and NASA Swift grant 80NSSC19K1639.
The Swift observations were triggered through the Swift GI program 80NSSC19K0316. SOUSA is supported by NASA's Astrophysics Data Analysis Program through grant NNX13AF35G. 
Some of the data presented herein were obtained at the W. M. Keck Observatory, which is operated as a scientific partnership among the California Institute of Technology, the University of California and the National Aeronautics and Space Administration. The Observatory was made possible by the generous financial support of the W. M. Keck Foundation. The Computational HEP program in The Department of Energy's Science Office of High Energy Physics provided simulation resources through Grant \#KA2401022. This research used resources of the National Energy Research Scientific Computing Center, a U.S. Department of Energy Office of Science User Facility operated under Contract No. DE-AC02-05CH11231.
D.-S.M., M.R.D., and C.D.M. are supported by Discovery Grants from the Natural Sciences and Engineering Research Council of Canada.
D.-S.M. was supported in part by a Leading Edge Fund from the Canadian Foundation for Innovation (project No. 30951).
M.R.D. was supported in part by the Canada Research Chairs Program, the Canadian Institute for Advanced Research (CIFAR), and the Dunlap Institute at the University of Toronto.
D.J.S. acknowledges support by NSF grants AST-1821987, 1821967, 1908972 and from the Heising-Simons Foundation under grant \#2020-1864. 
S.G.-G. acknowledges support by FCT under Project CRISP PTDC/FIS-AST-31546 and Project UIDB/00099/2020.
S.C.K., Y.L., and H.S.P. acknowledge support by KASI under the R\&D program (Project No. 2022-1-868-04) supervised by the Ministry of Science and ICT.
H.S.P. was supported in part by the National Research Foundation of Korea (NRF) grant funded by the Korean government (MSIT, Ministry of Science and ICT; No. NRF-2019R1F1A1058228).
P.J.B. acknowledges support from the Swift GI program 80NSSC19K0316.
S.V., Y.D., and K.A.B. acknowledge support by NSF grants AST-1813176 and AST-2008108.
C.M. acknowledges support by NSF grant AST-1313484.
R.L.B. acknowledges support by NASA through Hubble Fellowship grant \#51386.01 awarded by the Space Telescope Science Institute, which is operated by the Association of  Universities for Research in Astronomy, Inc., for NASA, under contract NAS 5-26555.
A.G.-Y's research is supported by the EU via ERC grant No. 725161, the ISF GW excellence center, an IMOS space infrastructure grant and BSF/Transformative and GIF grants, as well as the André Deloro Institute for Advanced Research in Space and Optics, the Schwartz/Reisman Collaborative Science Program and the Norman E. Alexander Family M Foundation ULTRASAT Data Center Fund, Minerva and Yeda-Sela; A.G.-Y. is the incumbent of The Arlyn Imberman Professorial Chair.
L.G. acknowledges financial support from the Spanish Ministerio de Ciencia e Innovaci\'on (MCIN), the Agencia Estatal de Investigaci\'on (AEI) 10.13039/501100011033, and the European Social Fund (ESF) "Investing in your future" under the 2019 Ram\'on y Cajal program RYC2019-027683-I and the PID2020-115253GA-I00 HOSTFLOWS project, and from Centro Superior de Investigaciones Cient\'ificas (CSIC) under the PIE project 20215AT016.
G.P. acknowledges support by ANID -- Millennium Science Initiative -- ICN12\_009 and by FONDECYT Regular 1201793.
J.A. is supported by the Stavros Niarchos Foundation (SNF) and the Hellenic Foundation for Research and Innovation (H.F.R.I.) under the 2nd Call of ``Science and Society'' Action Always strive for excellence -- ``Theodoros Papazoglou’' (Project Number: 01431).

\vspace{5mm}

\software{SNooPy \citep{Burns2011aj}, Castro \citep{Almgren2010apj, Zingale2018jphc}, Sedona \citep{Kasen2006bapj}, SNAP (\url{https://github.com/niyuanqi/SNAP}), IRAF}

\appendix
\restartappendixnumbering

\section{Optical Color Evolution After The Infant Phase}\label{apx:color}

The color evolution of \uname\ after the infant-phase is characterized by phase transitions at $-$4.6, 10.4, and 26.0 days since $B$-band maximum, marked by the second, third, and fourth vertical dotted lines in Figure~\ref{fig:optcolor}, respectively. These epochs are roughly aligned with the primary peak, onset of secondary rise, and secondary peak of the $i$-band light curve. 
These mark four clear phases of \tas\ color evolution first described by \citet{Moon2021apj}.
\begin{enumerate}
    \item Between the first and second color transition epochs, during the $i$-band primary rise, the \bv\ color of \uname\ evolves blueward by 0.4 mag while the \vi\ color evolves redward by 0.4 mag before both colors become stalled prior to the second color transition epoch. The blueward \bv\ color evolution during this phase is consistent with that of the ``early-red'' group of \tase\ \citep{Stritzinger2018apj}, dominated by normal events. During this phase, the \bv\ color is thought to evolve blueward as a result of increased heating from \ni56\ within the ejecta as it is revealed by the SN expansion \citep{Hoeflich2017apj, Piro&Nakar2014apj}. 
    While SNe have rarely been observed with \vi\ color in such early epochs, the redward \vi\ color evolution of \uname\ during this phase appears similar to those of SN~2004D \citep{Patat1996mnras} and KSP-OT-201509b \citep{Moon2021apj}, two normal \tase\ from the over- and sub-luminous extremes, respectively, attributed to the evolution of spectral features in the $i$ band \citep{Moon2021apj}, e.g., \caii\ \citep[][see Figure 1 therein]{Parrent2012apj}.
    \item By the second color transition epoch, near the $i$-band primary peak, both \bv\ and \vi\ colors have reversed their direction of evolution, evolving redward and blueward, respectively. The redward \bv\ color evolution has been attributed to the development of Fe-peak absorption features in the $B$ band and the blueward \vi\ color evolution to the continued temperature increase as deeper deposits of \ni56\ within the ejecta continue to be exposed by the SN expansion \citep{Moon2021apj}.
    \item While the \bv\ color sustains redward evolution by 1.0 mag until the fourth color transition epoch, the \vi\ color, after evolving blueward by 1.0 mag, changes direction again at the third color transition epoch, coinciding with the onset of the secondary rise in $i$ band. The $i$-band secondary rise, which is due to the increased line opacity in the $B$ and $V$ bands from the recombination of \feiii\ \citep{Kasen2006apj}, causes the \vi\ color to evolve redward by 1.2 mag until the fourth color transition epoch.
    \item After the fourth color transition epoch, near the $i$-band secondary peak, both \bv\ and \vi\ colors evolve linearly bluewards at rates of 0.013 and 0.009 mag per day, respectively, as the SN enters the ``Lira law phase'' ($\gtrsim$ 30 days post-peak).
    During this phase, the intrinsic \bv\ colors of \tase\ evolve blueward linearly, following the Lira law \citep{Phillips1999aj}.
\end{enumerate}

The dotted line in Figure~\ref{fig:optcolor} (top panel) represents the Lira law evolution of
the \bv\ color of \tase\ from \citet{Burns2014apj}, which provides a Lira law slope of $-$0.011 mag per day for \sbv\ = 0.797.
The difference of $\sim$ 0.003 mag per day between the observed \bv\ slope of \uname\ in the Lira law phase and that of the Lira law is consistent with the observed range of scatter in the \bv\ slopes of \tase\ in the Lira law phase \citep[$\sim$ 0.004 mag per day;][]{Burns2014apj}.
Thus, we find that a simple vertical shift corresponding to a reddening of $E(B-V)$ = 0.09 mag, assuming $R_V=3.1$, is sufficient to match the Lira law to the observed color 
evolution of \uname.
This reddening is consistent with the Galactic extinction expected towards the direction of the source, supporting a negligible amount of extragalactic extinction for \uname\  
as expected from its location in the halo of NGC~3923.

\section{Light Curve Classification}\label{apx:class}

\begin{figure}[t!]
\epsscale{\scl}
\plotone{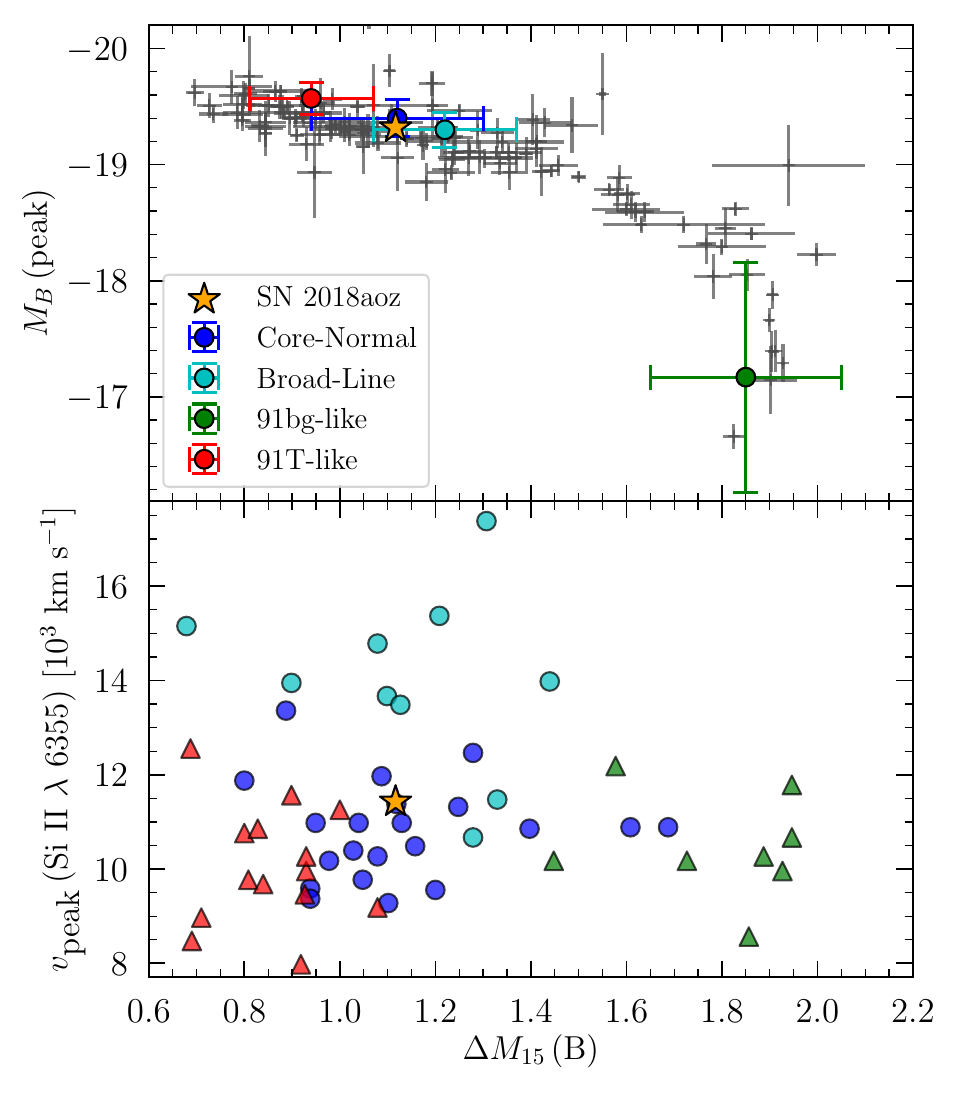}
\caption{(Top) Comparison of \dm15\ and \mb\ of \uname\ (orange star; Paper I) measured in rest frame with those of other \tase\ \citep[grey crosses,][]{Burns2018apj}. The colored circles represent the average values for the four main subtypes of \tase\ \citep{Parrent2014apss}: CN (blue), BL (cyan), 91bg-like (green), and 91T-like (red). (Bottom) Comparison of \dm15\ and peak \siii\ velocity of \uname\ (orange star) with those of other \tase\ \citep[colored circles;][]{Parrent2014apss}. The colors represent the same subtypes of \tase\ as in the top panel.
\label{fig:class-parrent}}
\end{figure}

\begin{figure*}[t!]
\plotone{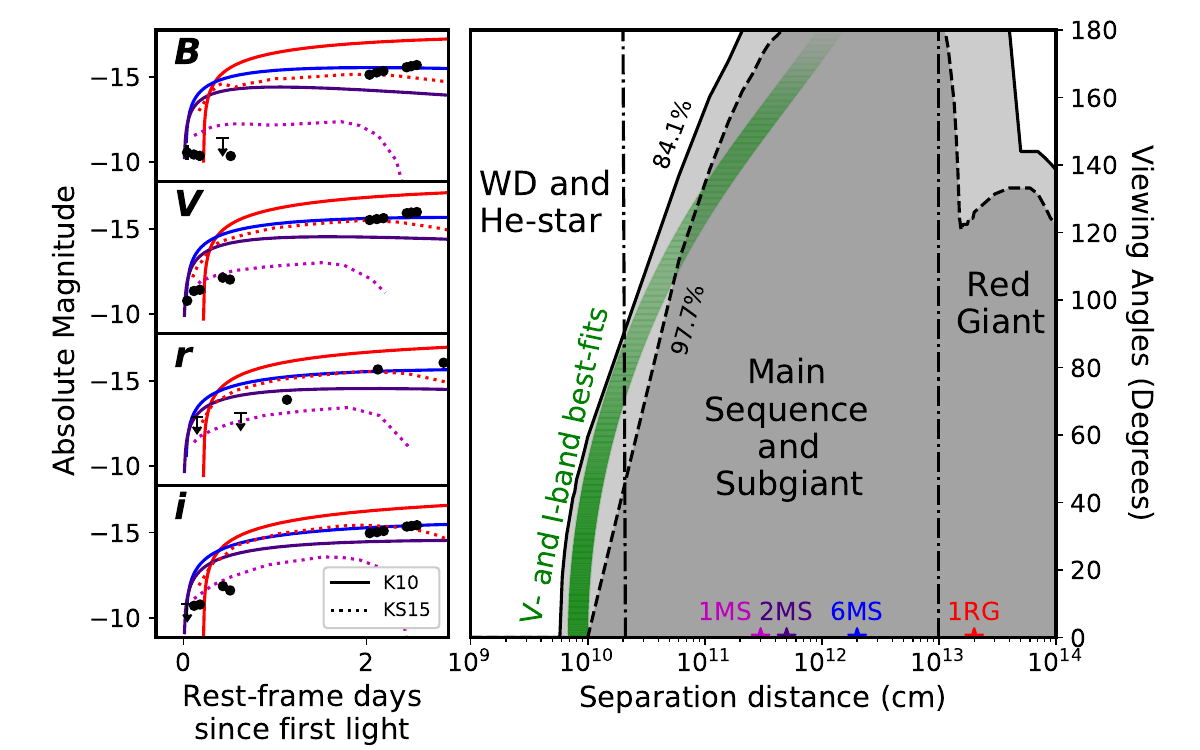}
\caption{(Left) Same as Figure~\ref{fig:kasmcVi}, but including the $B$-band light curve (top). (Right) Same as Figure~\ref{fig:kasmcVi}, but the solid and dashed lines representing the 84.1\% and 97.7\% confidence levels are calculated for the full $BVri$-band light curves, including that of the infant-phase $B$-band plateau which is affected by $B$-band suppression.
\label{fig:kasmcBVi}}
\end{figure*}

The classification of \uname\ as a normal \tas\ that is intermediate between the CN and BL subtypes (Section~\ref{sec:class}) is supported by its  near-peak light curves as follows.
Figure~\ref{fig:class-parrent} (top panel) compares the peak $B$-band absolute magnitude and \dm15\ light curve parameters of \uname\ to those of other \tase\ from the CN, BL, 91bg-like, and 91T-like subtypes.
The parameters of \uname\ are consistent with both CN and BL subtypes of the normal \tase, while they are inconsistent with the sub-luminous/fast-declining 91bg-like and the over-luminous/slow-declining 91T-like events.
This confirms that \uname\ is normal, belonging either CN or BL, though the two subtypes are not clearly distinguished based on their light curves properties alone.
The bottom panel compares the peak \siii\ velocity and \dm15\ of \uname\ with those of other \tase\ from the same four subtypes as in the top panel.
As seen in the panel, the CN and BL subtypes are more easily separated by these two parameters.
The \siii\ velocity and \dm15\ of \uname\ are each consistent with both CN and BL subtypes, while the SN itself is located within the CN-subtype cluster.
This confirms that \uname\ exhibits properties in common with both CN and BL subtypes, consistent with its intermediate classification, though the SN may be more similar to CN/NV than BL/HV events overall.

We note that for a normal \tas, the pre-peak light curves of \uname\ rise exceptionally fast.
Population studies of \tase\ find rise times of 18--25 days from first light to $B$-band maximum in normal events by fitting power-laws with indices of 2 to the early light curves \citep{Riess1999aj, Aldering2000aj, Conley2006aj}, while rise times of 15--22 days are found when the power-law index is freely fitted \citep{Miller2020apj}. 
The 15.32-day rise time of \uname, measured using a power-law fit with a free index \citep{Ni2022natas}, is consistent with what has been found in normal events, but near the low end of the distribution.

\section{Effect of Infant-Phase B-band Suppression on Companion Constraints}\label{apx:kasBsup}

In our analyses in Section~\ref{sec:kasan}, we excluded the $B$-band light curve of \uname\ during 0--1 days since first light from the comparisons between the observed
brightness and the model prediction from ejecta-companion interactions.
Figure~\ref{fig:kasmcBVi} is the same as Figure~\ref{fig:kasmcVi}, 
but obtained by comparing the entire $BVri$ early light curves of \uname\ during 0--3 days,
including those of the $B$-band plateau (top-left panel).
The inclusion of the $B$-band plateau results in much stronger
constraints on the companion by disallowing separation distances 
from $\sim 10^{11}$ to $\sim 4\times 10^{13}$~cm for almost all viewing angles (Figure~\ref{fig:kasmcBVi}, right panel).
This substantially increases the likelihood of small companions (i.e., WD or He-star) for \uname, 
with main sequence and subgiant companions of $>$ 1~\msol\ 
now essentially excluded and those with smaller masses (i.e., M or K dwarfs) only 
allowed for a more limited range of viewing angles $\sim$ 90--180\degr.
These clearly show that, compared to the $Vri$ bands, the early $B$-band light curve of \uname\ over-constrains its companion to favour much smaller ones as a result of being suppressed by surface Fe-peak elements (Paper I).
Our results suggests that caution needs to be exercised in general when future observations probing the infant-phase evolution of \tase\ are used to constrain physical parameters of
progenitors and explosion mechanisms, 
such as the companion size based on comparing the model brightness and light curves.

\end{document}